\documentclass[preprint]{aastex62}
\usepackage{natbib}
\usepackage[utf8]{inputenc}
\usepackage{inputenc}
\usepackage{epstopdf}
\usepackage[graphicx]{realboxes}
\usepackage{graphicx}
\usepackage[figuresright]{rotating}

\usepackage{graphicx}
\usepackage{amssymb}
\usepackage{multirow}
\received{January 4, 2018}
\revised{January 5, 2018}
\accepted{}
\submitjournal{ApJ}

\title{Confirmed width-$E_{iso}$ and width-$L_{iso}$ relations in GRB: comparison with the Amati and Yonetoku relations}
\shortauthors{Peng et al.}

\begin{document}
\nocite{*}
\begin{sloppypar}
\title{Confirmed width-$E_{iso}$ and width-$L_{iso}$ relations in GRB: comparison with the Amati and Yonetoku relations}
\correspondingauthor{Peng Z. Y}
\email{pengzhaoyang412@163.com}
\author{Peng Z. Y}
\affil{College of Physics and Electronics, Yunnan Normal University, Kunming 650500, China}
\author{Yin Y}
\affil{Department of Physics, Liupanshui Normal College, Liupanshui 553004, China}
\author{Li t}
\affil{Division of Assets, Yunnan Normal University, Kunming 650500, China}
\author{Wu H}
\affil{College of Physics and Electronics, Yunnan Normal University, Kunming 650500,China}
\author{Wang, D, Z}
\affiliation{College of Physics and Electronics, Yunnan Normal University, Kunming 650500, People's Republic of China}

\begin{abstract}
  The well-known Amati and Yonetoku relations in gamma-ray bursts (GRBs) show the strong correlations between the rest-frame $\nu f_{\nu}$ spectrum peak energy, $E_{p,i}$ and the isotropic energy, $E_{iso}$ as well as isotropic peak luminosity, $L_{iso}$. Recently, Peng et al. (2019) showed that the cosmological rest-frame spectral width are also correlated with $E_{iso}$ as well as $L_{iso}$. In this paper, we select a sample including 141 BEST time-integrated F spectra and 145 BEST peak flux P spectra observed by the Konus-Wind with known redshift to recheck the connection between the spectral width and $E_{iso}$ as well as $L_{iso}$. We define six types of absolute spectral widths as the differences between the upper ($E_{2}$) and lower energy bounds ($E_{1}$) of the full width at 50\%, 75\%, 85\%, 90\%, 95\%, 99\% maximum of the $EF_{E}$ versus E spectra. It is found that all of the rest-frame absolute spectral widths are strongly positive correlated with $E_{iso}$ as well as $L_{iso}$ for the long burst for both the F and P spectra. All of the short bursts are the outliers for width-$E_{iso}$ relation and most of the short bursts are consistent with the long bursts for the width-$L_{iso}$ relation for both F and P spectra. Moreover, all of the location energy, $E_{2}$ and $E_{1}$, corresponding to various spectral widths are also positive correlated with $E_{iso}$ as well as $L_{iso}$. We compare all of the relations with the Amati and Yonetoku relations and find the width-$E_{iso}$ and width-$L_{iso}$ relations when the widths are at about 90\% maximum of the $EF_{E}$ spectra almost overlap with Amati relation and Yonetoku relation, respectively. The correlations of $E_{2}-E_{iso}$, $E_{1}-E_{iso}$ and $E_{2}-L_{iso}$, $E_{1}-L_{iso}$ when the location energies are at 99\% maximum of the $EF_{E}$ spectra are very close to the Amati and Yonetoku relations, respectively. Therefore, we confirm the existence of tight width-$E_{iso}$ and width-$L_{iso}$ relations for long bursts. We further show that the spectral shape is indeed related to $E_{iso}$ and $L_{iso}$. The Amati and Yonetoku relations are not necessarily the best relationships to relate the energy to the $E_{iso}$ and $L_{iso}$. They may be the special cases of the width-$E_{iso}$ and width-$L_{iso}$ relations or the  energy-$E_{iso}$ and energy-$L_{iso}$ relations.
\end{abstract}

\section{Introduction}
Thanks to the extreme brightness Gamma-ray bursts (GRBs) are thought as the most energetic explosions in the universe. The emission mechanisms of the GRB prompt emission are still unresolved. One of the methods is to analyze the GRB prompt emission, especially for the spectroscopy, which can provide us valuable clues to the underlying processes giving rise to the phenomenon. Until now, no physical model has been demonstrated to systematically fit the observed spectra though some attempts to fit physical models to the observational data have been made (e.g. Titarchuk et al., 2012; Oganesyan et al., 2019; Burgess et al., 2020; Ahlgren et al., 2019; Ahlgren et al., 2019). Instead, a single simple empirical Band model (Band et al. 1993) or the Cutoff power law (CPL) model can well represent these GRB spectra due to the overall spectral shape is similar. Through performing the time-revolved spectral analysis it was found that the prompt emission undergoes significant spectral evolution (e. g. Preece et al. 2000; Kaneko et al. 2006, Peng et al. 2009).

Most of researchers still used empirical Band or CPL model to describe the time-integrated or time-resolved spectrum. They also found some interesting correlations (spectrum-energy correlations) between the model parameters and the energy-related physical parameters of the overall GRB. For example, the well-known Amati relation shows the intrinsic peak energy $E_{p,i}$ in the $\nu f_{\nu}$ spectra correlates with the isotropic energy ($E_{iso}$) (Amati et al. 2002). Moreover, the $E_{p,i}$ is also correlated with isotropic peak luminosity ($L_{iso}$) (Yonetoku relation; Yonetoku et al. 2004), as well as the collimation-corrected energy ($E_{\gamma}$) (Ghirlanda relation, Ghirlanda et al. 2004).

Traditionally, GRBs can be roughly divided into short and long-duration classes based on $T_{90}$ where all the bursts are likely to be separated at about 2 s (Kouveliotou et al. 1993). In fact, $T_{90}$ depends on an energy range used for the calculation and it is highly affected by the number of selection effects (e.g. Minaev \& Pozanenko, 2020). Therefore, some authors  derived burst types using I (merger-origin), II (collapsar-origin), I/II (the type is uncertain) (Horv\'{a}th et al. 2010). A method similar to Horv\'{a}th et al. (2010) are put forward by Svinkin et al., (2016).  They divide the burst types as I $>$ 0.9-Type I, 0.1  $<$ I $<$ 0.9-Type I/II, I $<$ 0.1- Type II based on the short/hard burst indicator function I($T_{50}$, $HR_{32}$) (see Equation (5) in Horv\'{a}th et al. 2010). For different burst classes the spectrum-energy correlation properties may be different. It is shown that short burst classes does not follow the Amati relation (e.g. Qin et al. 2013). Until now, the nature of the Amati correlation is still unclear. Some authors think it may be connected with selection effects  (e.g. Band \& Preece 2005). More authors claimed that the ${E}_{p,i}-{E}_{{iso}}$ and/or ${E}_{p,i}-{E}_{\gamma }$ correlations were not significantly affected by the selection effects.

Recently, Axelsson \& Borgonovo (2015) used a relative spectral widths (the ratio of two energies) of the observed $\nu F_{\nu}$ spectrum to compare with those of known emission mechanisms. Peng et al. (2019, hereafter Paper I) further studied the properties of spectral width fitted by the BEST model with Fermi/GBM data. They have found there are correlated relationships between the relative spectral width and the isotropic energy ($E_{iso}$) as well as the isotropic peak luminosity ($L_{iso}$) with the Fermi GBM data. Moreover, the short bursts extend the correlations to the long ones, which seems to show the long and short burst share same physics origins.

However, the correlation between the relative spectral width and $E_{iso}$ as well $L_{iso}$ was not very significant for time-integrated spectra and it was not significant for the peak flux spectra. When they used the intrinsic absolute spectral width defined by the difference of two energies the correlations of width-energy and width-luminosity improved noticeably. Moreover, the two correlated relationships were also established for the peak energy spectra. Both Amati and Yonetoku relations are based on the peak energy of GRB spectra and the peak energy and the spectral widths are related to the spectral shape of GRB. Therefore, they further deduced that $E_{iso}$ and $L_{iso}$ might be connected with GRB spectral shape combined with the Amati and Yonetoku relation. They also suspected the other shape parameters $E_{2}$, $E_{1}$ (where $E_{2}$ and $E_{1}$ are the upper, lower energy bounds of the full width at half maximum of the $EF_{E}$ versus E spectra) also correlate with luminosity as well as energy. Which parameters is the relative better indicator of energy and luminosity? We shall investigate these issues in detail in our work. Besides these, there are some other issues are not resolved. Firstly, if the short burst is outliers of the width-energy relation or width-luminosity. Secondly, whether there are the connections between width-energy relation, width-luminosity and the Amati and Yonetoku relations. This issues motivate our investigations below. In section 2, we present a sample description and data analysis. The results of the analysis are given in section 3. Discussion and conclusions are presented in the last section.

\section{Sample selection and data analysis}
In this paper, we would like to investigate the correlation properties between the intrinsic absolute spectral width and the isotropic-equivalent radiated energy $E_{iso}$ as well as the isotropic-equivalent peak luminosity $L_{iso}$. Therefore, we need spectral data which can provide well-represented the observed spectral model data. Tsvetkova et al. (2017) just provide a such large spectral sample what we need. They have presented the results of a systematic study of GRBs with reliable redshift estimates detected in the triggered mode of the Konus-Wind (KW) experiment during the period from 1997 February to 2016 June. The sample consists of 150 GRBs (including 12 short/hard bursts) and represents the largest set of cosmological GRBs studied to date over a broad energy band (13 keV- 10 MeV). The wide energy range facilitate deriving the complete spectral shape directly from the KW data and the GRB  energetics can also be estimated using fewer extrapolations coupled with reliable redshift estimate, which is very important to obtain a complete spectral width for our work. These can provide an excellent testing ground for widely discussed correlations between rest-frame spectral hardness and  energetics (Tsvetkova et al. 2017). The KW catalogue provides two types of spectra, the time-integrated spectral fits (F spectra) and spectral fits at the brightest time bin (P spectra). They employed two different spectral models, Band GRB function (BAND) and the exponential cutoff power law (COMP) to fit the data, respectively. They also compare the two model and give the BEST models for these bursts.

In order to investigate in detail the connections between width-energy and width-luminosity relations and compare the Amati and Yonetoku relations we want to define six types of absolute spectral widths, $W_{ab,50}$, $W_{ab,75}$, $W_{ab,85}$, $W_{ab,90}$, $W_{ab,95}$, and $W_{ab,99}$ based on eq. 1.
\begin{equation}
    W_{ab} = log (E_{2}-E_{1})
	\label{}
\end{equation}
where $E_{1}$ and $E_{2}$ are the lower and upper energy bounds of the full width of 50\%, 75\%, 85\%, 90\%, 95\%, and 99\% maximum of the $EF_{E}$ versus E spectra, respectively.

Similar to Axelsson \& Borgonovo (2015) we also minimize selection effects due to parameter limit by requiring $\alpha$ $> -1.90$ and $\beta <-2.10$ to obtain better observed spectral width. Finally there are 141 F burst spectral widths (including 12 short bursts) and 145 P spectral widths (including 12 short bursts) in our sample under these restrictions. All of the spectra are fitted by the BEST models with curvature shapes (Band function and Compton). The number of the two models for the subset is listed in Table 1. The uncertainty for each burst spectral width is estimated by using Monte Carlo methods.

\begin{deluxetable}{lcccccccc}
\tablecaption{A list of sample size of the two models.\label{}}
\tablehead{
\colhead{} &\colhead{} & \colhead{F spectra} & \colhead{} & \colhead{} &\colhead{P spectra} & \colhead{} & \colhead{} & \colhead{} \\
\colhead{} &\colhead{long GRBs} & \colhead{short GRBs}& \colhead{entire bursts} & \colhead{long GRBs}& \colhead{short GRBs}& \colhead{entire bursts}
}
\startdata
 BAND  & 48          & 0         & 48         & 44      & 0            &44      \\
 COMP  & 81          & 12        & 93         & 89      & 12           &64      \\
 all   & 129         & 12        & 141        & 133     & 12           &145     \\
\enddata
\end{deluxetable}

\section{analysis results}

\subsection{The distribution of the burst absolute spectral width}
In this section, we first check the distributions of the different absolute spectral widths defined in previous section. Then we compare the distributions between the F spectra and the P spectra as well as the two type of bursts, long and short bursts. The various $W_{ab}$ distributions for the long and short bursts are demonstrated in Figure 1 and the characteristics are summarized in Table 2. We only analyze in detail the popular half width $W_{ab,50}$ distributions below.

We find from Figure 1 and Table 2 for the long burst of two types of spectra that (1) the widths range from 2.16$\pm$0.03 to 4.00$\pm$0.17 for the F spectra and from 2.16$\pm$0.03 to 4.00$\pm$0.21 for the P spectra; (2) for both the P spectra and the F spectra, the distribution peak at $<$ 3,  there is a very small fraction of bursts extending towards larger widths; and (3) the corresponding median values of $W_{ab,50}$ are 2.90$\pm$0.11 and 2.93$\pm$0.19 for the P spectra and the F spectra, respectively.

While for the short burst of two types of spectra (1) the widths range from 2.88$\pm$0.02 to 3.86$\pm$0.05 for the F spectra and from 2.88$\pm$0.02 to 3.85$\pm$0.02 for the P spectra; (2) the distribution peaks $>$ 3 for both spectra; (3) the corresponding median values of W are 3.16$\pm$0.05 and 3.22$\pm$0.08 for the F spectra and the P spectra, respectively. It is found that the widths of the F spectra are close to those of the P spectra for both of the long bursts and the short bursts.

When comparing the long and short bursts for two types of spectra from Figure 1 and Table 2 it is found that the peak value and the median of the short burst widths are evidently greater than those of the long burst for both spectra. This absolute width distributions are contrary to those of the relative spectral width of Paper I. Whether or not only the median value is different for the two types of bursts is the question.  We do a K-S test to check this and find that the significance probabilities and D values are 5.39$\times$10$^{-3}$, 0.50 and 8.55$\times$10$^{-3}$, 0.48 for the F and the P spectra, respectively. Therefore there are significant differences between the two types of GRBs for both the F spectra and the P spectra. However, there is substantial overlap between the long burst and short burst (see, Figure 1 and Table 2). Therefore the distributions of the two types of bursts are perfectly compatible when taking into account the variances of the distributions.

We also compare the width distributions between the F spectra and the P spectra. For the short burst, the probability (0.99) and D value (0.17) reveal that the distribution of the short burst between the F and the P spectra is the same. The K-S test gives a significance probability 0.58 and D value 0.10, which also shows that the distribution of the long burst between the F spectra and the P spectra is also the same.

The distributions of $W_{ab,75}$, $W_{ab,85}$, $W_{ab,90}$, $W_{ab,95}$, $W_{ab,99}$ are also shown in Table 2 and Figures 2 and 6. Naturally the median and peak width decrease. The K-S test also show that there are significant differences between the two types of GRBs for both the F spectra and the P spectra and the significance increases as the width decreases. However, the distributions between the F spectra and the P spectra are the same for both long and short bursts.

\begin{figure}
  \includegraphics[width=0.24\textwidth]{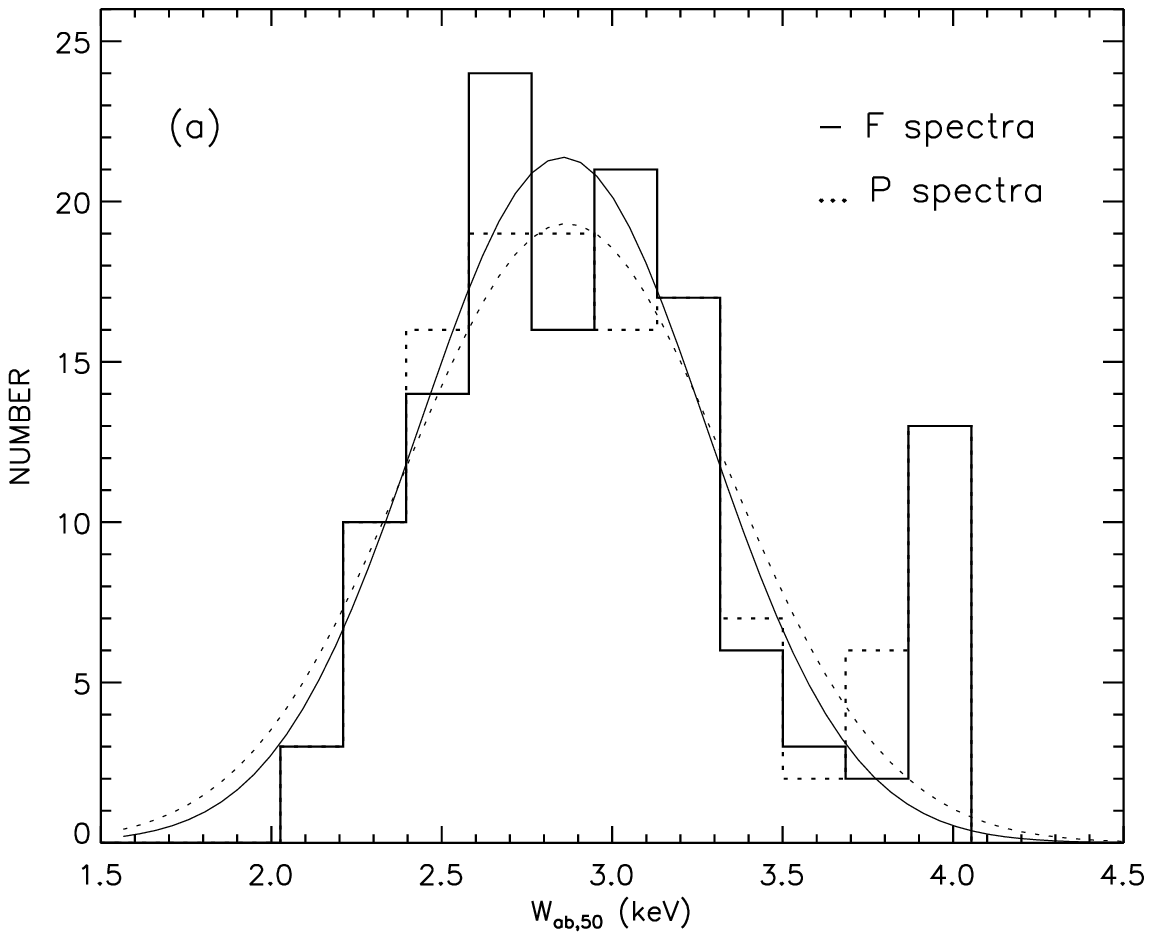}
  \includegraphics[width=0.24\textwidth]{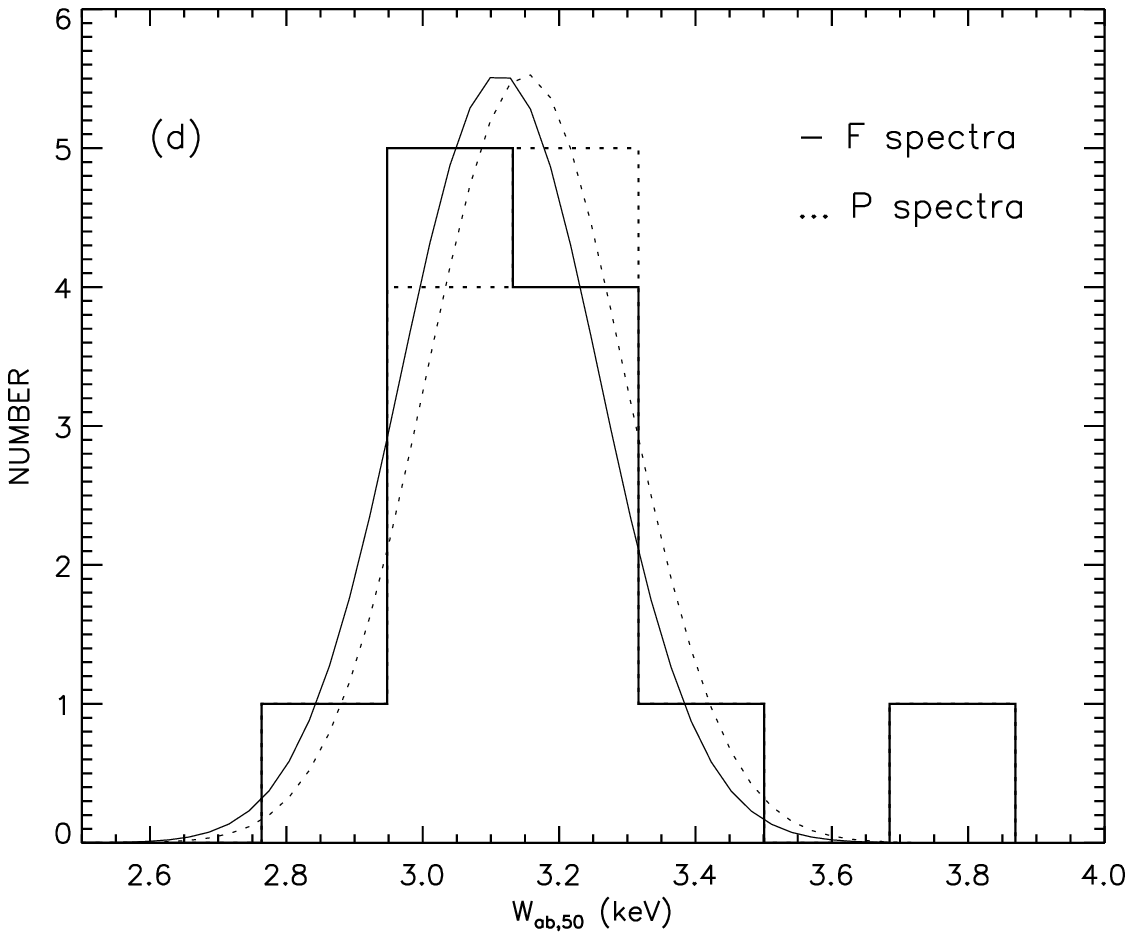}
  \includegraphics[width=0.24\textwidth]{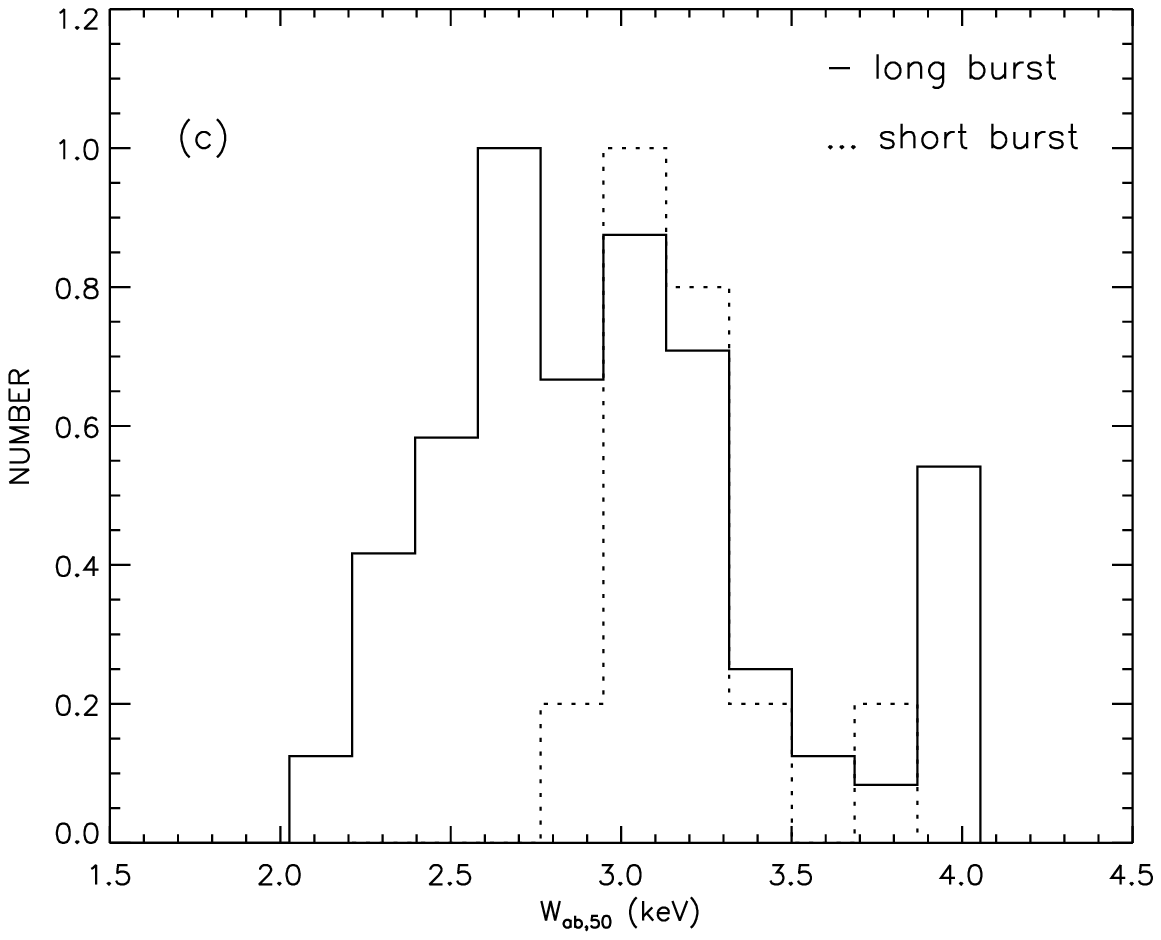}
  \includegraphics[width=0.24\textwidth]{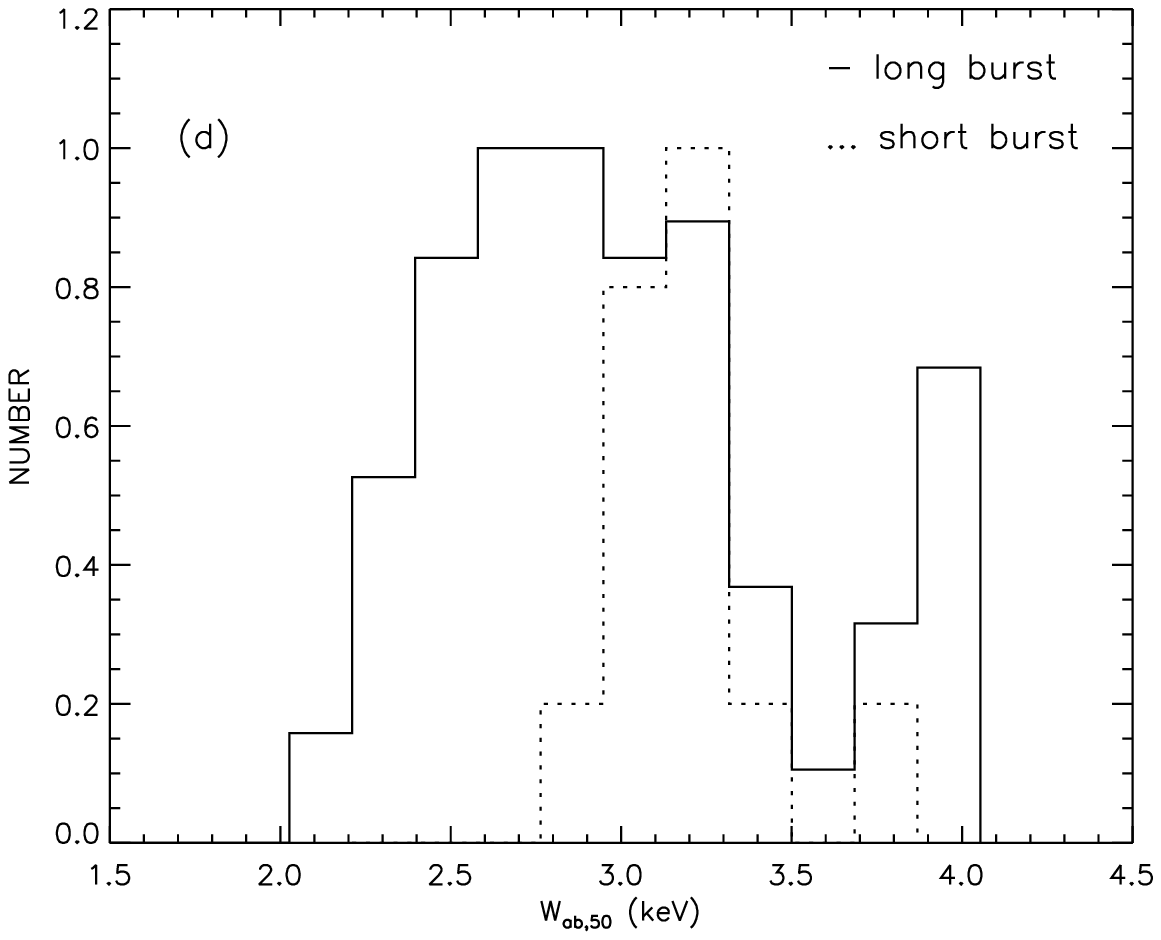}
  \caption{Distributions of the spectral width $W_{ab,50}$ for the long (a) and the short burst set (b) and the comparison of the spectral width distribution between the long and short burst for the P spectra (c) and the F spectra (d).\label{fig:f1}}
\end{figure}

\begin{figure}
  \includegraphics[width=0.24\textwidth]{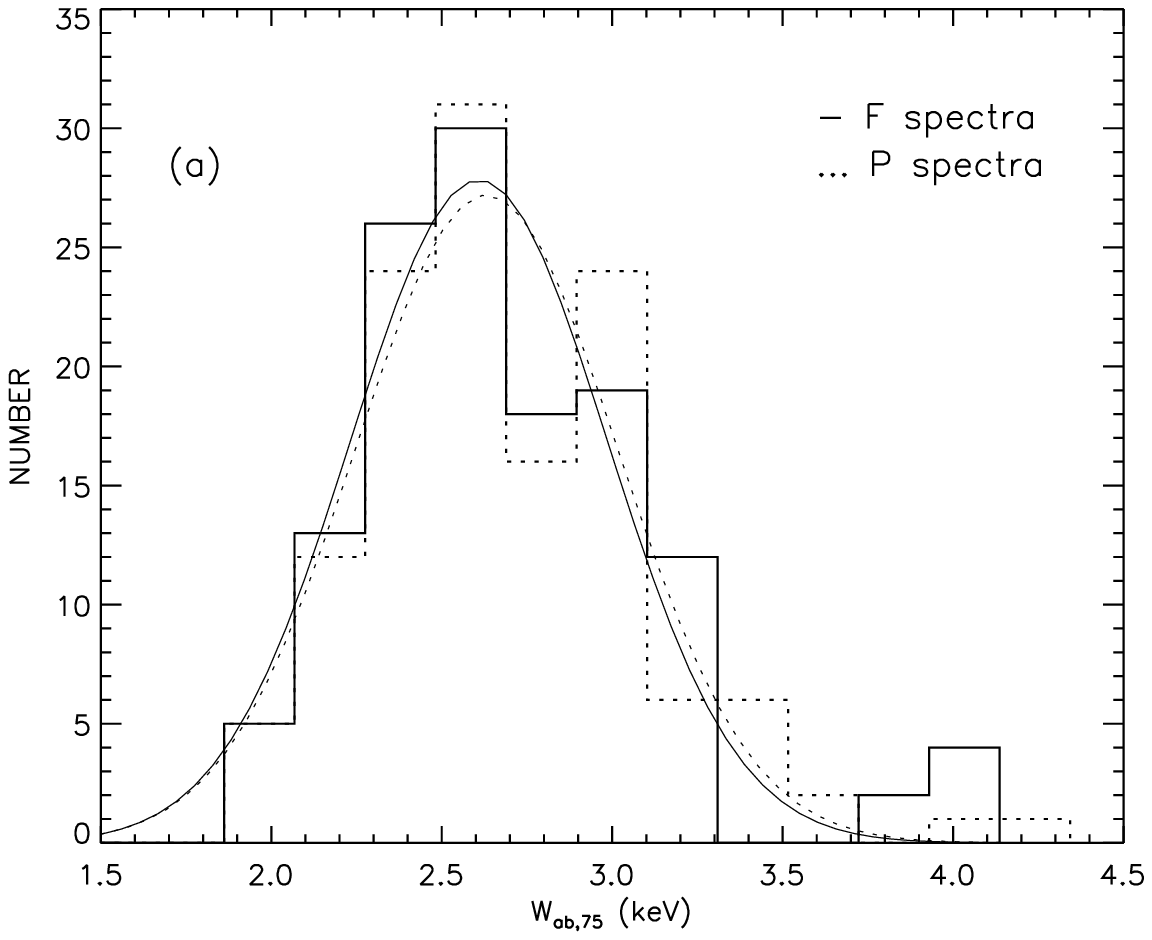}
  \includegraphics[width=0.24\textwidth]{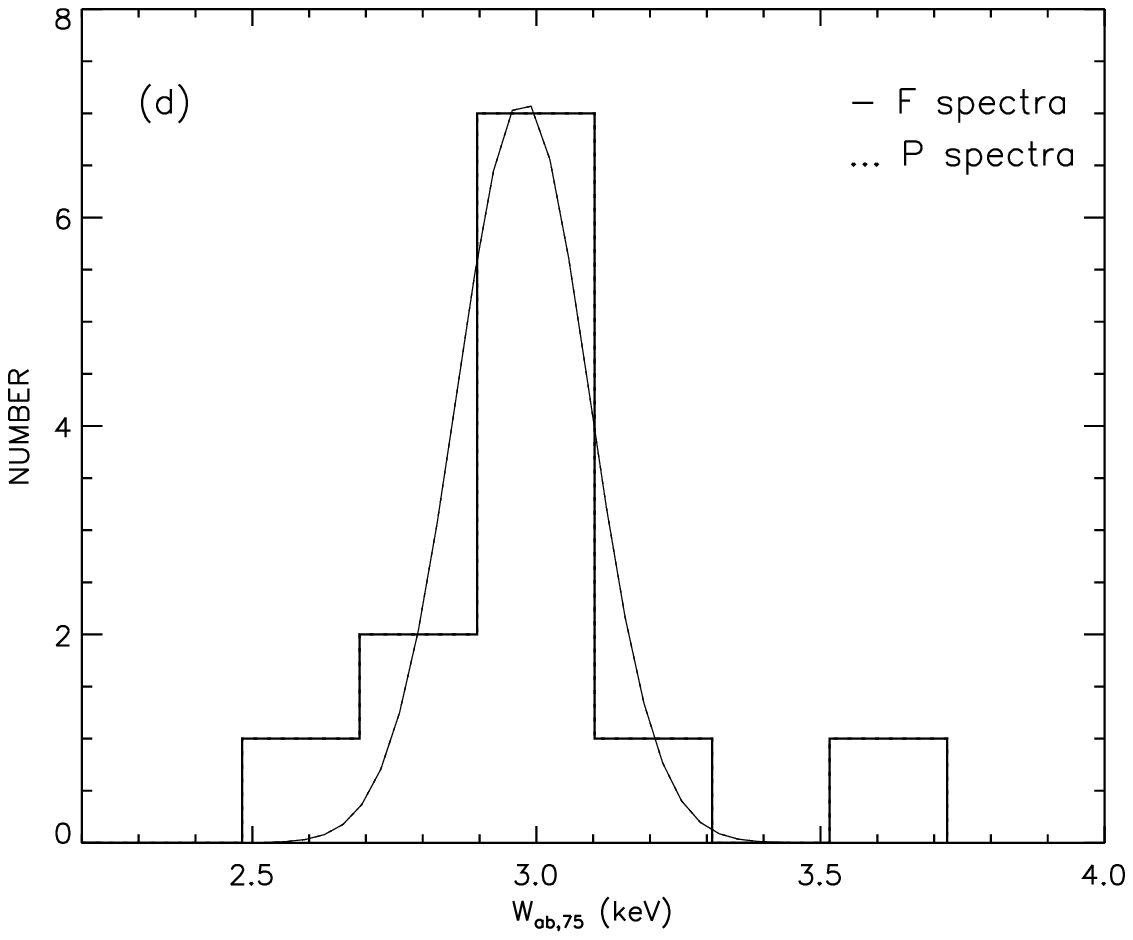}
  \includegraphics[width=0.24\textwidth]{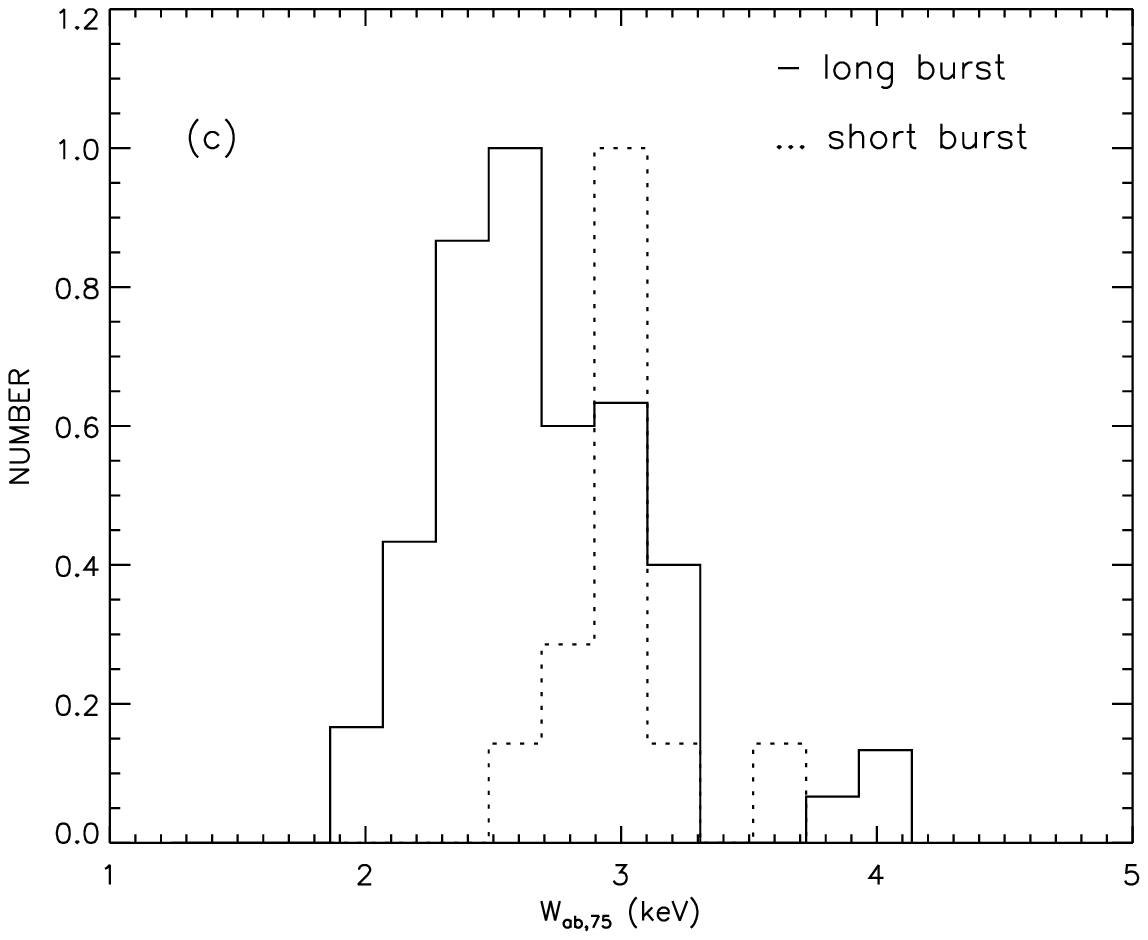}
  \includegraphics[width=0.24\textwidth]{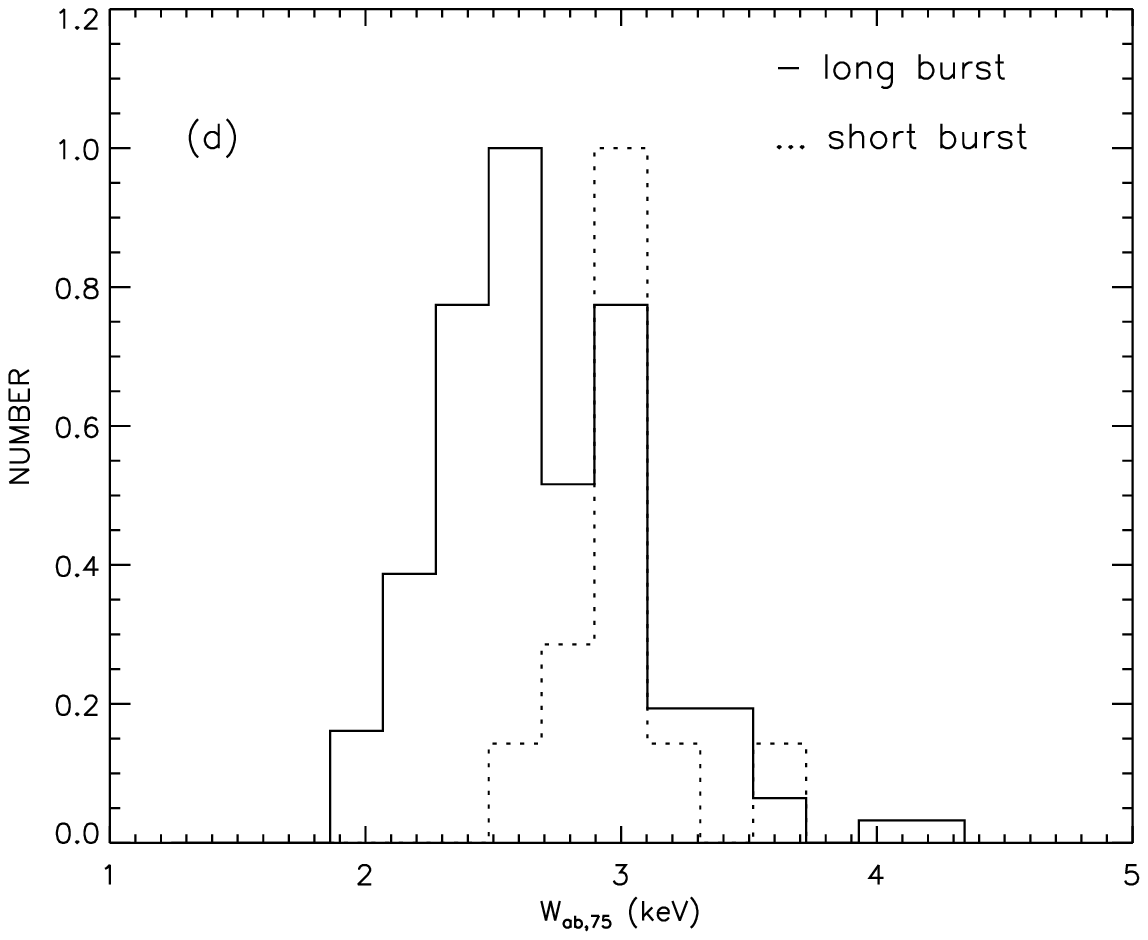}
  \caption{Distributions of the spectral width $W_{ab,75}$ for the long (a) and the short burst set (b) and the comparison of the spectral width distribution between the long and short burst for the P spectra (c) and the F spectra (d).\label{fig:f1}}
\end{figure}

\begin{figure}
  \includegraphics[width=0.24\textwidth]{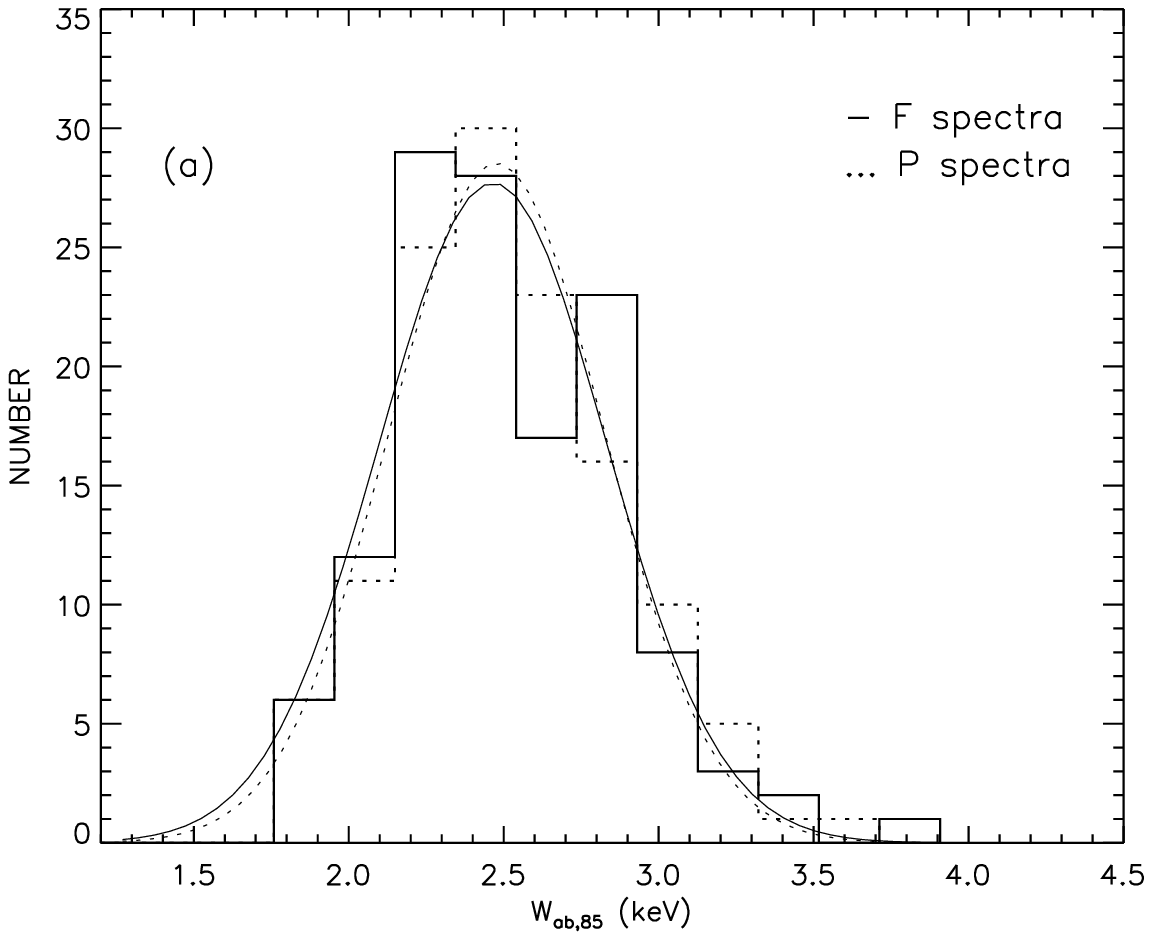}
  \includegraphics[width=0.24\textwidth]{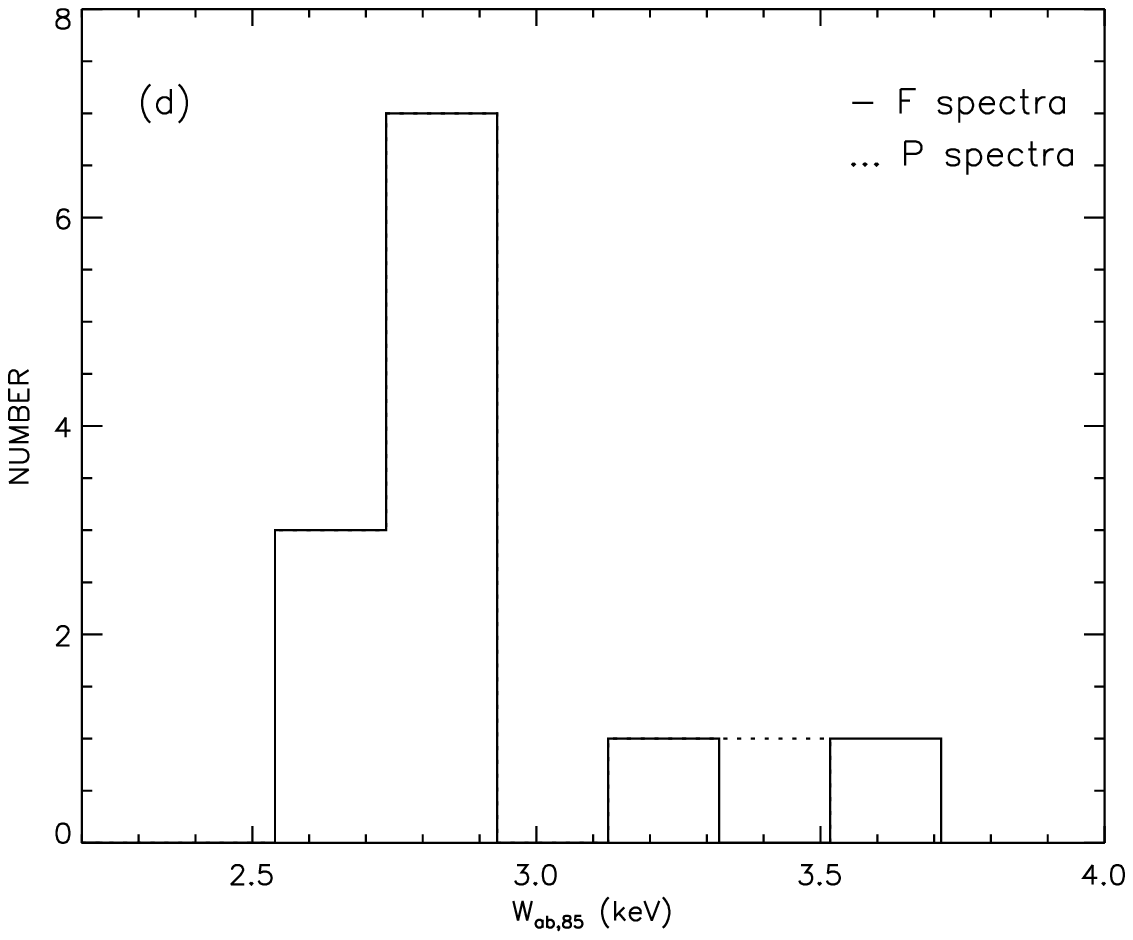}
  \includegraphics[width=0.24\textwidth]{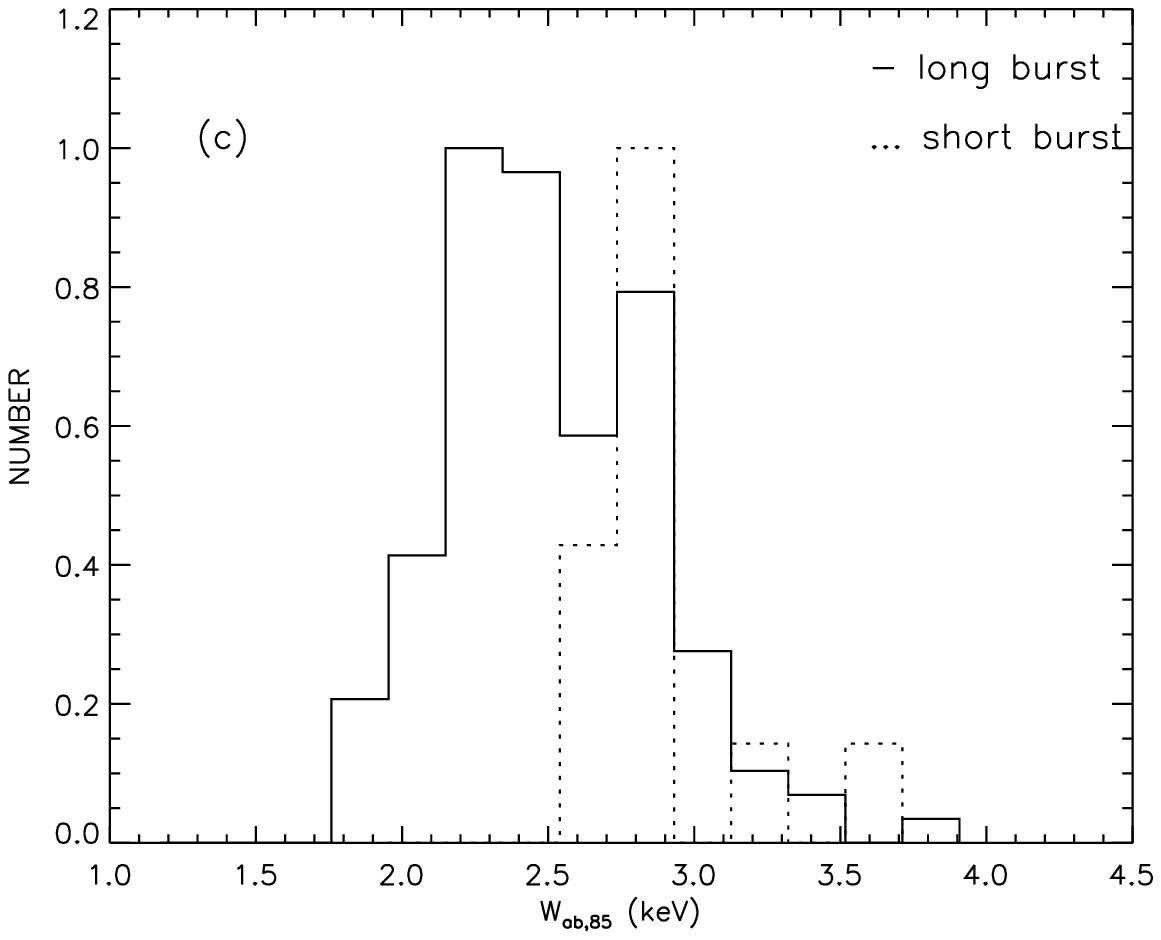}
  \includegraphics[width=0.24\textwidth]{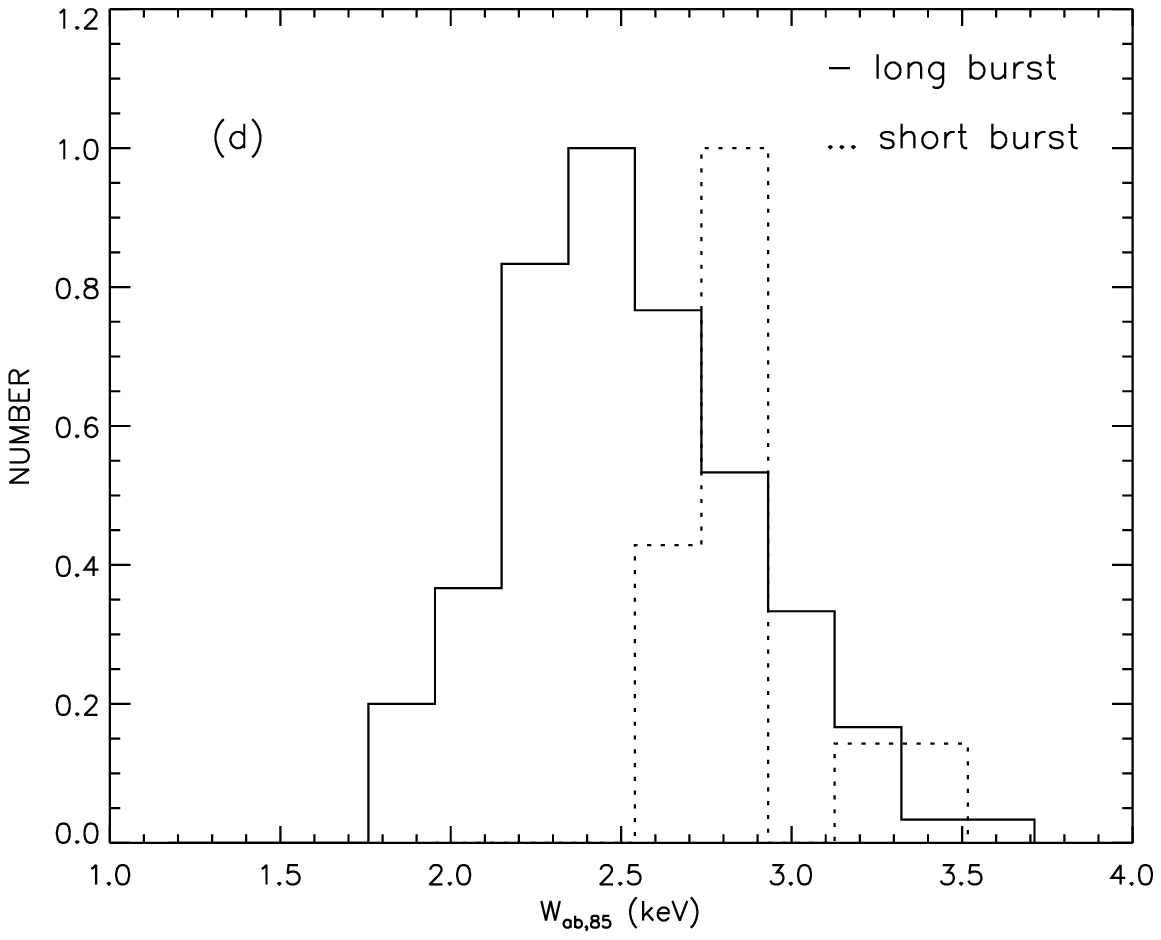}
  \caption{Distributions of the spectral width $W_{ab,85}$ for the long (a) and the short burst set (b) and the comparison of the spectral width distribution between the long and short burst for the P spectra (c) and the F spectra (d).\label{fig:f1}}
\end{figure}

\begin{figure}
  \includegraphics[width=0.24\textwidth]{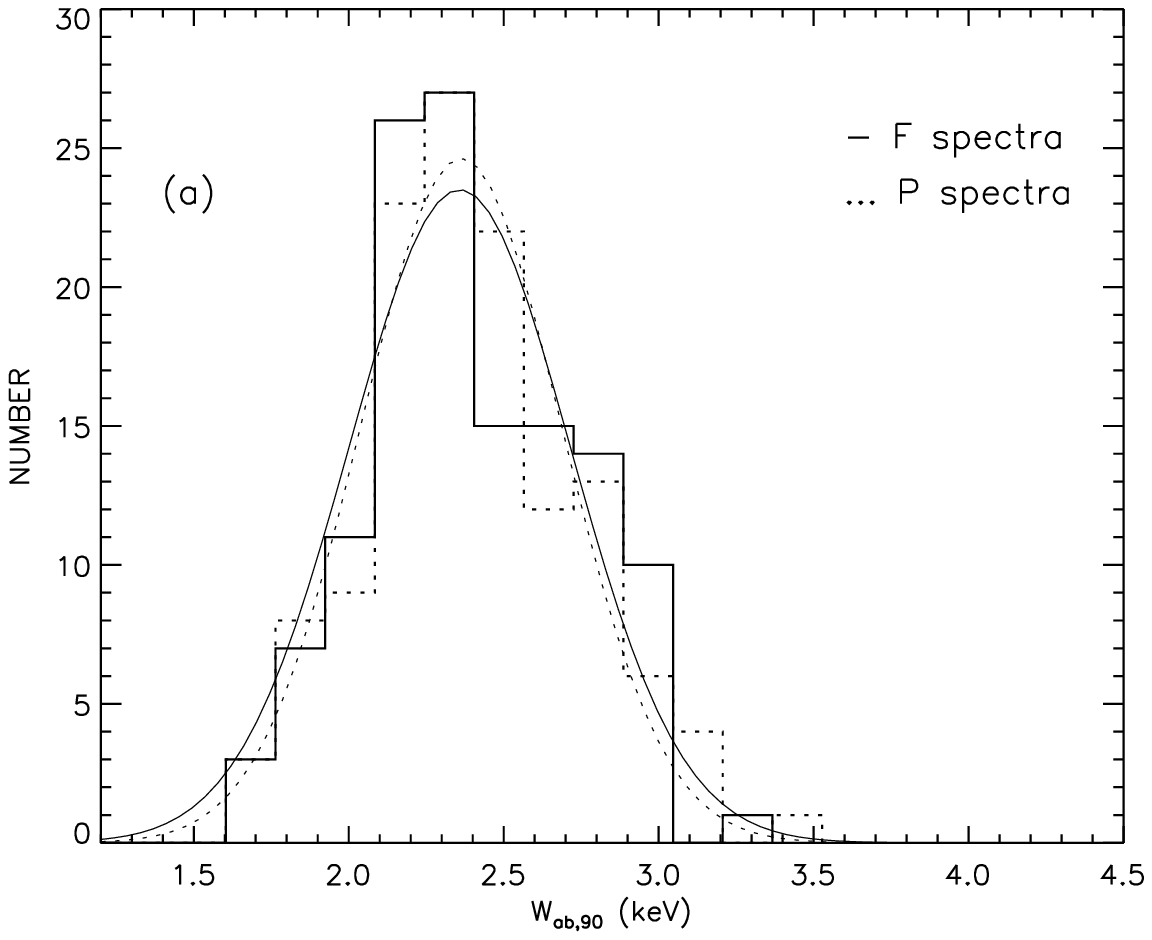}
  \includegraphics[width=0.24\textwidth]{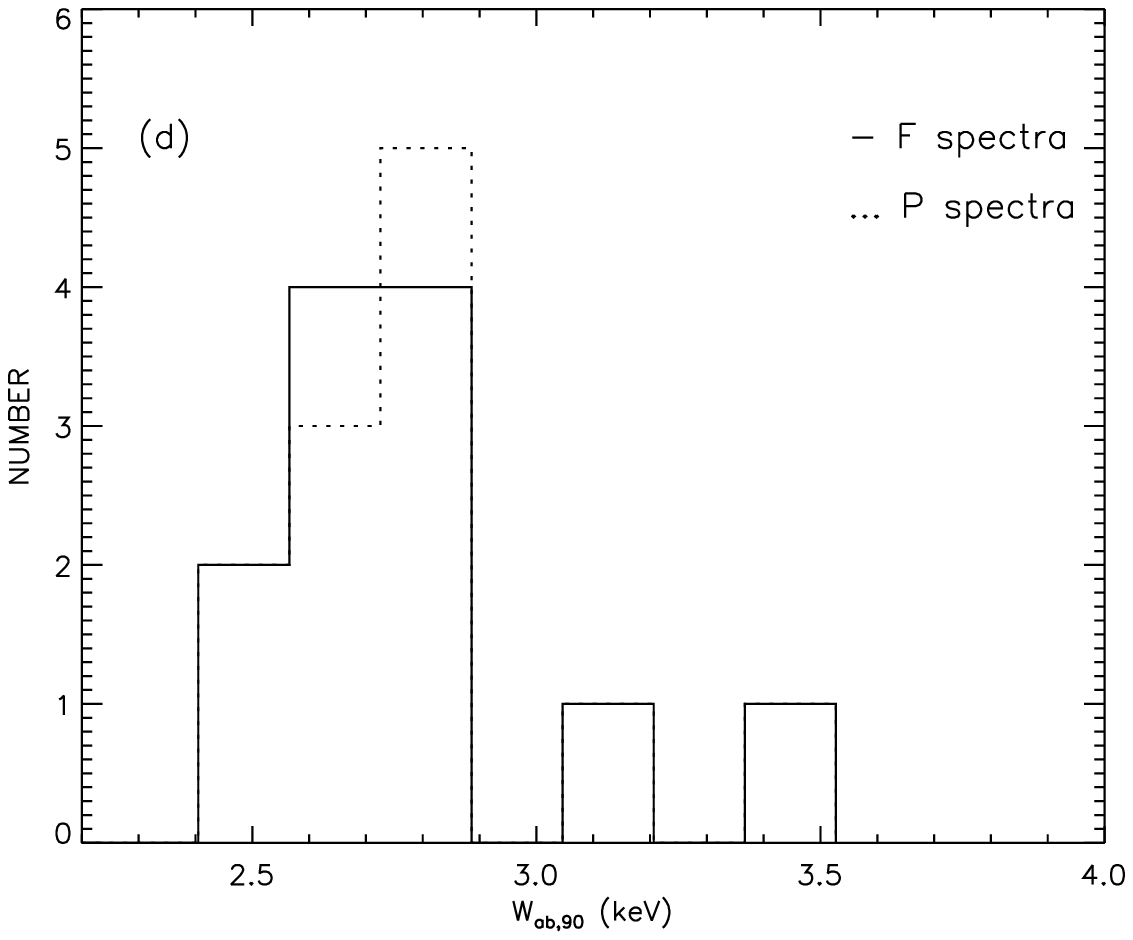}
  \includegraphics[width=0.24\textwidth]{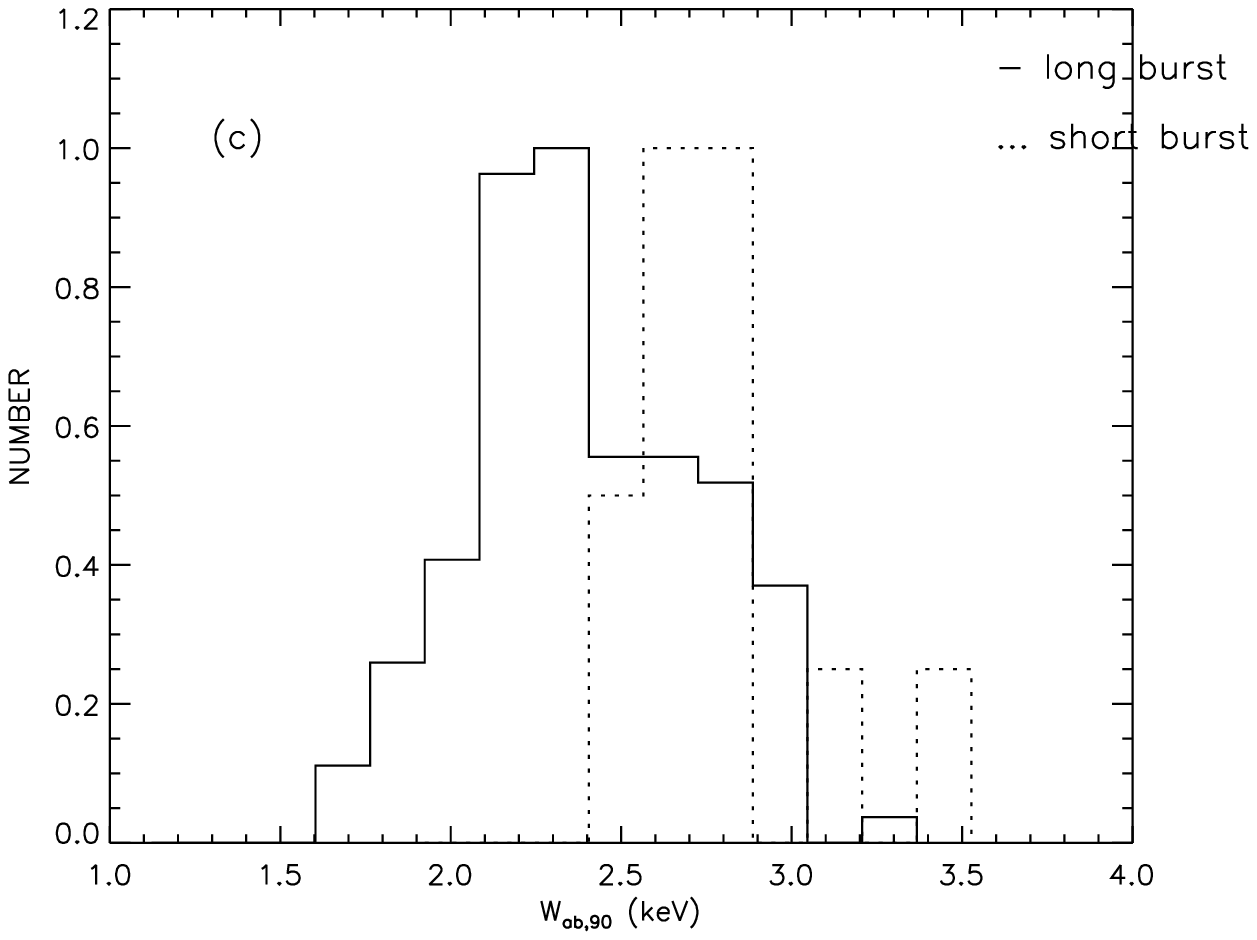}
  \includegraphics[width=0.24\textwidth]{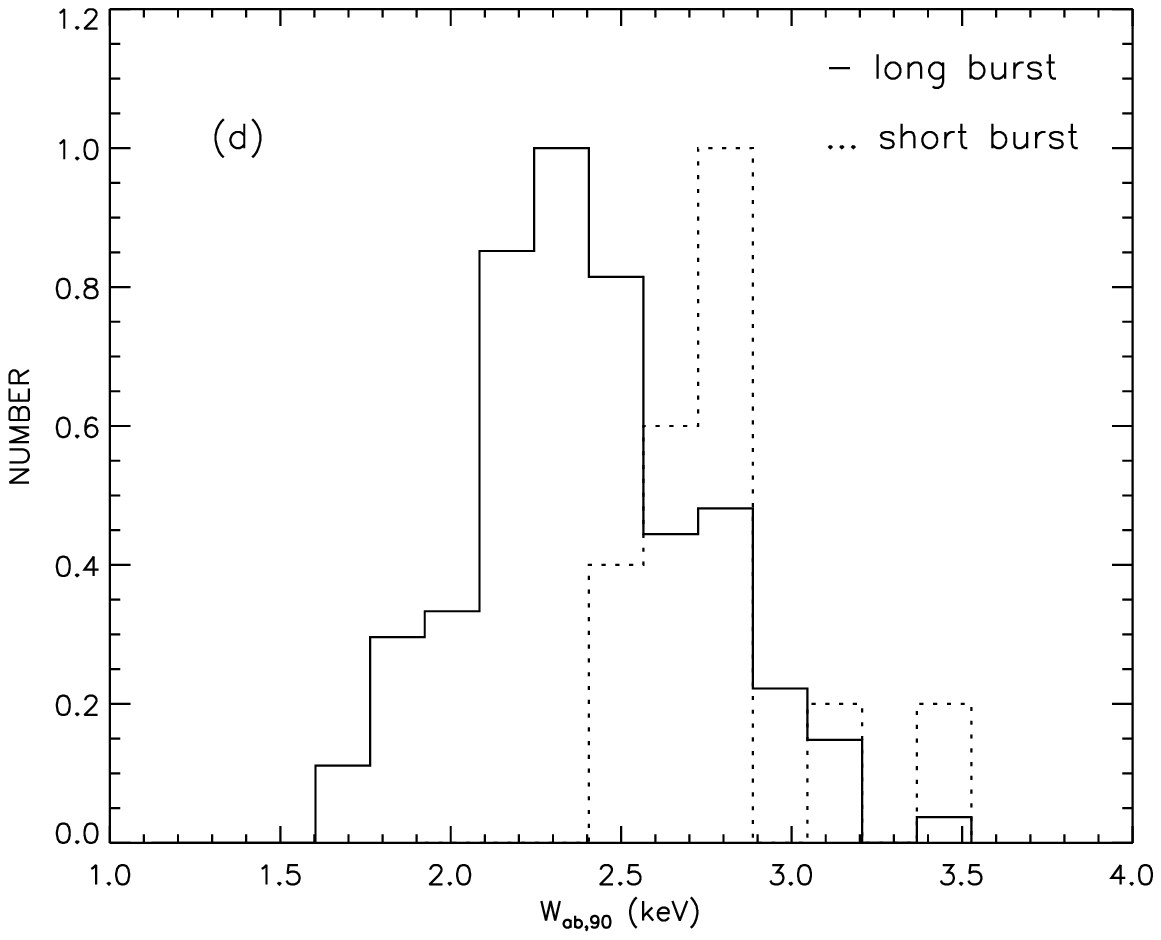}
  \caption{Distributions of the spectral width $W_{ab,90}$ for the long (a) and the short burst set (b) and the comparison of the spectral width distribution between the long and short burst for the P spectra (c) and the F spectra (d).\label{fig:f1}}
\end{figure}

\begin{figure}
  \includegraphics[width=0.24\textwidth]{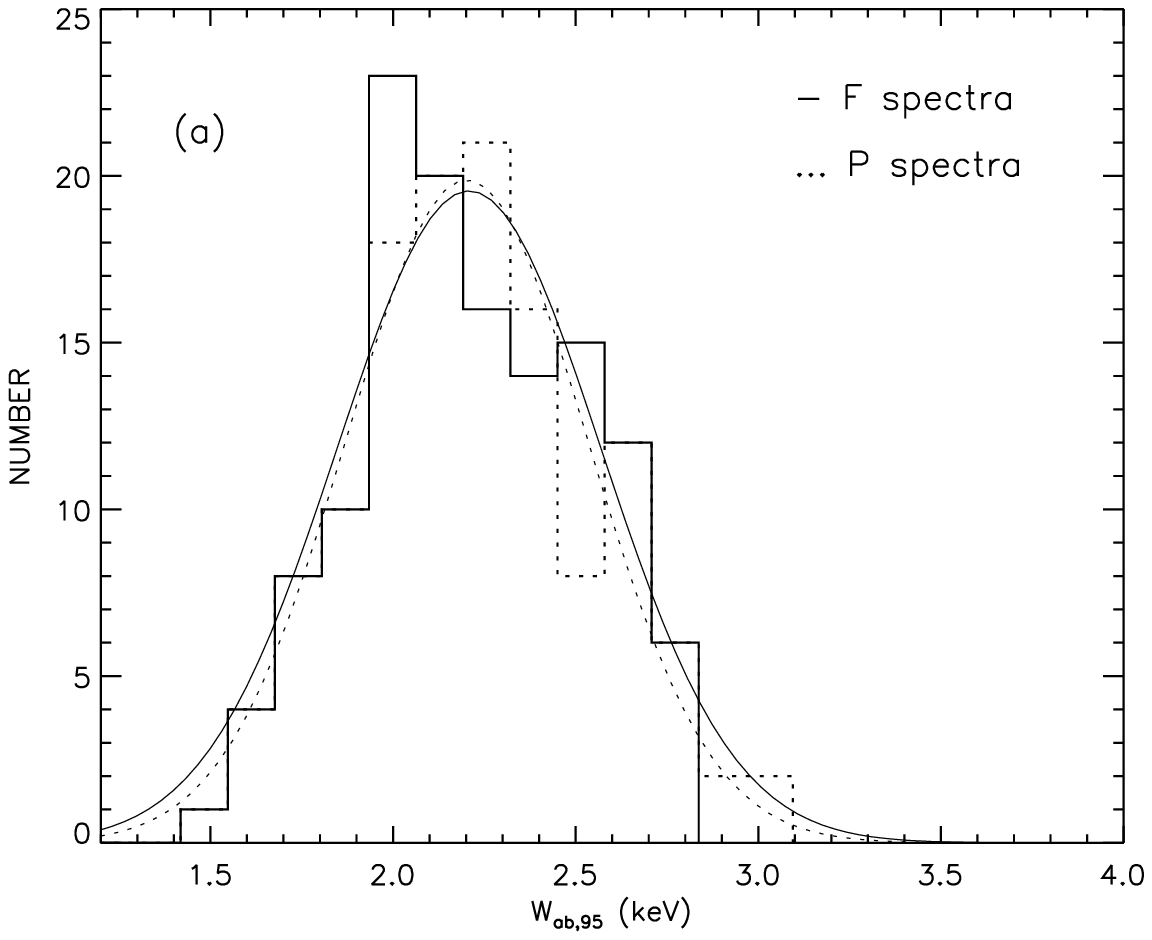}
  \includegraphics[width=0.24\textwidth]{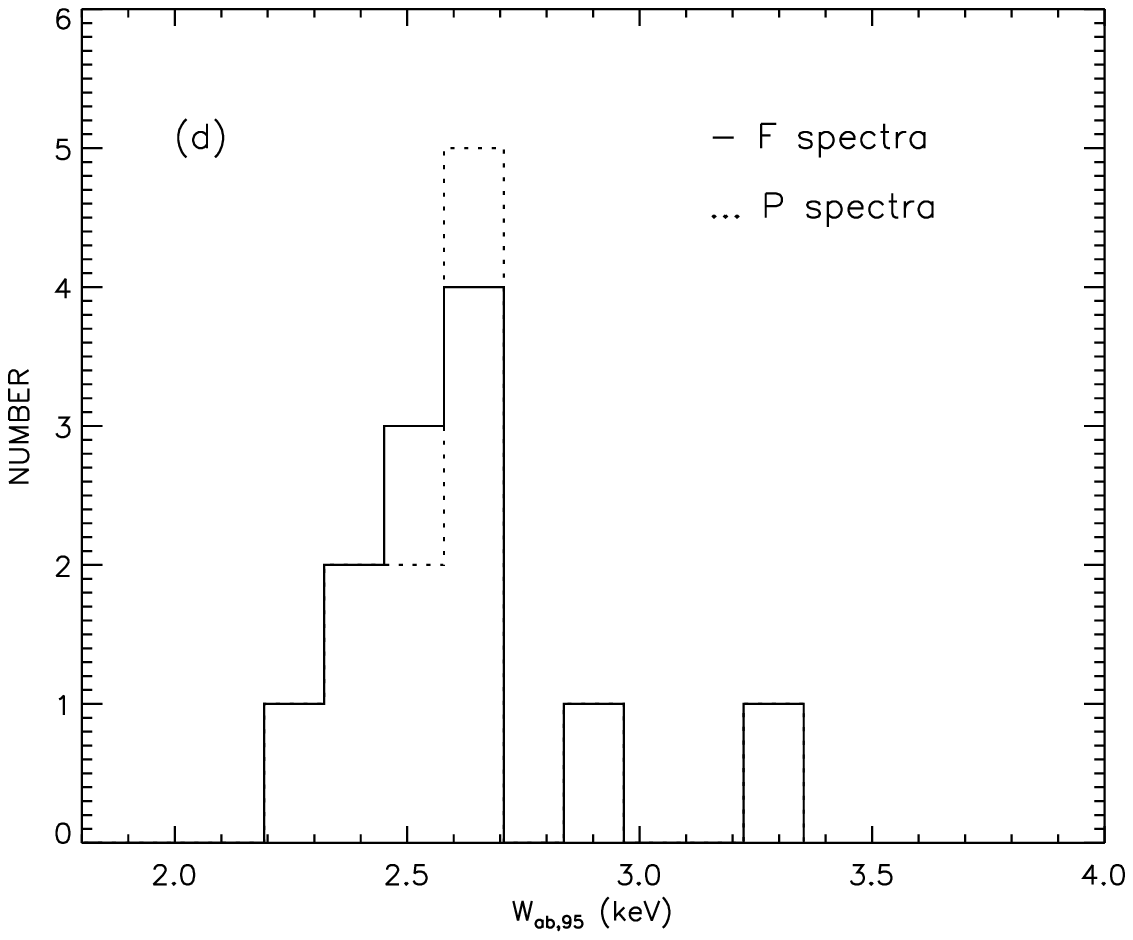}
  \includegraphics[width=0.24\textwidth]{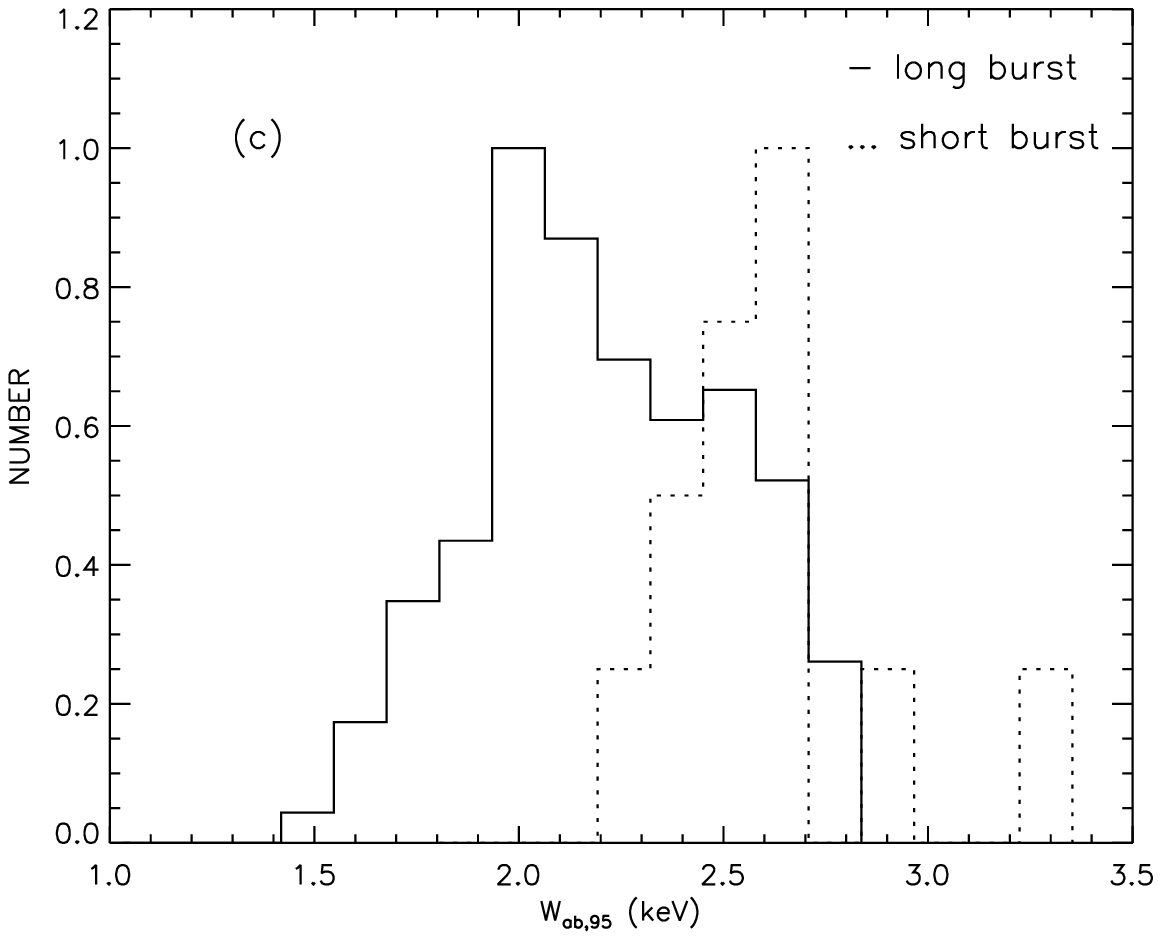}
  \includegraphics[width=0.24\textwidth]{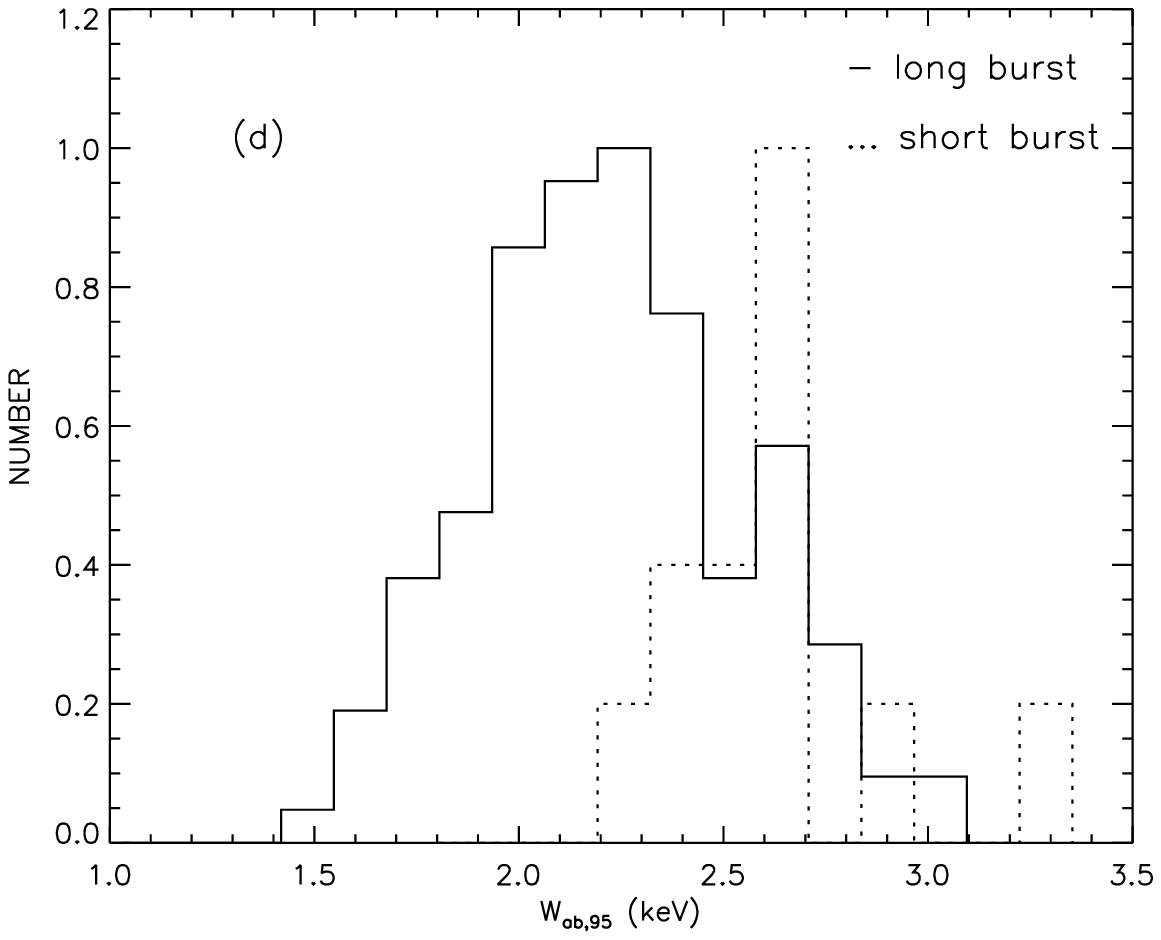}
  \caption{Distributions of the spectral width $W_{ab,95}$ for the long (a) and the short burst set (b) and the comparison of the spectral width distribution between the long and short burst for the P spectra (c) and the F spectra (d).\label{fig:f1}}
\end{figure}

\begin{figure}
  \includegraphics[width=0.24\textwidth]{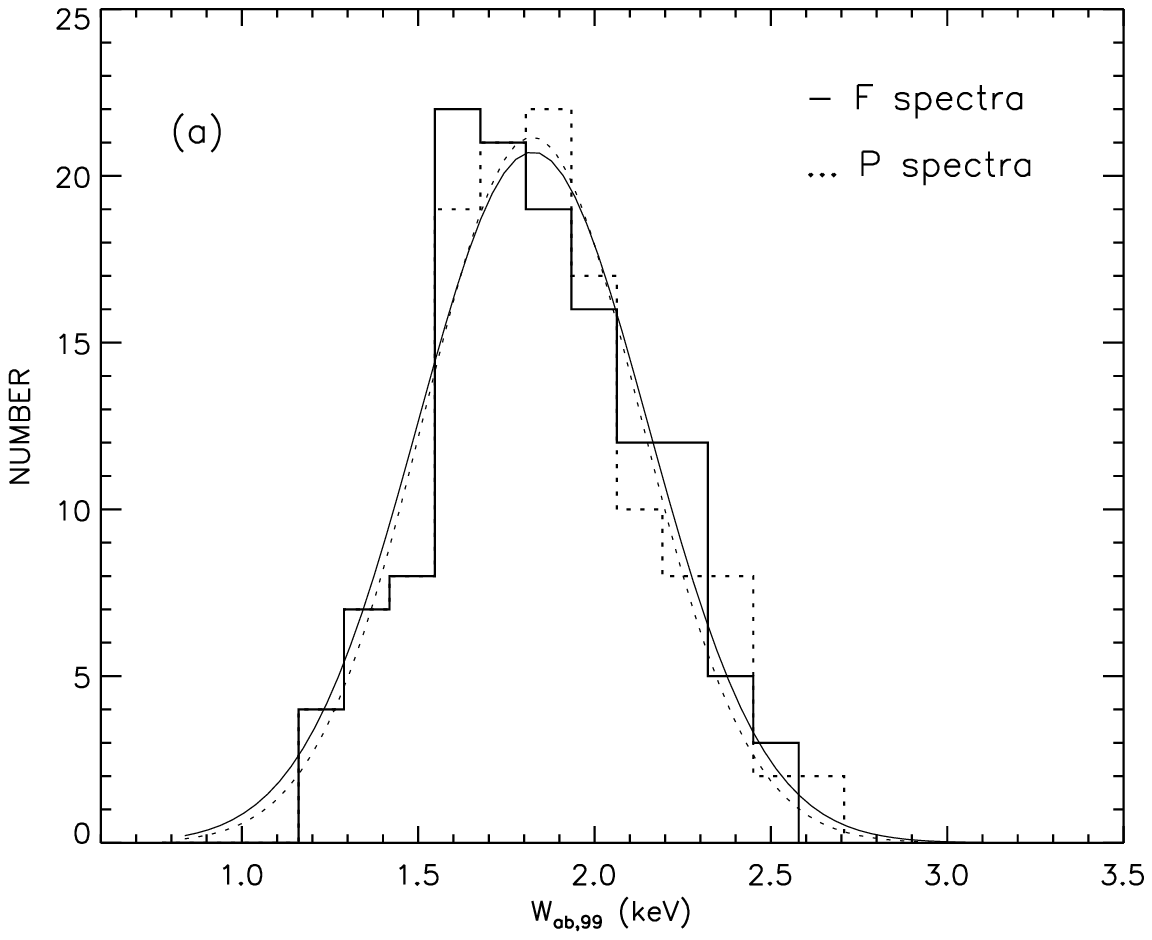}
  \includegraphics[width=0.24\textwidth]{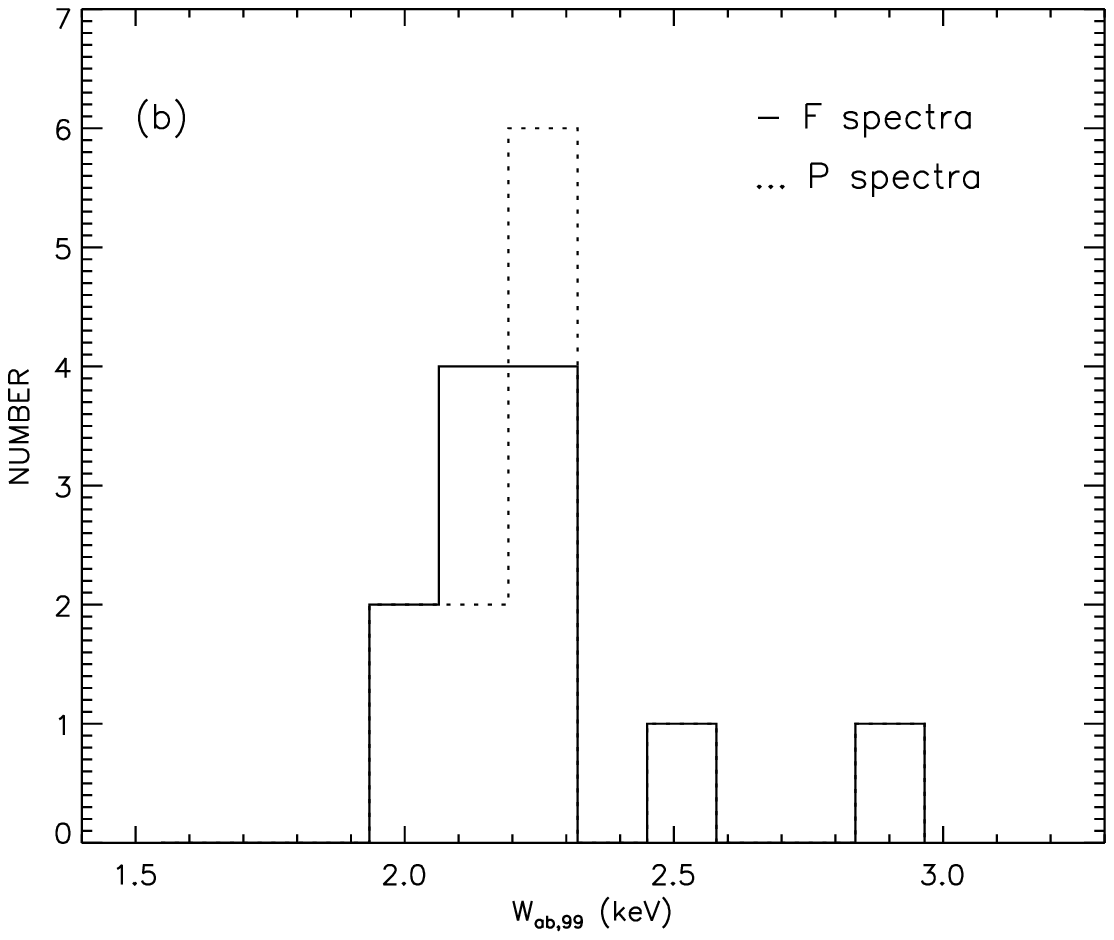}
  \includegraphics[width=0.24\textwidth]{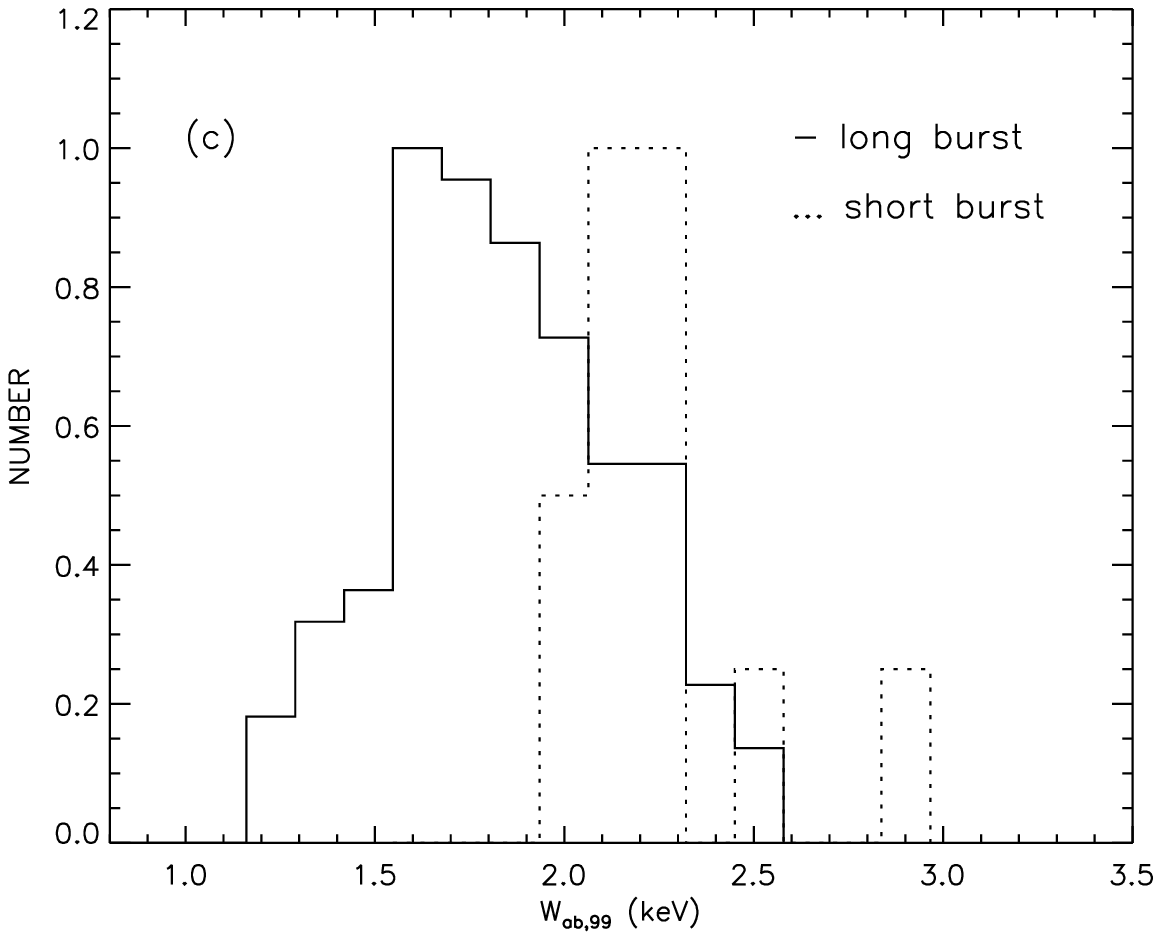}
  \includegraphics[width=0.24\textwidth]{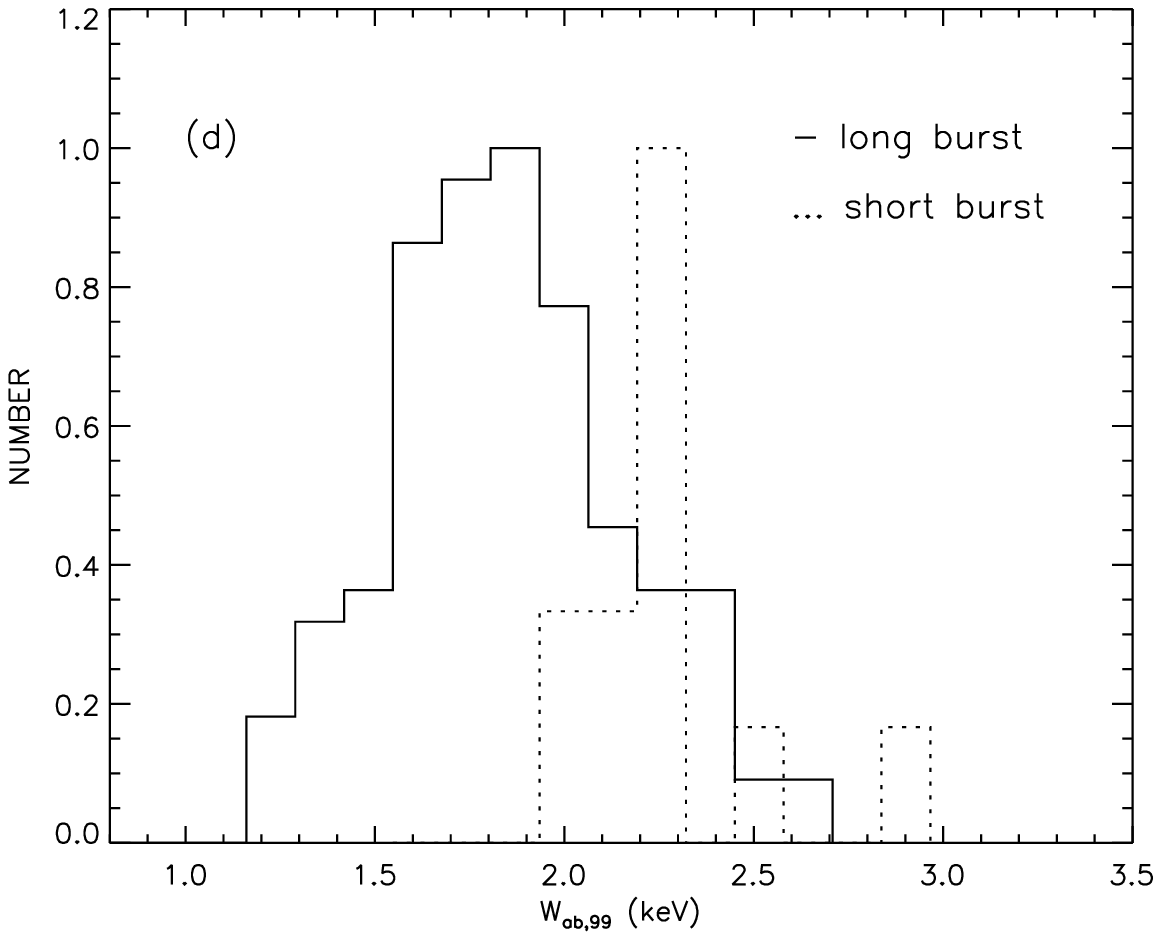}
  \caption{Distributions of the spectral width $W_{ab,99}$ for the long (a) and the short burst set (b) and the comparison of the spectral width distribution between the long and short burst for the P spectra (c) and the F spectra (d).\label{fig:f1}}
\end{figure}

\startlongtable
\begin{deluxetable}{lcccccccc}
\tablecaption{Characteristics of various spectral width distributions of the two types of spectra, separated for the long and the short GRBs.\label{}}
\tablehead{
\colhead{} &\colhead{} & \colhead{F Spectra} & \colhead{} & \colhead{} &\colhead{P Spectra} & \colhead{} & \colhead{} & \colhead{} \\
\colhead{} &\colhead{Long GRBs} & \colhead{Short GRBs}& \colhead{Entire Bursts} & \colhead{Long GRBs}& \colhead{Short GRBs}& \colhead{Entire Bursts}
}
\startdata
    Number     & 129               & 12              & 141               &  133               & 12                &145 \\
               &                   &                 &      $W_{ab,50}$  &                    &                   &                \\
	Median     & 2.90 $\pm$0.11  & 3.16$\pm$0.05 & 2.97$\pm$0.055   &  2.93$\pm$0.19   & 3.22$\pm$0.08   & 2.96 $\pm$0.30\\
    Minimum    & 2.16 $\pm$0.03  & 2.88$\pm$0.02 & 2.16$\pm$0.03   &  2.16$\pm$0.03   & 2.88$\pm$0.02   & 2.16 $\pm$0.03\\
    Maximin    & 4.00 $\pm$0.17  & 3.86$\pm$0.05 & 4.00$\pm$0.17   &  4.00$\pm$0.21   & 3.83$\pm$0.04   & 4.00 $\pm$0.21\\
               &                   &                 &      $W_{ab,75}$  &                    &                   &                \\
    Median     & 2.59 $\pm$0.33  & 2.97$\pm$0.05 & 2.63$\pm$0.02  &  2.64$\pm$0.03   & 3.02$\pm$0.10   & 2.68 $\pm$0.13\\
    Minimum    & 1.93 $\pm$0.10  & 2.69$\pm$0.02 & 1.93$\pm$0.097   &  1.93$\pm$0.10   & 2.69$\pm$0.02   & 1.93 $\pm$0.10\\
    Maximin    & 4.00 $\pm$0.40  & 3.66$\pm$0.05 & 4.00$\pm$0.40   &  4.32$\pm$0.46   & 3.63$\pm$0.05   & 4.32$\pm$0.46\\
               &                   &                 &      $W_{ab,85}$  &                    &                   &                \\
    Median     & 2.45 $\pm$0.09  & 2.84$\pm$0.06 & 2.49$\pm$0.04   &  2.47$\pm$0.01   & 2.89$\pm$0.10   & 2.54 $\pm$0.03\\
    Minimum    & 1.79 $\pm$0.08  & 2.56$\pm$0.02 & 1.79$\pm$0.08   &  1.79$\pm$0.08   & 2.56$\pm$0.02   & 1.80 $\pm$0.08\\
    Maximin    & 3.74 $\pm$0.36  & 3.53$\pm$0.05 & 3.74$\pm$0.36   &  3.60$\pm$0.29   & 3.50$\pm$0.05   & 3.60 $\pm$0.29\\
               &                   &                 &      $W_{ab,90}$  &                    &                   &                \\
    Median     & 2.35 $\pm$0.07  & 2.75$\pm$0.05 & 2.39$\pm$0.07   &  2.37$\pm$0.08   & 2.79$\pm$0.10   & 2.41 $\pm$0.03\\
    Minimum    & 1.70 $\pm$0.06  & 2.47$\pm$0.02 & 1.70$\pm$0.06   &  1.70$\pm$0.06   & 2.47$\pm$0.02   & 1.70 $\pm$0.06\\
    Maximin    & 3.30 $\pm$0.26  & 3.44$\pm$0.05 & 3.44$\pm$0.05   & 3.50$\pm$0.31    & 3.41$\pm$0.05   & 3.50 $\pm$0.31\\
               &                   &                 &      $W_{ab,95}$  &                    &                   &                \\
    Median     & 2.19 $\pm$0.03  & 2.59$\pm$0.05 & 2.22$\pm$0.01   &  2.20$\pm$0.03   & 2.64$\pm$0.10   & 2.25 $\pm$0.09\\
    Minimum    & 1.54$\pm$0.05   & 2.31$\pm$0.02 & 1.54$\pm$0.05   &  1.54$\pm$0.05   & 2.31$\pm$0.02   & 1.54 $\pm$0.05\\
    Maximin    & 2.83 $\pm$0.01  & 3.28$\pm$0.05 & 3.28$\pm$0.05   &  3.03$\pm$0.14   & 3.25$\pm$0.04   & 3.25 $\pm$0.04\\
               &                   &                 &      $W_{ab,99}$  &                    &                   &                \\
    Median     & 1.83 $\pm$0.07  & 2.23$\pm$0.06 & 1.86$\pm$0.01   &  1.84$\pm$0.01   & 2.28$\pm$0.10   & 1.86 $\pm$0.01\\
    Minimum    & 1.18$\pm$0.05  & 1.96$\pm$0.02  & 1.18$\pm$0.05   &  1.18$\pm$0.05   & 1.96$\pm$0.02   & 1.18 $\pm$0.05\\
    Maximin    & 2.47$\pm$0.01  & 2.93$\pm$0.05  & 2.93$\pm$0.05   &  2.68$\pm$0.13   & 2.89 $\pm$0.05  & 2.89 $\pm$0.05\\
\enddata
\end{deluxetable}

\subsection{Confirmed correlated relationships of the $Width-E_{iso}$ and $Width-L_{iso}$}
In this section, we first reanalyze the correlations between the absolute spectral width in the rest frame $W_{ab,i}$=$W_{ab}+\log(1+z)$ and the $E_{iso}$ as well as $L_{iso}$. The case of the most popular spectral width at half (50\%) maximum $W_{ab,50,i}$ is first considered. We calculate the Spearman rank-order correlation coefficients ($\rho$) and the associated null-hypothesis (chance) probabilities or P values to check the existence of correlations. In fact, astronomical data are often uncertain with errors that are heteroscedastic (Robotham \& Obreschkow 2015). The correlation is often characterized by the "intrinsic" scatter for the two subsamples. A Hyper-fit package can derive the general likelihood function to be maximised to recover the best fitting model if a set of D-dimensional data points can be described by a (D-1)-dimensional plane with intrinsic scatter (Robotham \& Obreschkow 2015). We can use the Hyper-fit package to perform the linear model fitting (equation (2)) with both statistical errors in $x=log \frac{E_{iso}}{10^{51} erg}$ or $x=log \frac{L_{iso}}{10^{51} erg s^{-1}}$ and $y=W_{ab,50,i}$ values to estimate the best fitting lines. The associated residuals and intrinsic scatters are then obtained. The slope and intercept as well as the intrinsic scatter $\sigma _{int}$ are estimated by using Hyper-fit package to perform the linear model fitting (equation (2)) with both statistical errors in X and Y values. The correlation coefficients ($\rho$), P values, the slope (a), intercept (b), dispersion of the points around the best-fit relations ($\sigma$) are summarized in Table 3.

\begin{equation}
    y = a \times x + b
	\label{}
\end{equation}

\begin{figure*}
  \includegraphics[width=0.48\textwidth]{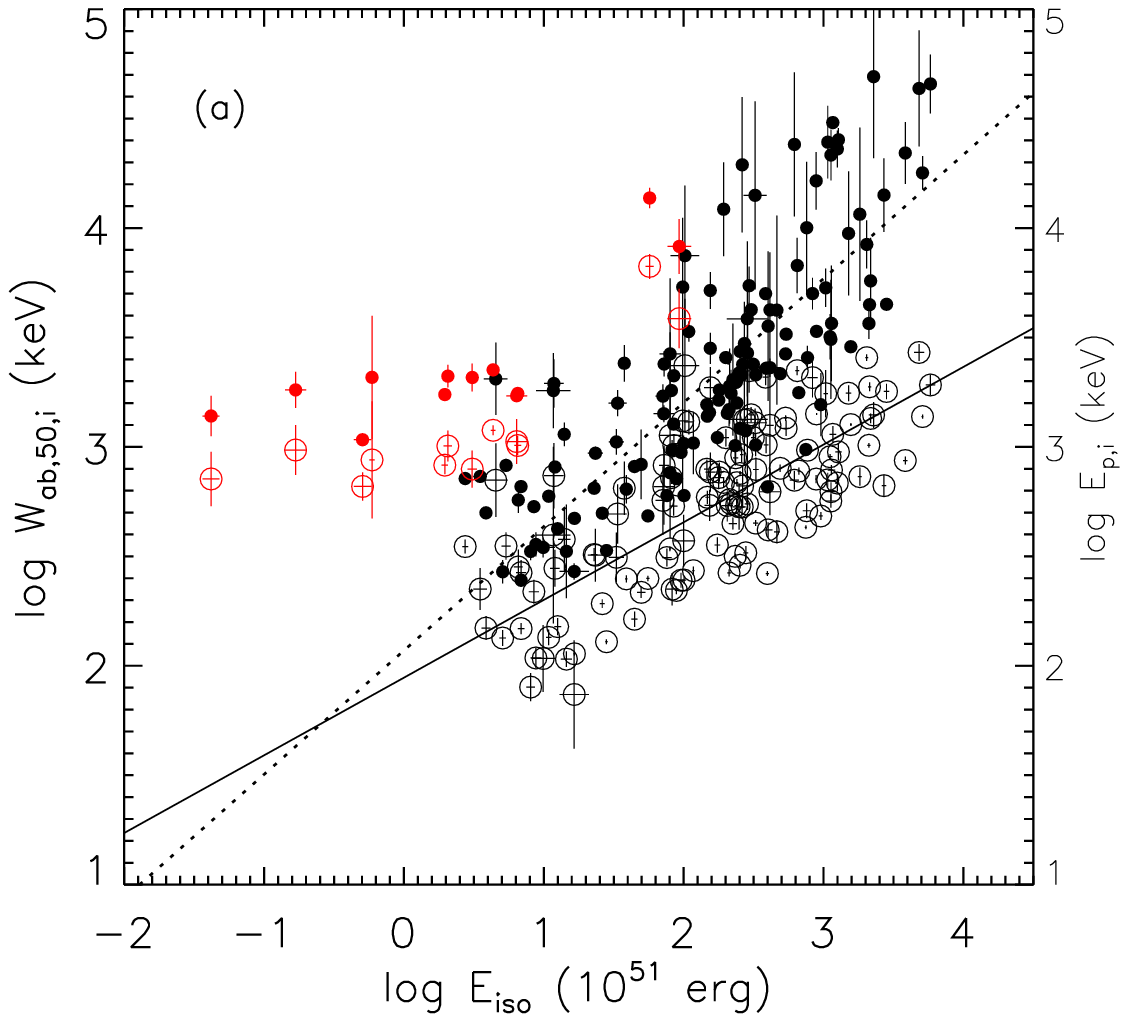}
  \includegraphics[width=0.48\textwidth]{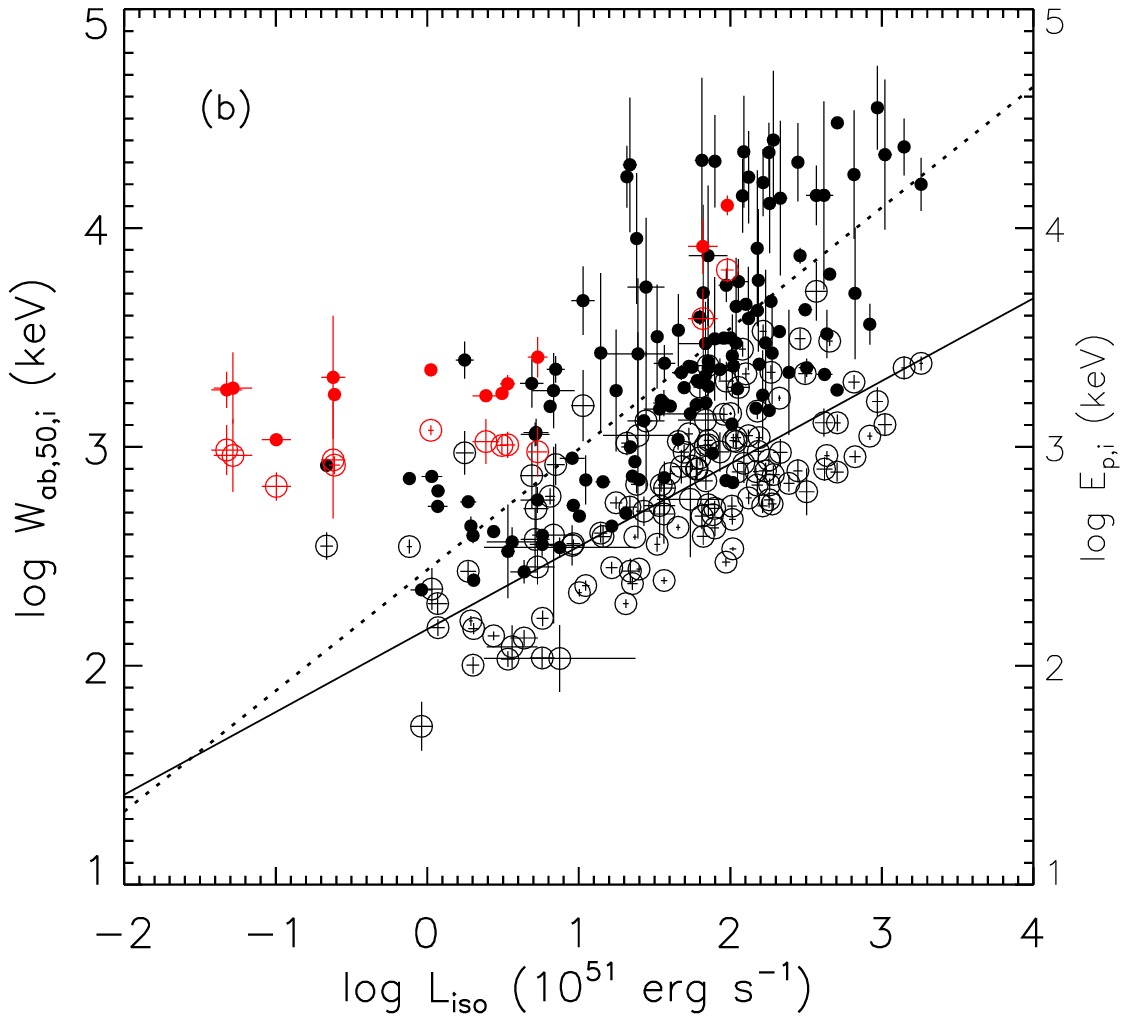}
  \caption{$W_{ab,50,i}$ vs. $E_{iso}$ (a) and $W_{ab,50,i}$ vs. $L_{iso}$ (b), along with the Amati and Yonetoku relations. The black filled circles denote $W_{ab,i}-E_{iso}$ and $W_{ab,i}-L_{iso}$ relations for long bursts and black open circles represent the Amati and Yonetoku relations for long bursts, respectively. while the red filled circles and red open circles stand for the short bursts for the same relations. The dashed lines are the best-fitting lines of $W_{ab,50,i}$-$E_{iso}$ and $W_{ab,50,i}$-$L_{iso}$ and the solid lines are the best-fitting lines of the Amati and Yonetoku relations.}
 \label{fig:example_figure}
 \end{figure*}

Figure 7 and Table 3 demonstrate the relationships between $W_{ab,50,i}$ versus $E_{iso}$ and $W_{ab,50,i}$ versus $L_{iso}$. The short GRBs are shown in red filled circles, while long GRBs are denoted in black filled circles. We first check the relationships between $W_{ab,50,i}$ and $E_{iso}$ for the F spectra. For the case of F spectra a strong correlation exists between them for entire sample (including long and short bursts) with the Spearman rank-order correlation coefficients $\rho$=0.72 (P = 2.25 $\times$ $10^{-23}$). The short bursts clearly deviate from the correlations of the long bursts in Figure 7, which can be also seen from the fact that the correlations of the long burst is $\rho$=0.80 (P = 8.75 $\times$ $10^{-30}$). This correlation of the long burst is much more significant than the entire bursts. The short bursts are the outliers for the $W_{ab,50,i}-E_{iso}$ relation. This property is similar to the Amati relation.

Then we examine the the relationships between $W_{ab,50,i}$ and $L_{iso}$ for the P spectra. The corresponding Spearman rank-order correlation coefficients are $\rho$=0.69 (P $<$0.0001) and $\rho$=0.70 (P$<$0.0001) for the entire bursts and long bursts, which also reveal the very strong correlations for $W_{ab,50,i}- L_{iso}$. In addition, this shows that the short and long bursts are not well separated and the short bursts almost follow the same correlated relationships as the long bursts for the case of $W_{ab,50,i}- L_{iso}$. This property is consistent with the Yonetoku relation.

However, the dispersions of the $W_{ab,50,i}-E_{iso}$ (0.300) for the F spectra is also slightly greater than the Amati relation (0.227) obtained with the same F spectral data. Likewise, the scatters of $W_{ab,50,i}- L_{iso}$ (0.346) for the P spectra are also greater than the Yonetoku relation (0.243) with the same P spectral data.

\begin{figure*}
  \includegraphics[width=0.48\textwidth]{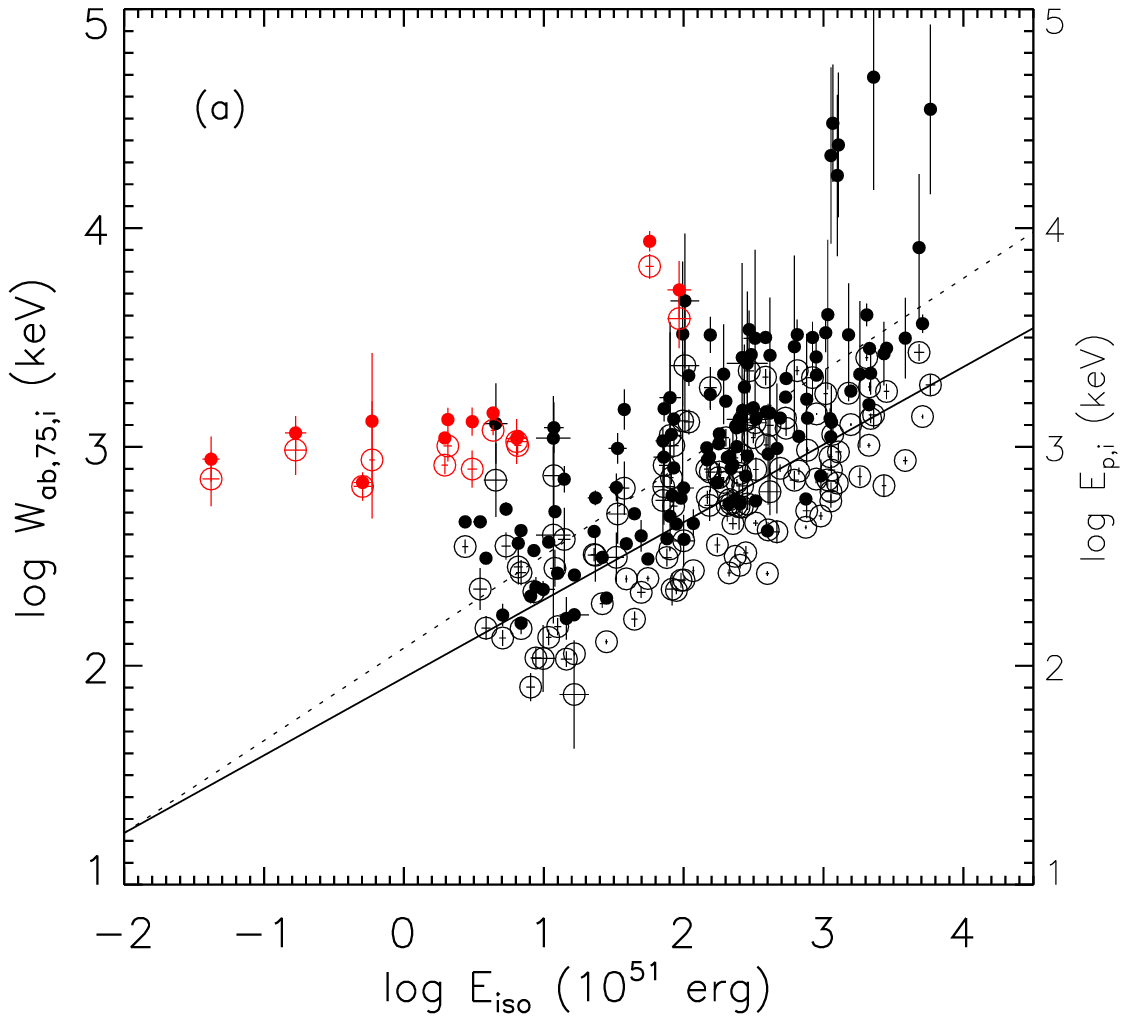}
  \includegraphics[width=0.48\textwidth]{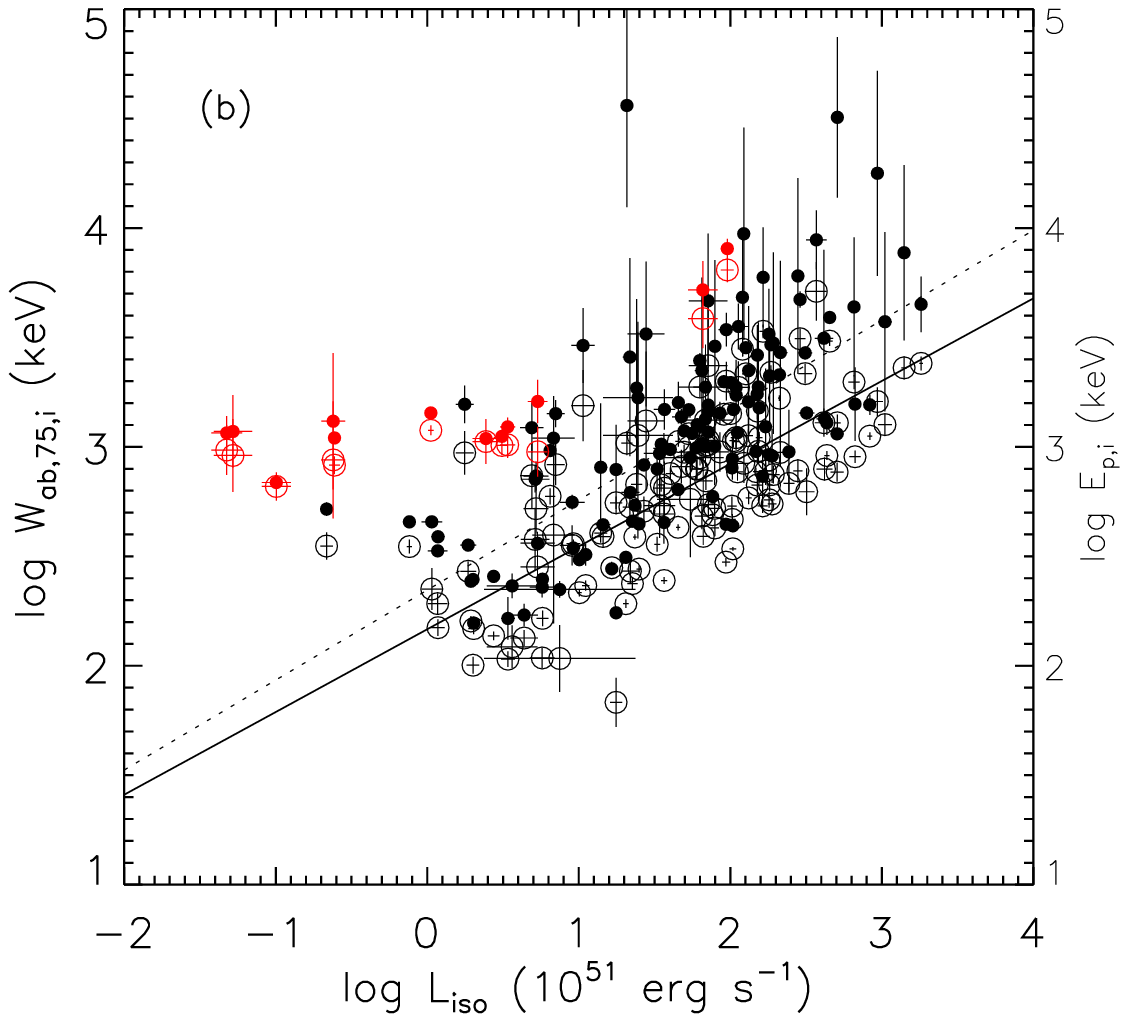}
  \caption{$W_{ab,75,i}$ vs. $E_{iso}$ for the F spectra (a) and $W_{ab,75,i}$ vs. $L_{iso}$ for the P spectra (b), along with the Amati and Yonetoku relations. All the symbols are same as Figure 7.}
\label{fig:example_figure}
\end{figure*}

So we try to investigate the possible reason of the larger scatters. We suspect that the wider spectra cause larger scatter by inspection from the Figure 7. In fact, the upper energy bounds of the full width half maximum $E_{2}$ defined in equations 1 are less reliable due to the fact that the detector effective area and the flux fall rapidly with increasing energy, which can be seen from the larger errors of the wide spectral bursts.

\begin{figure*}
  \includegraphics[width=0.48\textwidth]{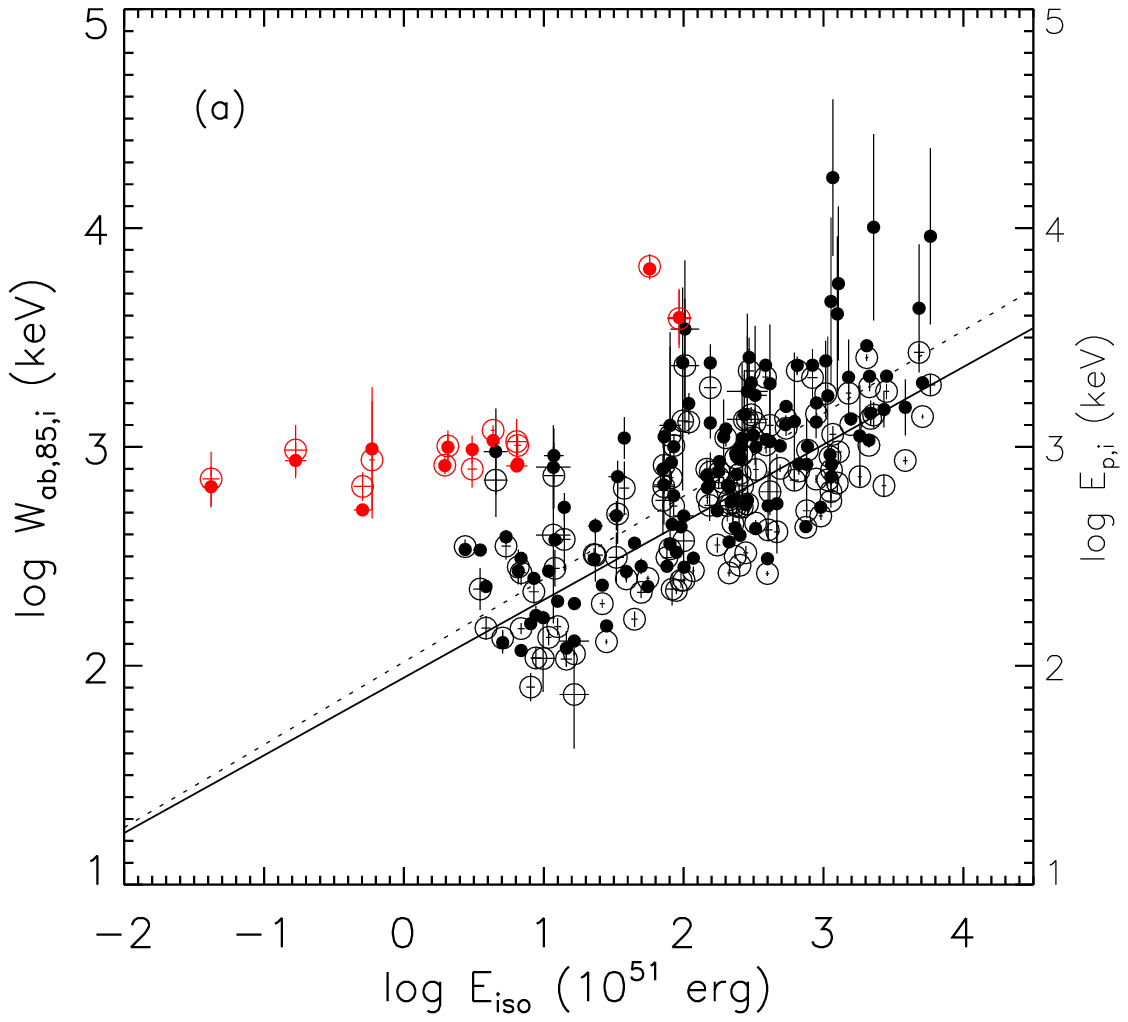}
  \includegraphics[width=0.48\textwidth]{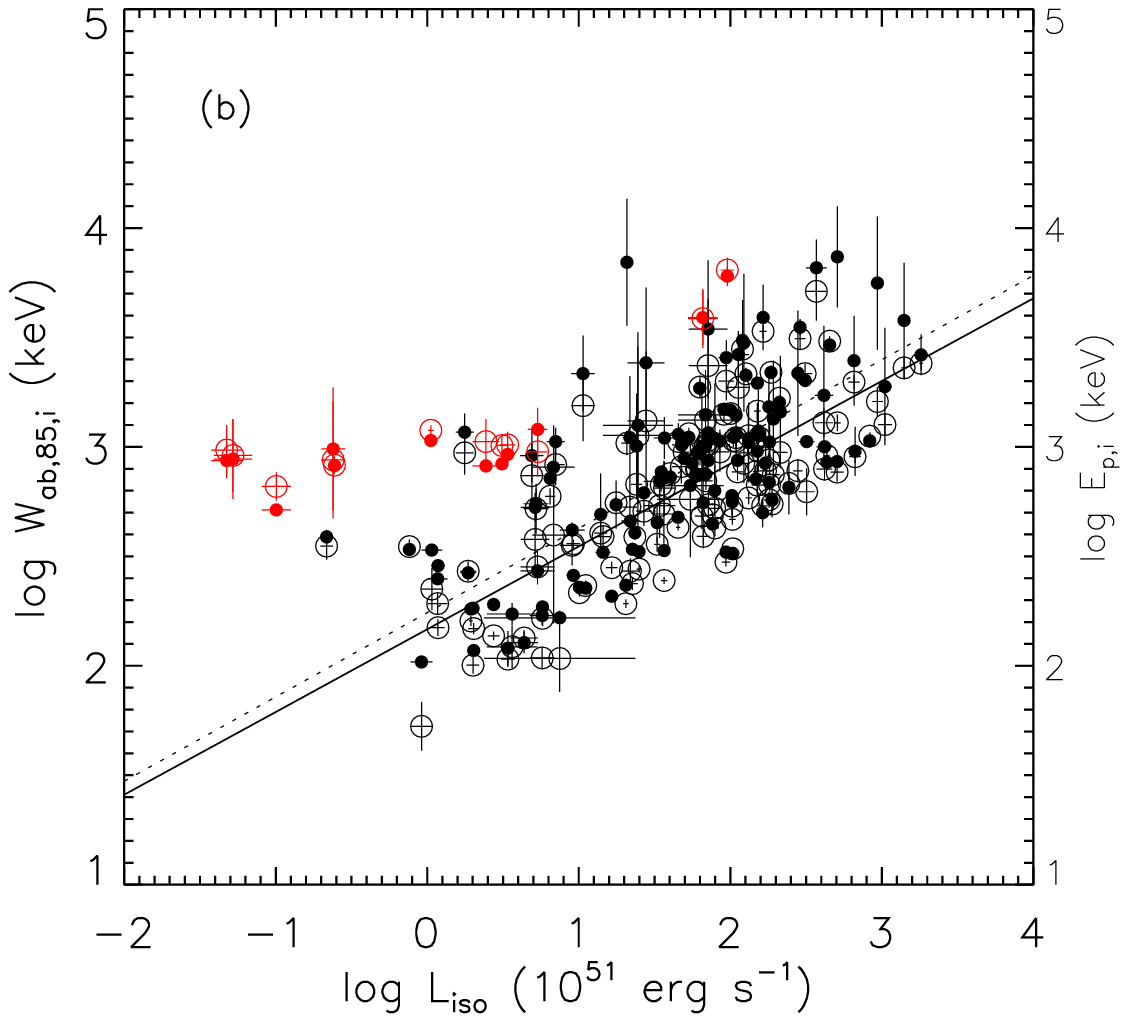}
  \caption{$W_{ab,85,i}$ vs. $E_{iso}$ for the F spectra (a) and $W_{ab,85,i}$ vs. $L_{iso}$ for the P spectra (b), along with the Amati and Yonetoku relations. All the symbols are same as Figure 7.}
  \label{fig:example_figure}
\end{figure*}

Therefore we check the correlations between other much smaller absolute spectral widths defined in previous section in the rest frame and the $E_{iso}$ as well as $L_{iso}$. The diagrams and correlated properties constructed for different spectral width
$W_{ab,75,i}$, $W_{ab,85,i}$, $W_{ab,90,i}$, $W_{ab,95,i}$, and $W_{ab,99,i}$ with $E_{iso}$ and $L_{iso}$ are presented in Figures 8-12 and Table 4.
It is found that the smallest correlation coefficient among the correlation parameter pairs arrives at 0.64 (p$<$0.0001).
These show all of the absolute spectral widths in the rest frame are strongly correlated with $E_{iso}$ as well as $L_{iso}$. Short bursts are evident outliers for width-$E_{iso}$ for the F spectra. For the width-$L_{iso}$ short and long GRBs are not well separated.

When comparing the correlated properties between all of the absolute spectral width in the rest frame and $E_{iso}$ as well as $L_{iso}$ from Table 4 and Figures 7-12 we find that: (1)the correlation coefficient and significance of the $W_{ab,50,i}$ $-E_{iso}$ is the largest and most significant among the six correlated parameter pairs; (2)the correlation coefficients, significances, and the scatters of correlations decrease with the widths decrease. However, the variability of Spearman rank-order correlation coefficients for $W_{ab,i}$ $-E_{iso}$ and $W_{ab,i}$ $-L_{iso}$ is very different. Note that the variability here is the amount between the smallest and the largest item in the data set. The correlations for the $W_{ab,i}$ $-E_{iso}$ have a large variability ($\Delta \rho \sim$ 0.14, but the amplitude of variability $\Delta \rho \sim$ 0.07 for $W_{ab,i}$ $-L_{iso}$) for the P spectra is very small (see, Table 4). These seem to show that width-$L_{iso}$ for the peak flux spectra are more stabler than that of time-integrated spectra.

\begin{figure*}
  \includegraphics[width=0.48\textwidth]{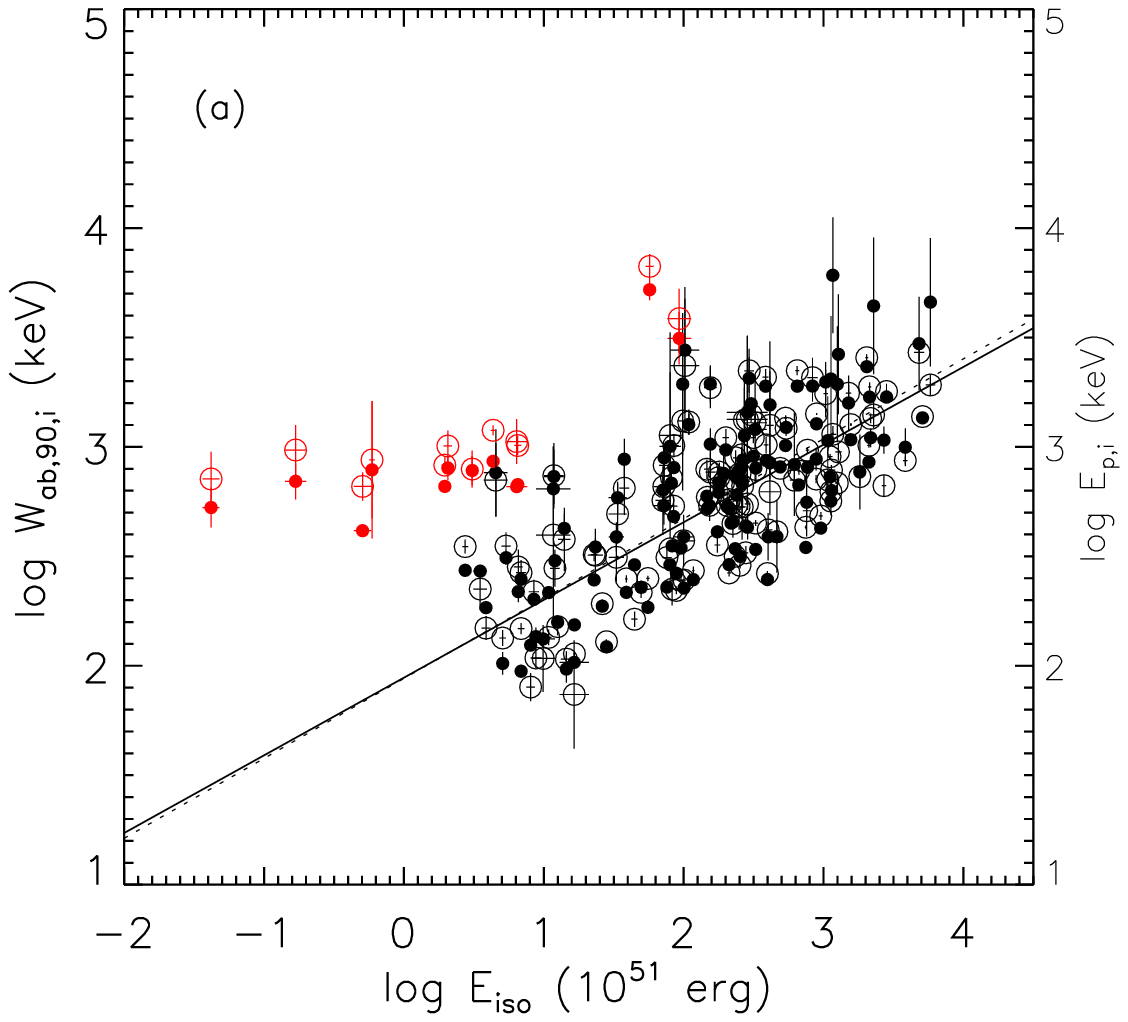}
  \includegraphics[width=0.48\textwidth]{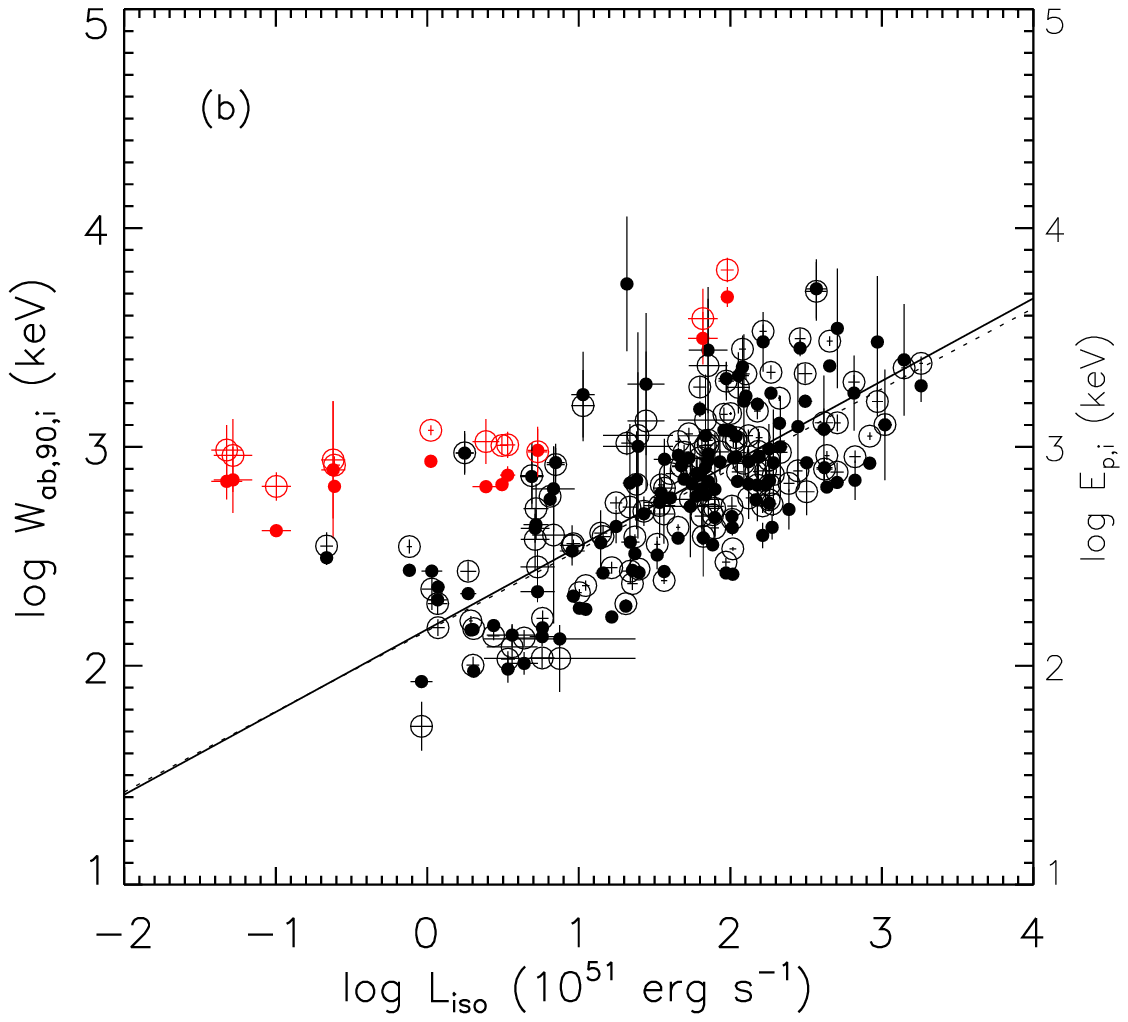}
  \caption{$W_{ab,90,i}$ vs. $E_{iso}$ for the F spectra (a) and $W_{ab,90,i}$ vs. $L_{iso}$ for the P spectra (b), along with the Amati and Yonetoku relations. All the symbols are same as Figure 7.}
\label{fig:example_figure}
\end{figure*}

\startlongtable
\begin{deluxetable}{ccccccccccc}
\tablecaption{Correlation analysis results of $W_{ab,50,i}$  and $E_{iso}$ as well as $L_{iso}$.
\label{}}
\tablehead{
\colhead{Correlation} &\colhead{Number} &\colhead{$\rho_{E}$} &\colhead{$P_{E}$} &\colhead{$\rho_{L}$} &\colhead{$P_{L}$} &\colhead{$a_{L}$} &\colhead{$b_{L}$} &\colhead{$\sigma_{int,L}$}}
\startdata %
\hline
$W_{ab,50,i,F}$ vs. $E_{iso}$ & 141 & 0.72 &  $<$0.0001 & 0.80 & $<$0.0001  & 0.57$\pm$ 0.04 & 2.07$\pm$ 0.09& 0.30$\pm$ 0.022\\
$W_{ab,50,i,P}$ vs. $L_{iso}$ & 145 &0.68  &  $<$0.0001 & 0.71 & $<$0.0001 & 0.55$\pm$0.05 & 2.44$\pm$0.08 & 0.346$\pm$0.027\\
\enddata
\tablecomments{The subscript F, P, E, L denote the F spectra, P spectra, the entire burst set, and the long burst set, respectively.}
\end{deluxetable}

\begin{figure*}
  \includegraphics[width=0.48\textwidth]{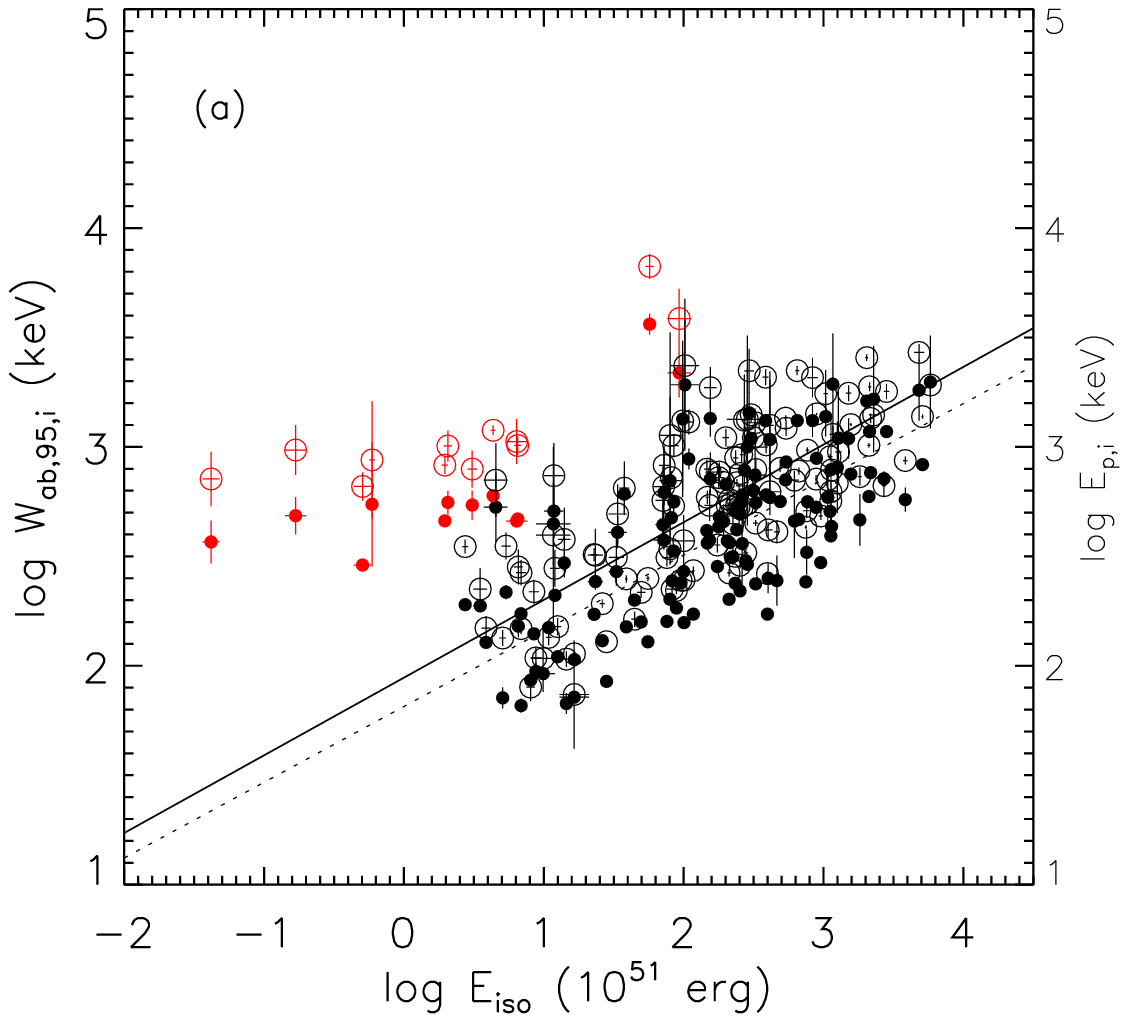}
  \includegraphics[width=0.48\textwidth]{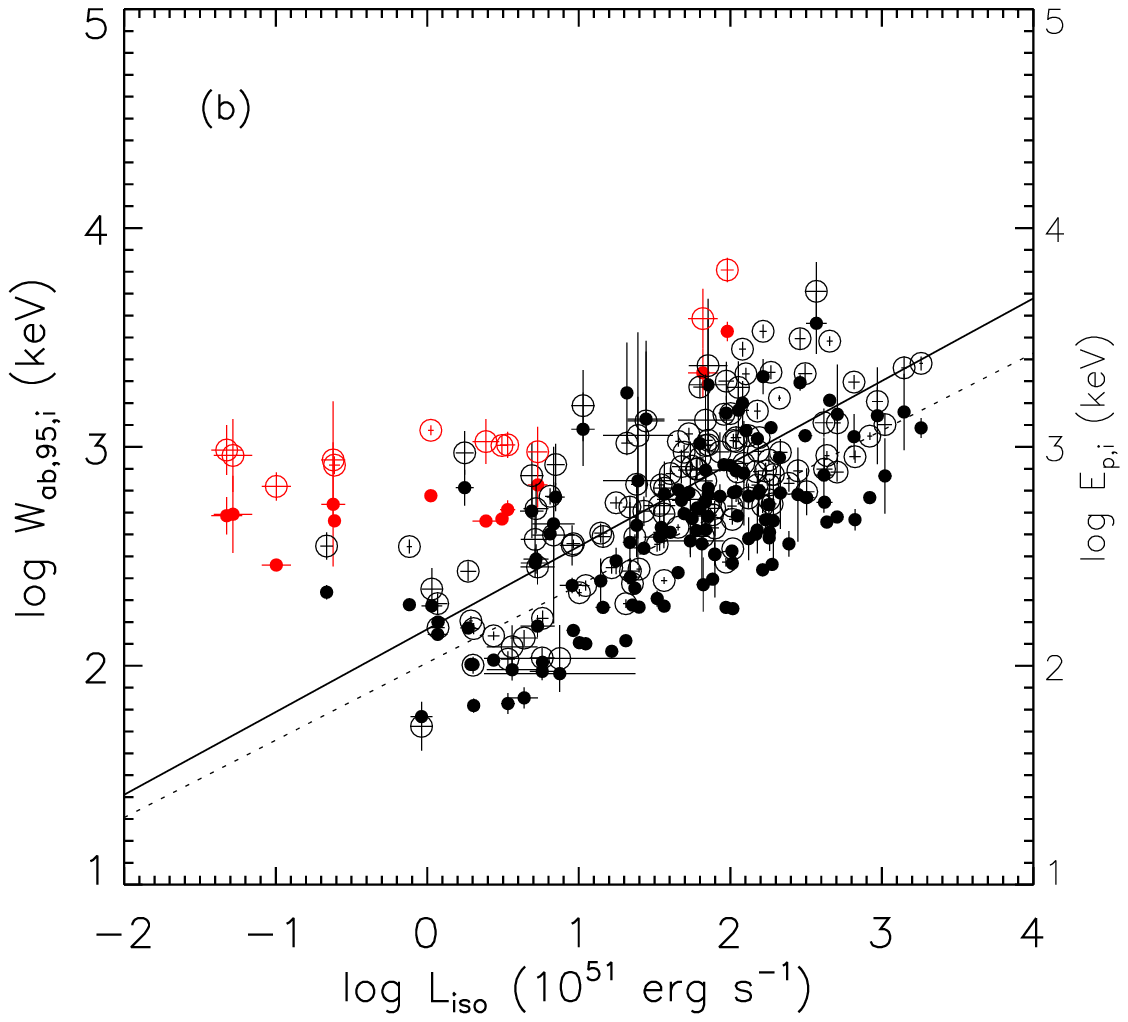}
 \caption{$W_{ab,95,i}$ vs. $E_{iso}$ for the F spectra (a) and $W_{ab,95,i}$ vs. $L_{iso}$ for the P spectra (b), along with the Amati and Yonetoku relations. All the symbols are same as Figure 7.}
\label{fig:example_figure}
\end{figure*}

\begin{figure*}
  \includegraphics[width=0.48\textwidth]{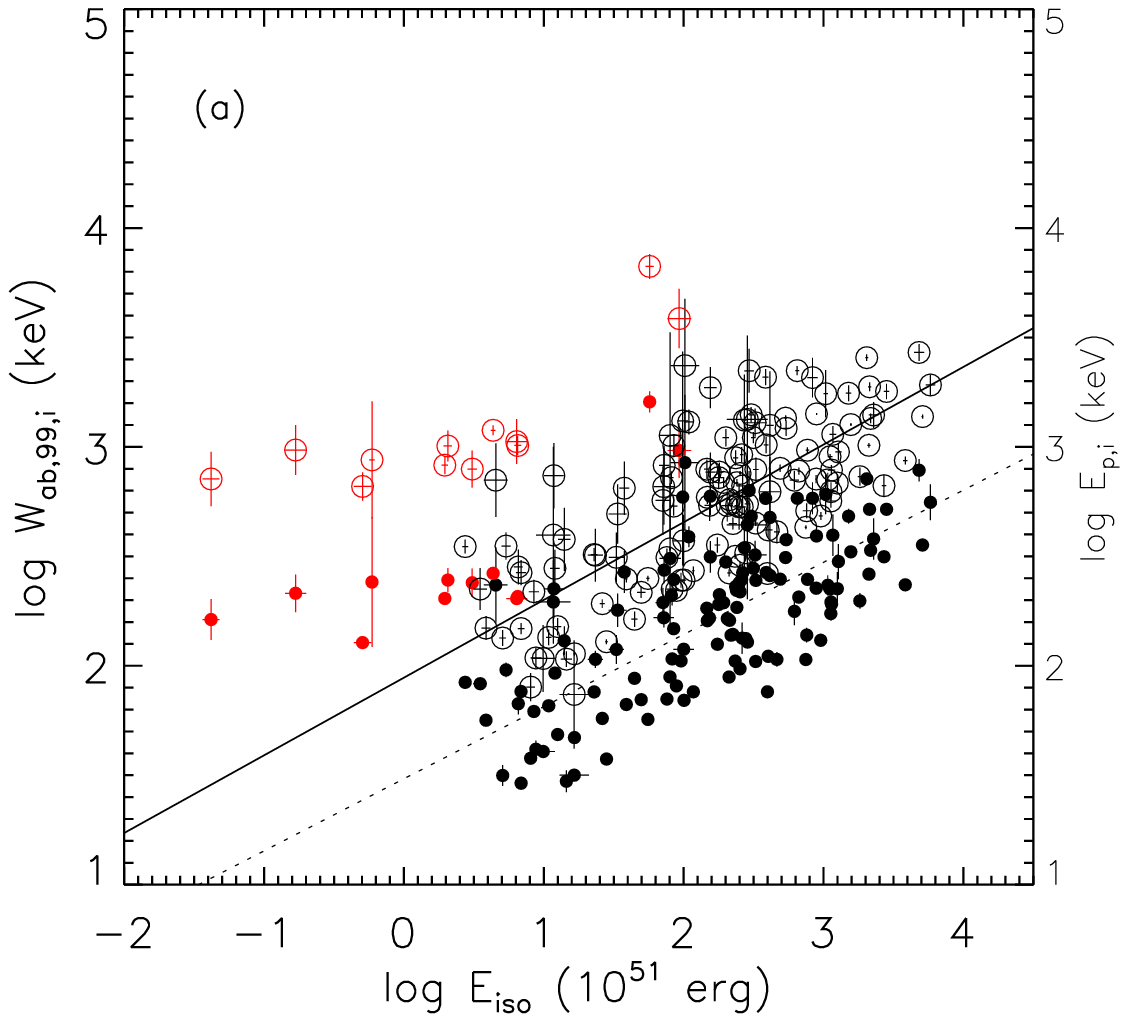}
  \includegraphics[width=0.48\textwidth]{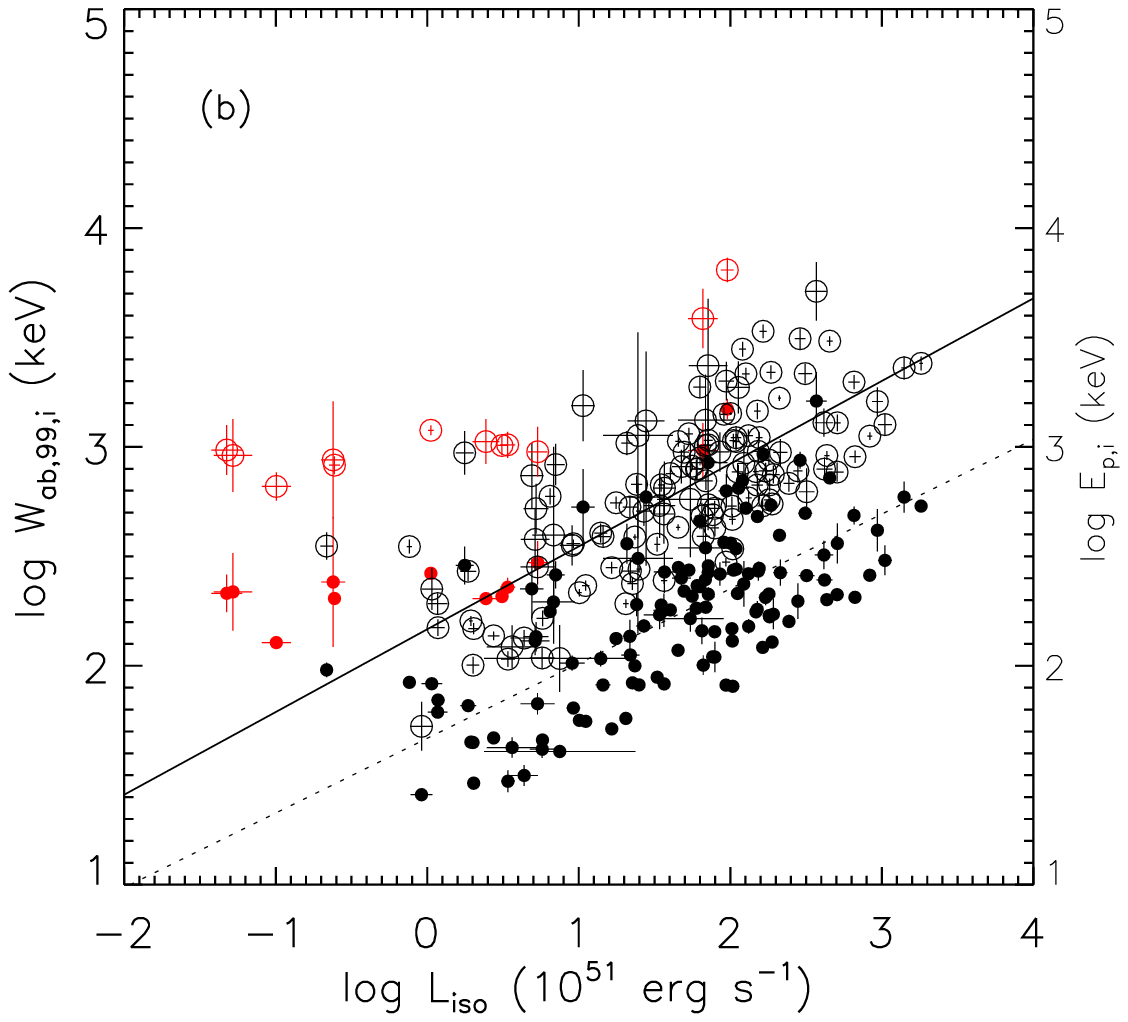}
  \caption{$W_{ab,99,i}$ vs. $E_{iso}$ for the F spectra (a) and $W_{ab,99,i}$ vs. $L_{iso}$ for the P spectra (b), along with the Amati and Yonetoku relations. All the symbols are same as Figure 7.}
\label{fig:example_figure}
\end{figure*}

\subsection{Comparison of $width-E_{iso}$ and $width-L_{iso}$ with Amati and Yonetoku relations}
Since the correlated properties of $W_{ab,i}-E_{iso}$ and $W_{ab,i}- L_{iso}$ pairs are very similar to the well-known Amati relation
(Amati et al. 2002) and Yonetoku relation (Yonetoku et al. 2004) we would like to compare the correlation properties of the rest-frame width-$E_{iso}$ and rest-frame width-$L_{iso}$ with the Amati and Yonetoku relations. For the sake of comparison Figures 7-12 and Table 4 also give the correlated properties of $E_{p,i}-E_{iso}$ and $E_{p,i}- L_{iso}$ for both the F and P spectra.

It can be seen form Figure 7 and Table 4, for the case of the F spectra, the correlation of $W_{ab,50,i}-E_{iso}$ ($\rho=$0.80,
P$<$0.0001) is much stronger than Amati relation ($\rho=$0.69, P$<$0.0001). However, the dispersions of the $W_{ab,i}-E_{iso}$ (0.300 $\pm$ 0.022) is also slightly greater than the Amati relation (0.227 $\pm$ 0.016) obtained with the same spectral data. Nevertheless, it is found that the two correlated properties ($\rho=$0.71, P$<$0.0001 for $W_{ab,50,i}-L_{iso}$) are almost consist with Yonetoku relation ($\rho=$0.69, P$<$0.0001) when comparing the P spectra. But the dispersions of the $W_{ab,i}- L_{iso}$ (0.35 $\pm$ 0.03) for the F spectra and P spectra are also slightly greater than the Yonetoku relation (0.24 $\pm$ 0.02) obtained with the same spectral data.

We then compare other width$-E_{iso}$ and width$-L_{iso}$ with Amati relation and Yonetoku relation, respectively. It is very interesting that, for both F spectra and P spectra, we can also find from Figures 8-12 that rest-frame absolute spectral width approach $E_{p,i}$ with the decrease of the widths until the $W_{ab,90,i}$ almost overlapping with $E_{p,i}$. Meantime, the best regression line of $W_{ab,90,i}-E_{iso}$ and $W_{ab,90,i}-L_{iso}$ also almost overlap with those of Amati and Yonetoku relations, respectively. Moreover, it can be observed from Table 4 that the derived slopes, intercept and intrinsic scatter of $W_{ab,90,i}-E_{iso}$ and $W_{ab,90,i}-L_{iso}$ are also very close to those of the Amati and Yonetoku relations, respectively. It is also can be seen that the $W_{ab,95,i}-E_{iso}$, $W_{ab,95,i}-L_{iso}$ deviate from the Amati and Yonetoku relations and the correlation coefficients and the significances also decrease when the absolute spectral width decreases.

These striking similarities between $W_{ab,90,i}-E_{iso}$ and the Amati relation as well as $W_{ab,90,i}-L_{iso}$ and Yonetoku relation seem to show that the Amati and Yonetoku relations are the special cases of the width-$E_{iso}$ and width-$L_{iso}$. In addition, it also can be seen that the correlations of Amati relation is not the strongest among the width-$E_{iso}$ relations when the scatter is very close. While the Yonetoku relation also does not show the most significant correlation properties comparing with the width-$L_{iso}$.

\startlongtable
\begin{deluxetable}{cccccccc}
\tablecaption{Correlation analysis results of six absolute spectral widths in the rest frame and $E_{iso}$ as well as $L_{iso}$.}
\tablehead{\colhead{Correlation}  & \colhead{number}  &\colhead{$\rho$} &\colhead{$P$} &\colhead{a} &\colhead{b} &\colhead{$\sigma$}}
\startdata
\hline
$W_{ab,50,i,F}-E_{iso}$   &   129  & 0.80  &   $<$0.0001     & 0.57$\pm$0.04   & 2.07 $\pm$0.09  & 0.30$\pm$ 0.02\\
$W_{ab,75,i,F}-E_{iso}$   &   129  & 0.75  &   $<$0.0001     & 0.42$\pm$0.03   & 2.08 $\pm$0.07  & 0.240$\pm$0.02 \\
$W_{ab,85,i,F}-E_{iso}$   &   129  & 0.72  &   $<$0.0001     & 0.38$\pm$0.03   & 2.02 $\pm$0.07  & 0.23$\pm$0.02\\
$W_{ab,90,i,F}-E_{iso}$   &   129  & 0.71  &   $<$0.0001     & 0.37$\pm$0.03   & 1.94 $\pm$ 0.07  & 0.23$\pm$0.02\\
$W_{ab,95,i,F}-E_{iso}$   &   129  & 0.69  &   $<$0.0001     & 0.35$\pm$0.03   & 1.81 $\pm$0.06  & 0.23$\pm$0.02\\
$W_{ab,99,i,F}-E_{iso}$   &   129  & 0.66  &   $<$0.0001     & 0.33$\pm$0.03   & 1.48 $\pm$0.06  & 0.23$\pm$0.02\\
$E_{p,i,F}-E_{iso}$       &   129  & 0.69  &   $<$0.0001     & 0.36$\pm$0.03   & 1.95$\pm$0.07 & 0.23$\pm$0.02\\
$W_{ab,50,i,P}-L_{iso}$   &   133  & 0.71  & $<$0.0001     & 0.55$\pm$0.05  & 2.44 $\pm$0.08  & 0.35$\pm$0.03\\
$W_{ab,75,i,P}-L_{iso}$   &   133  & 0.69  & $<$0.0001     & 0.41$\pm$0.04  & 2.35 $\pm$0.06  & 0.26$\pm$0.02 \\
$W_{ab,85,i,P}-L_{iso}$   &   133  & 0.68  & $<$0.0001     & 0.39$\pm$0.03  & 2.24 $\pm$0.06  & 0.25$\pm$0.02\\
$W_{ab,90,i,P}-L_{iso}$   &   133  & 0.67  & $<$0.0001     & 0.37$\pm$0.03  & 2.16 $\pm$0.06  & 0.25$\pm$0.02\\
$W_{ab,95,i,P}-L_{iso}$   &   133  & 0.65  & $<$0.0001     & 0.35$\pm$0.03  & 2.01 $\pm$0.06  & 0.25$\pm$0.02\\
$W_{ab,99,i,P}-L_{iso}$   &   133  & 0.64  & $<$0.0001     & 0.34$\pm$0.03  & 1.67 $\pm$0.05  & 0.25$\pm$0.02\\
  $E_{p,i,P}-L_{iso}$     &   133  & 0.69  & $<$0.0001     & 0.38$\pm$0.03  & 2.17 $\pm$0.05  & 0.24$\pm$0.02\\
\enddata
\tablecomments{The F and P correspond to the F spectra and P spectra, respectively.}
\end{deluxetable}

\subsection{Is the $E_{p,i}$ a best energy correlated with  $E_{iso}$ and $L_{iso}$?}
As we known $E_{2}$, $E_{1}$, and $E_{p}$ are the location energy of $EF_{E}$ spectra. The absolute spectral width is the difference between location energy $E_{2}$ and $E_{1}$ and it must be relate to $E_{2}$ and $E_{1}$. For the wide spectra the absolute width should be strongly correlate with $E_{2}$. While the absolute width should be also correlated with $E_{1}$ for the narrower absolute spectral width. The corresponding Spearman correlation coefficients $\rho$ and P are shown in Table 5 and the absolute spectral widths indeed correlated with both $E_{2}$ and $E_{1}$.

As a consequence we suspect there must be correlations between rest-frame $E_{2,i}$, $E_{1,i}$ and $E_{iso}$ as well as $L_{iso}$. So we check the correlations between the two location energies $E_{2,i}$ and $E_{1,i}$ in various spectral widths and the $E_{iso}$ as well as $L_{iso}$. The correlations between them are shown in Table 6 and Figures 13-24. It is found that both $E_{1,i}$ and $E_{2,i}$ in various spectral widths are also strongly correlated with $E_{iso}$ as well as $L_{iso}$. The short bursts for $E_{2,i}-E_{iso}$ and $E_{1,i}-E_{iso}$ deviate from the long bursts for both F spectra and P spectra. While a few short bursts for $E_{2,i}-L_{iso}$ and $E_{1,i}-L_{iso}$ deviate from the long bursts for both F spectra and P spectra. The correlated properties are very similar to the correlations between various absolute spectral widths and $E_{iso}$ and $L_{iso}$ above. As the widths decrease the both $E_{2,i}$ and $E_{1,i}$ approach to $E_{p,i}$ (see, Figures 13-24, Tables 6 and 7). The correlation coefficients, significances, intrinsic scatters also decrease and the best regression lines of $E_{2,99,i}-E_{iso}$, $E_{2,99,i}-L_{iso}$, $E_{1,99,i}-E_{iso}$ and $E_{1,99,i}-L_{iso}$ almost overlap with those of Amati and Yonetoku relation, respectively.

In addition, we find there are several correlations between location energy ($E_{2,i}$, $E_{1,i}$) and $-E_{iso}$, such as $E_{2,95,i,F}- E_{iso}$, $E_{2,90,i,F}- E_{iso}$, are stronger than that of the Amati relation. Likewise, we also find there are several correlations of  $E_{2,i}-L_{iso}$, $E_{1,i}-L_{iso}$, such as $E_{2,85,i,P}- L_{iso}$, $E_{1,95,i,P}-L_{iso}$, are stronger than that of the Yonetoku relation. Therefore the Amati and Yonetoku relations are not most significant correlated relationships among the relations of $E_{2,i}-E_{iso}$, $E_{1,i}-E_{iso}$, $E_{2,i}-L_{iso}$, $E_{1,i}-L_{iso}$. These seem to also show that Amati and Yonetoku relations are only the special correlations between location energy in the GRB spectra and $E_{iso}$ as well as $L_{iso}$.

\begin{deluxetable}{cccccc}
\tablecaption{Correlation analysis results of six location energies $E_{2}$, $E_{1}$ and the corresponding six absolute spectral widths for the case of F spectra.}
\tablehead{\colhead{Correlation}   &\colhead{$\rho$} &\colhead{$P$} &\colhead{$\rho$} &\colhead{P}}
\startdata
\hline
  $E_{2,50}-W_{ab,50}$  & 1.00    &   $<$ 0.0001            &  1.00 & $<$0.0001  \\
  $E_{1,50}-W_{ab,50}$  & 0.69    &   $<$ 0.0001  &  0.71 & $<$ 0.0001 \\
  $E_{2,75}-W_{ab,75}$  & 0.99    &   $<$ 0.0001           &  0.99 & $<$ 0.0001 \\
  $E_{1,75}-W_{ab,75}$  & 0.81    &   $<$ 0.0001  &  0.85 & $<$ 0.0001\\
  $E_{2,85}-W_{ab,85}$  & 0.99    &   $<$ 0.0001            &  1.00 & $<$ 0.0001 \\
  $E_{1,85}-W_{ab,85}$  & 0.86    &   $<$ 0.0001  &  0.91 & $<$ 0.0001\\
  $E_{2,90}-W_{ab,90}$  & 0.99    &   $<$ 0.0001            &  0.99 & $<$ 0.0001  \\
  $E_{1,90}-W_{ab,90}$  & 0.89    &   $<$ 0.0001            &  0.93 & $<$ 0.0001\\
  $E_{2,95}-W_{ab,95}$  & 0.98    &   $<$ 0.0001            &  0.99 & $<$ 0.0001  \\
  $E_{1,95}-W_{ab,95}$  & 0.92    &   $<$ 0.0001            &  0.96 & $<$ 0.0001 \\
  $E_{2,99}-W_{ab,99}$  & 0.97    &   $<$ 0.0001           &  0.98 & $<$ 0.0001 \\
  $E_{1,99}-W_{ab,99}$  & 0.95    &   $<$ 0.0001            &  0.97 & $<$ 0.0001 \\
\enddata
\end{deluxetable}

\startlongtable
\begin{deluxetable}{cccccccc}
\tablecaption{Correlation analysis results of six location energies $E_{2}$ in the rest frame and $E_{iso}$ as well as $L_{iso}$.}
\tablehead{\colhead{Correlation}  & \colhead{number}  &\colhead{$\rho$} &\colhead{$P$} &\colhead{a} &\colhead{b} &\colhead{$\sigma$}}
\startdata
\hline
$E_{2,50,i,F}-E_{iso}$   &   129  & 0.80  & $<$0.0001     & 0.59$\pm$0.04   & 2.08 $\pm$0.09  & 0.33$\pm$0.02\\
$E_{2,75,i,F}-E_{iso}$   &   129  & 0.76  & $<$0.0001     & 0.48$\pm$0.04   & 2.08 $\pm$0.09  & 0.30$\pm$0.02\\
$E_{2,85,i,F}-E_{iso}$   &   129  & 0.73  & $<$0.0001     & 0.39$\pm$0.03   & 2.15 $\pm$0.07  & 0.24$\pm$0.02\\
$E_{2,90,i,F}-E_{iso}$   &   129  & 0.72  & $<$0.0001     & 0.38$\pm$0.03   & 2.13 $\pm$0.07  & 0.23$\pm$0.02\\
$E_{2,95,i,F}-E_{iso}$   &   129  & 0.70  & $<$0.0001     & 0.36$\pm$0.03   & 2.09 $\pm$0.06  & 0.23$\pm$0.02\\
$E_{2,99,i,F}-E_{iso}$   &   129  & 0.68  & $<$0.0001     & 0.35$\pm$0.03   & 2.02 $\pm$0.06  & 0.23$\pm$0.02\\
$E_{p,i,F}-E_{iso}$      &   129  & 0.69  & $<$0.0001     & 0.36$\pm$0.03   & 1.95$\pm$0.07   & 0.23$\pm$0.02\\
$E_{2,50,i,P}-L_{iso}$   &   133  & 0.72  & $<$0.0001     & 0.56$\pm$0.05  & 2.49 $\pm$0.08  & 0.36$\pm$0.03\\
$E_{2,75,i,P}-L_{iso}$   &   133  & 0.70  & $<$0.0001     & 0.45$\pm$0.04  & 2.43 $\pm$0.07  & 0.32$\pm$0.02\\
$E_{2,85,i,P}-L_{iso}$   &   133  & 0.70  & $<$0.0001     & 0.40$\pm$0.03  & 2.40 $\pm$0.06  & 0.26$\pm$0.02\\
$E_{2,90,i,P}-L_{iso}$   &   133  & 0.69  & $<$0.0001     & 0.39$\pm$0.03  & 2.36 $\pm$0.06  & 0.25$\pm$0.02\\
$E_{2,95,i,P}-L_{iso}$   &   133  & 0.68  & $<$0.0001     & 0.38$\pm$0.03  & 2.31 $\pm$0.06  & 0.25$\pm$0.02\\
$E_{2,99,i,P}-L_{iso}$   &   133  & 0.68  & $<$0.0001     & 0.37$\pm$0.03  & 2.24 $\pm$0.05  & 0.24$\pm$0.02\\
  $E_{p,i,P}-L_{iso}$    &   133  & 0.69  & $<$0.0001     & 0.38$\pm$0.03  & 2.17 $\pm$0.05  & 0.24$\pm$0.02\\
\enddata
\tablecomments{The F and P correspond to the F spectra and P spectra, respectively.}
\end{deluxetable}

\startlongtable
\begin{deluxetable}{cccccccc}
\tablecaption{Correlation analysis results of six location energy $E_{1}$ in the rest frame and $E_{iso}$ as well as $L_{iso}$.}
\tablehead{\colhead{Correlation}  & \colhead{number}  &\colhead{$\rho$} &\colhead{$P$} &\colhead{a} &\colhead{b} &\colhead{$\sigma$}}
\startdata
\hline
$E_{1,50,i,F}-E_{iso}$   &   129  & 0.67  & $<$0.0001     & 0.41$\pm$0.04   & 1.16 $\pm$0.09  & 0.28$\pm$0.02\\
$E_{1,75,i,F}-E_{iso}$   &   129  & 0.69  & $<$0.0001     & 0.39$\pm$0.03   & 1.46 $\pm$0.08  & 0.26$\pm$0.02\\
$E_{1,85,i,F}-E_{iso}$   &   129  & 0.69  & $<$0.0001     & 0.38$\pm$0.03   & 1.59 $\pm$0.08  & 0.25$\pm$0.02\\
$E_{1,90,i,F}-E_{iso}$   &   129  & 0.69  & $<$0.0001     & 0.38$\pm$0.03   & 1.67 $\pm$0.07  & 0.24$\pm$0.02\\
$E_{1,95,i,F}-E_{iso}$   &   129  & 0.69  & $<$0.0001     & 0.37$\pm$0.03   & 1.76 $\pm$0.07  & 0.24$\pm$0.02\\
$E_{1,99,i,F}-E_{iso}$   &   129  & 0.69  & $<$0.0001     & 0.36$\pm$0.03   & 1.87 $\pm$0.07  & 0.23$\pm$0.02\\
$E_{p,i,F}-E_{iso}$      &   129  & 0.69  & $<$0.0001     & 0.36$\pm$0.03   & 1.95 $\pm$0.07  & 0.23$\pm$0.02\\
$E_{1,50,i,P}-L_{iso}$   &   133  & 0.72  & $<$0.0001     & 0.42$\pm$0.03  & 1.49 $\pm$0.06  & 0.25$\pm$0.02\\
$E_{1,75,i,P}-L_{iso}$   &   133  & 0.71  & $<$0.0001     & 0.41$\pm$0.03  & 1.75 $\pm$0.06  & 0.26$\pm$0.02 \\
$E_{1,85,i,P}-L_{iso}$   &   133  & 0.71  & $<$0.0001     & 0.40$\pm$0.03  & 1.86 $\pm$0.06  & 0.25$\pm$0.02\\
$E_{1,90,i,P}-L_{iso}$   &   133  & 0.70  & $<$0.0001     & 0.40$\pm$0.03  & 1.92 $\pm$0.06  & 0.25$\pm$0.02\\
$E_{1,95,i,P}-L_{iso}$   &   133  & 0.70  & $<$0.0001     & 0.39$\pm$0.03  & 2.00 $\pm$0.06  & 0.24$\pm$0.02\\
$E_{1,99,i,P}-L_{iso}$   &   133  & 0.69  & $<$0.0001     & 0.38$\pm$0.03  & 2.10 $\pm$0.06  & 0.23$\pm$0.02\\
  $E_{p,i,P}-L_{iso}$    &   133  & 0.69  & $<$0.0001     & 0.38$\pm$0.03  & 2.17 $\pm$0.05  & 0.24$\pm$0.02\\
\enddata
\tablecomments{The F and P correspond to the F spectra and P spectra, respectively.}
\end{deluxetable}

\begin{figure*}
  \includegraphics[width=0.48\textwidth]{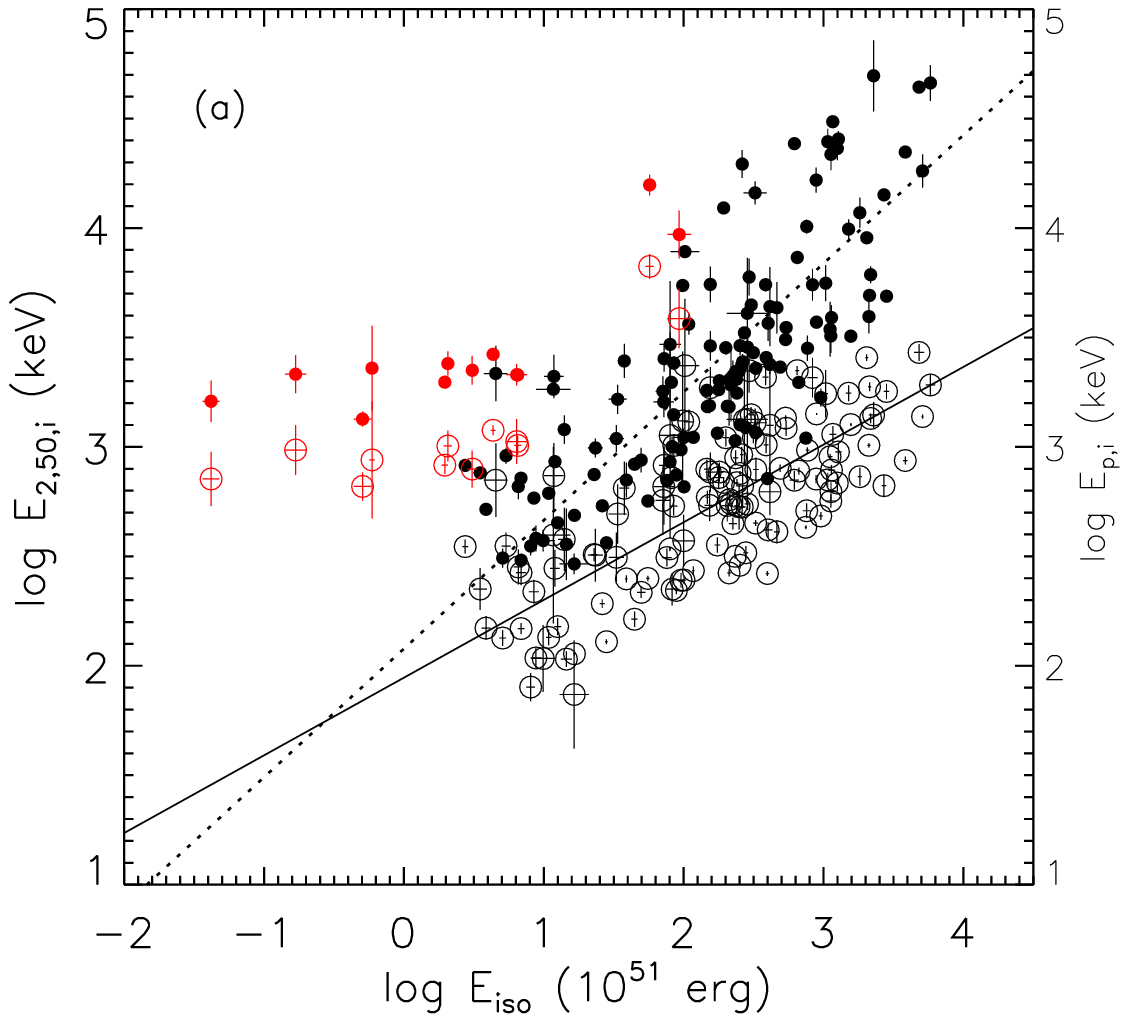}
  \includegraphics[width=0.48\textwidth]{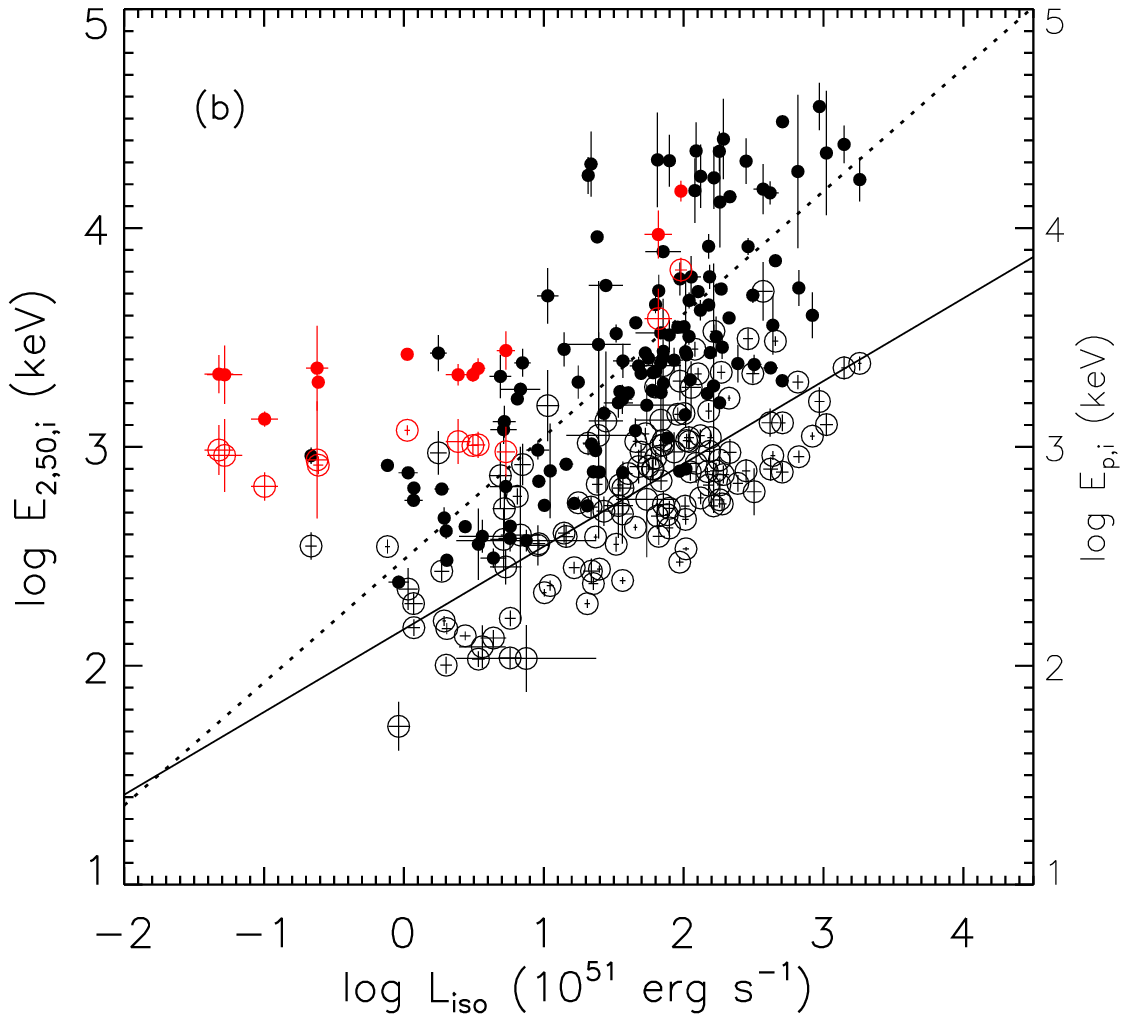}
  \caption{$E_{2,50,i}$ vs. $E_{iso}$ for the F spectra (a) and P spectra (b) and $E_{2,50,i}$ vs. $L_{iso}$ for the F spectra (c) and P spectra (d), along with the Amati and Yonetoku relations for the F and P spectra. All the symbols are same as Figure 7.}
\label{fig:example_figure}
\end{figure*}

\begin{figure*}
  \includegraphics[width=0.48\textwidth]{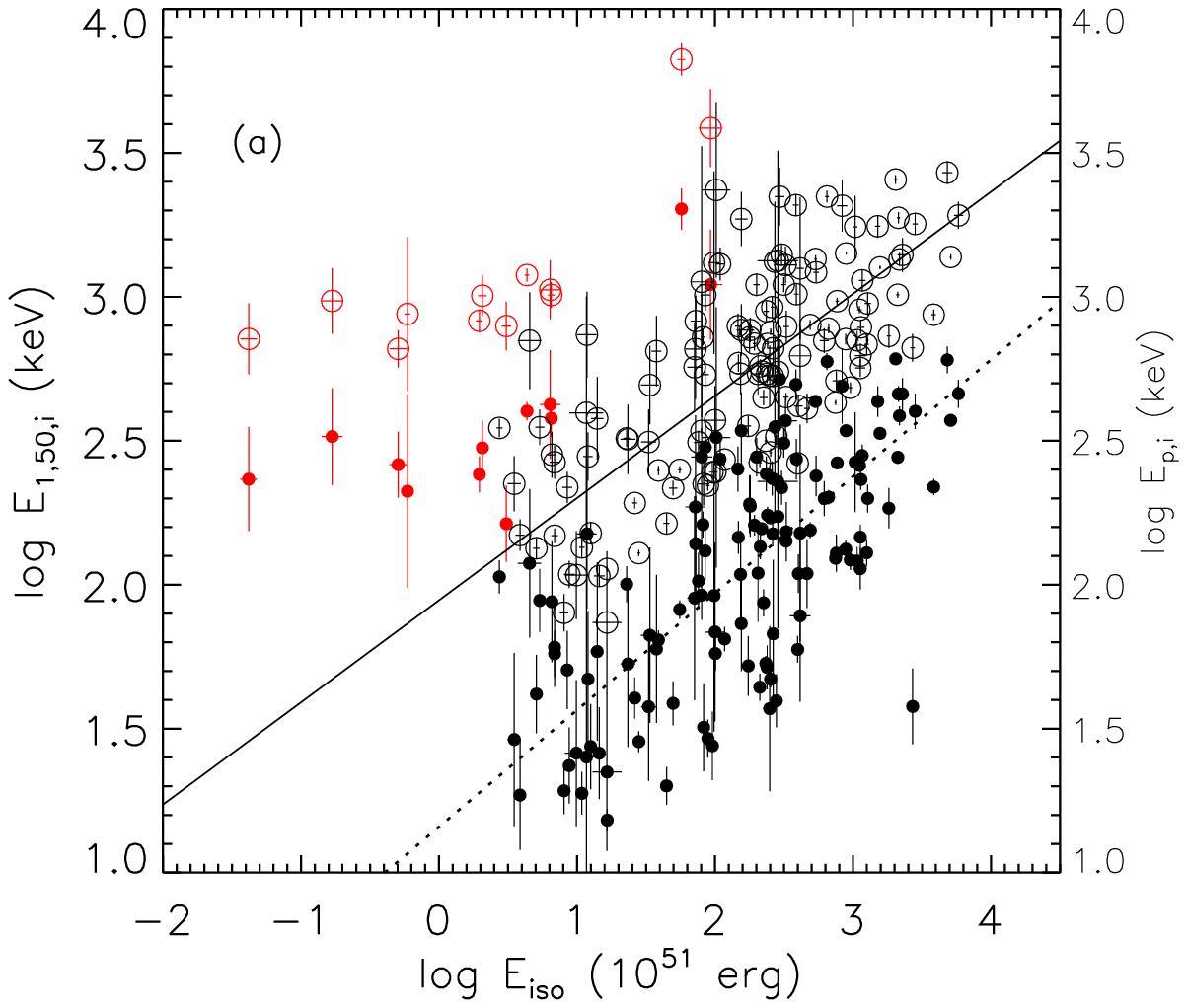}
  \includegraphics[width=0.48\textwidth]{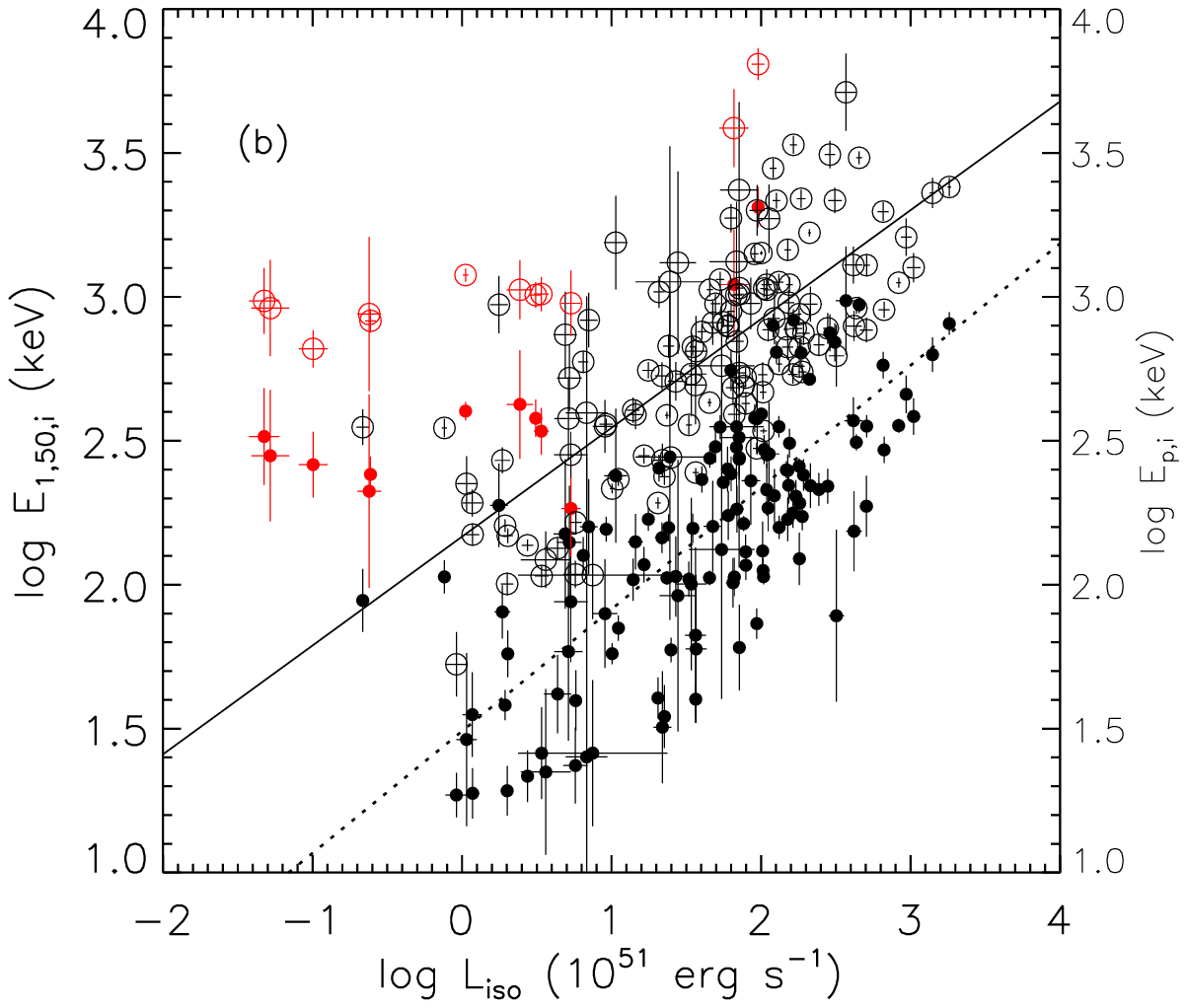}
  \caption{$E_{1,50,i}$ vs. $E_{iso}$ for the F spectra (a) and P spectra (b) and $E_{1,50,i}$ vs. $L_{iso}$ for the F spectra (c) and P spectra (d), along with the Amati and Yonetoku relations for the F and P spectra. All the symbols are same as Figure 7.}
\label{fig:example_figure}
\end{figure*}

\begin{figure*}
  \includegraphics[width=0.48\textwidth]{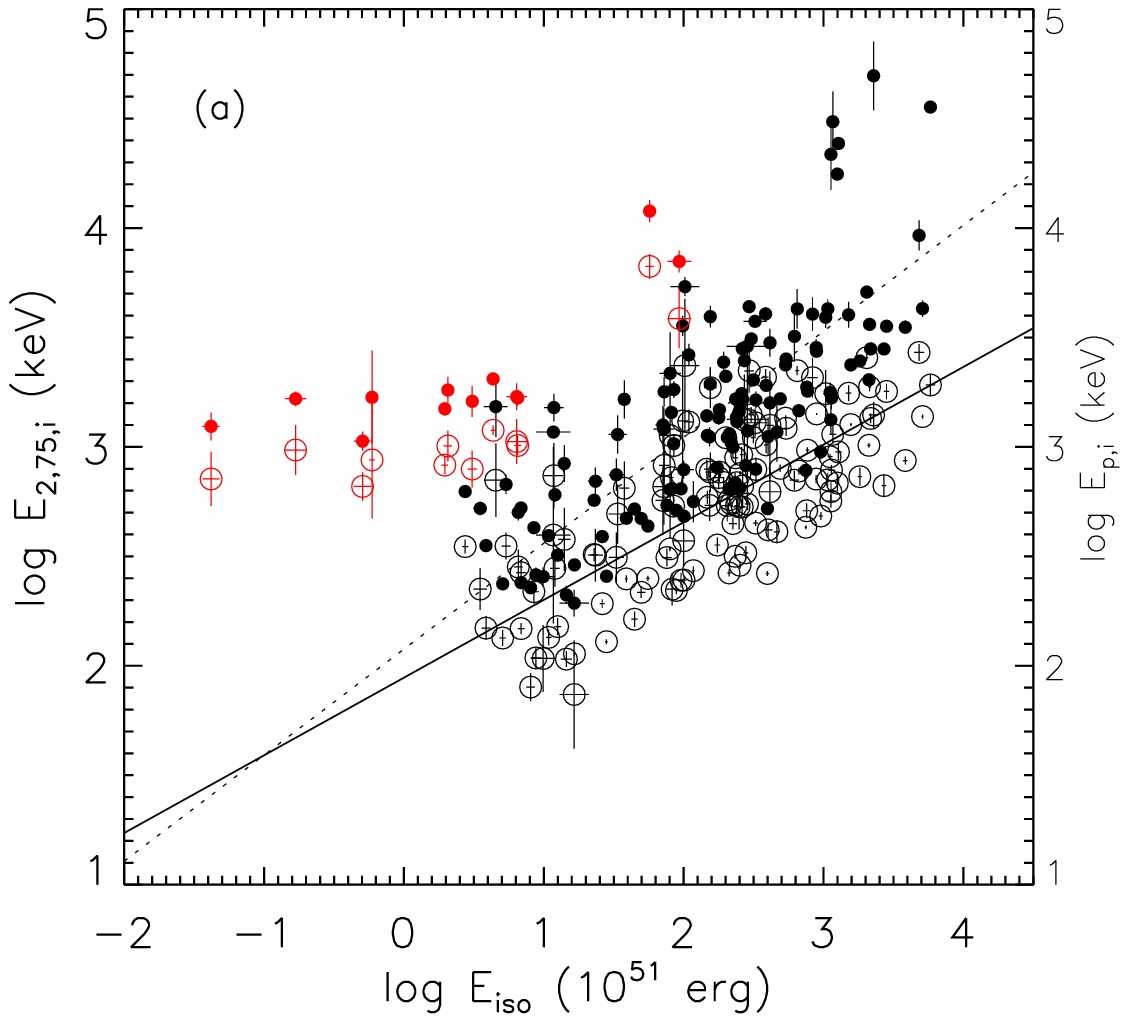}
  \includegraphics[width=0.48\textwidth]{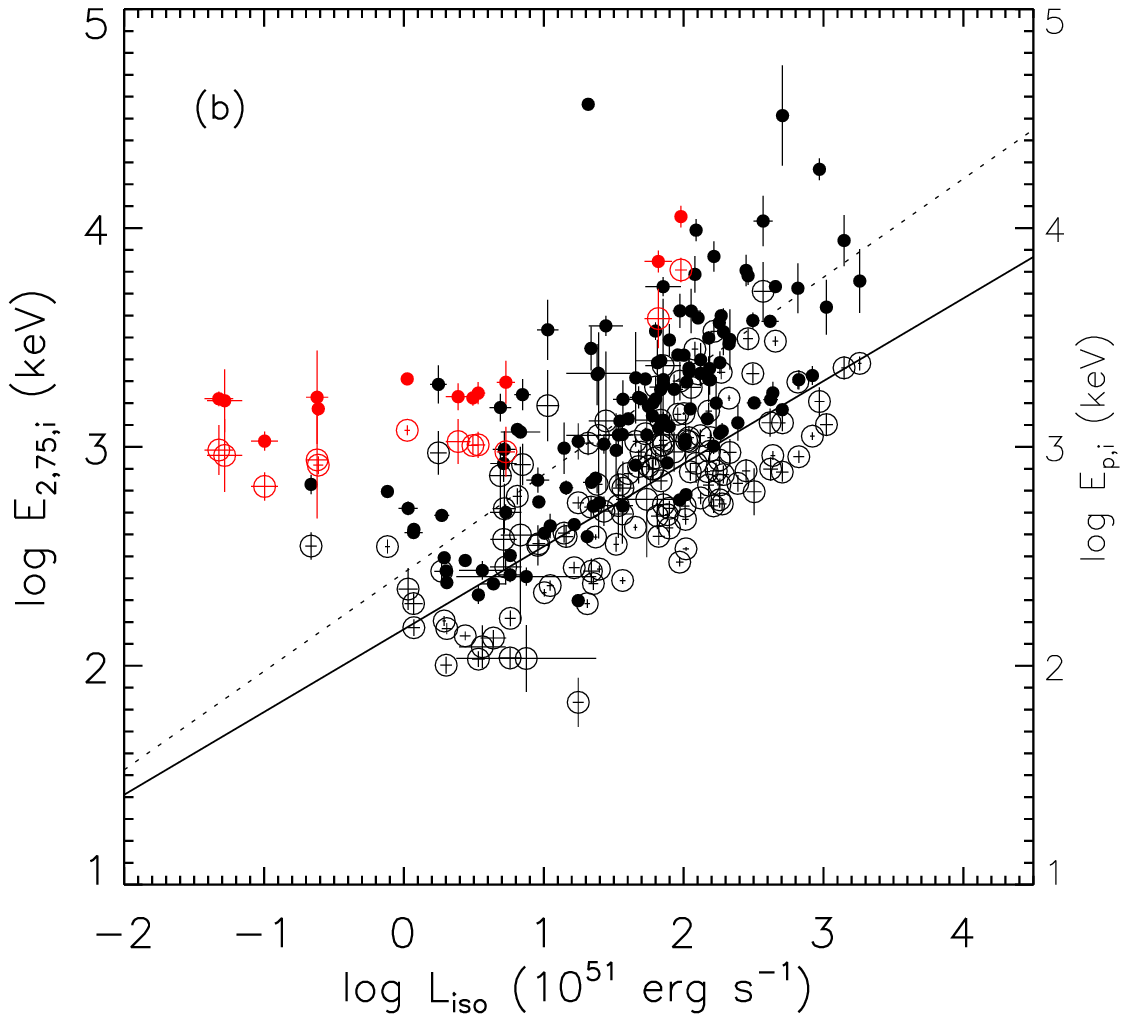}
  \caption{$E_{2,75,i}$ vs. $E_{iso}$ for the F spectra (a) and P spectra (b) and $E_{2,75,i}$ vs. $L_{iso}$ for the F spectra (c) and P spectra (d), along with the Amati and Yonetoku relations for the F and P spectra. All the symbols are same as Figure 7.}
\label{fig:example_figure}
\end{figure*}

\begin{figure*}
  \includegraphics[width=0.48\textwidth]{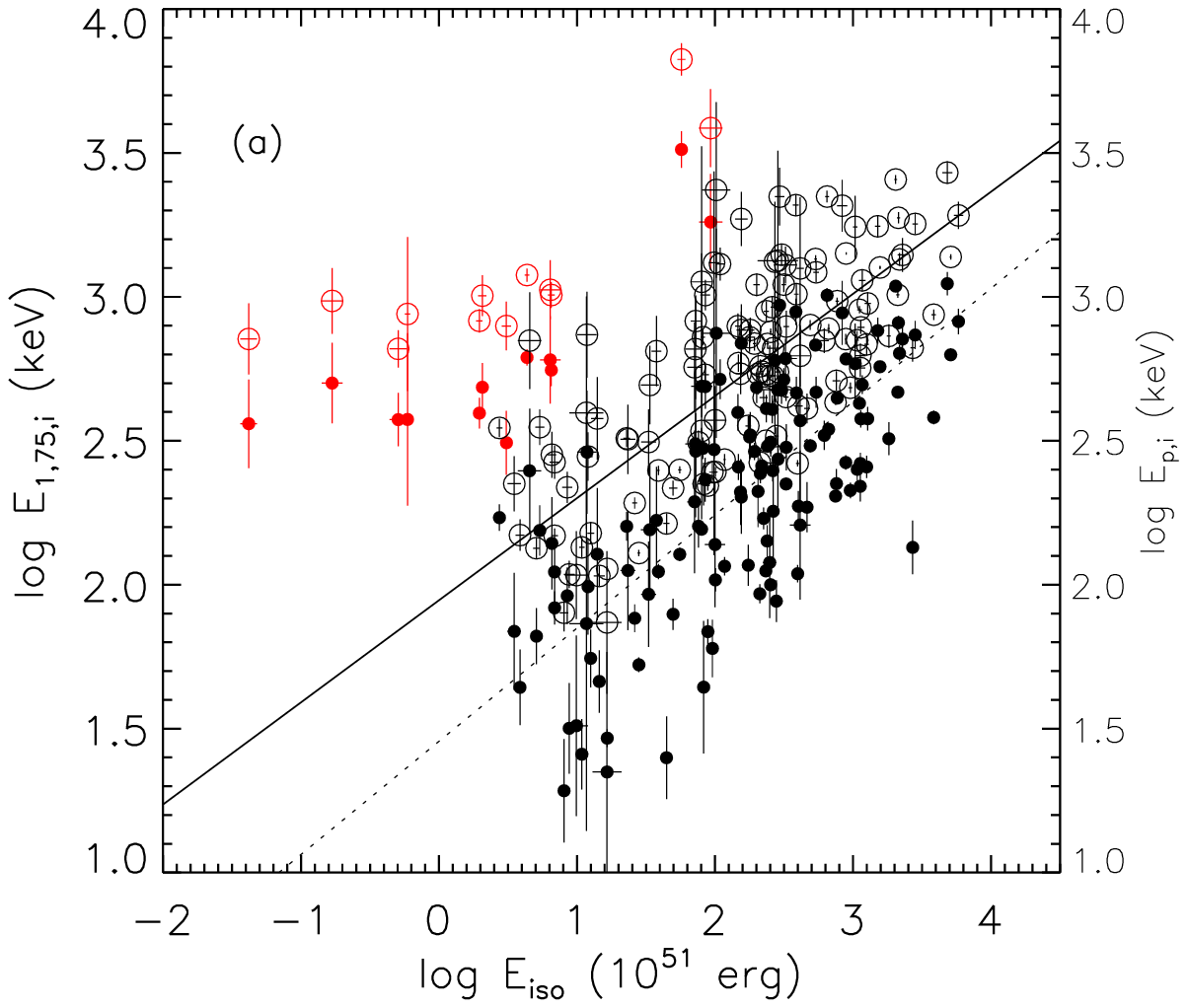}
  \includegraphics[width=0.48\textwidth]{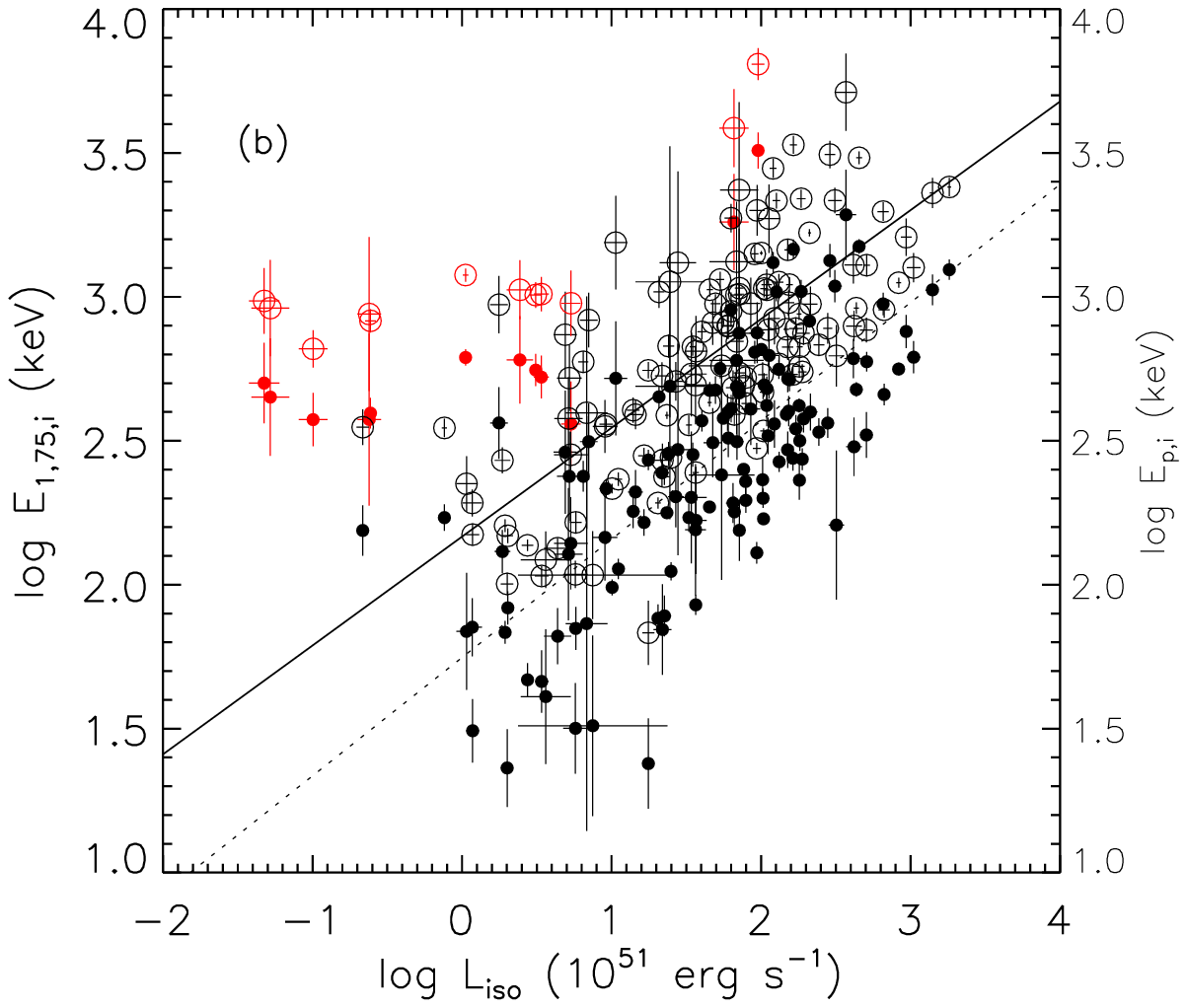}
  \caption{$E_{1,75,i}$ vs. $E_{iso}$ for the F spectra (a) and P spectra (b) and $E_{1,75,i}$ vs. $L_{iso}$ for the F spectra (c) and P spectra (d), along with the Amati and Yonetoku relations for the F and P spectra. All the symbols are same as Figure 7.}
\label{fig:example_figure}
\end{figure*}

\begin{figure*}
  \includegraphics[width=0.48\textwidth]{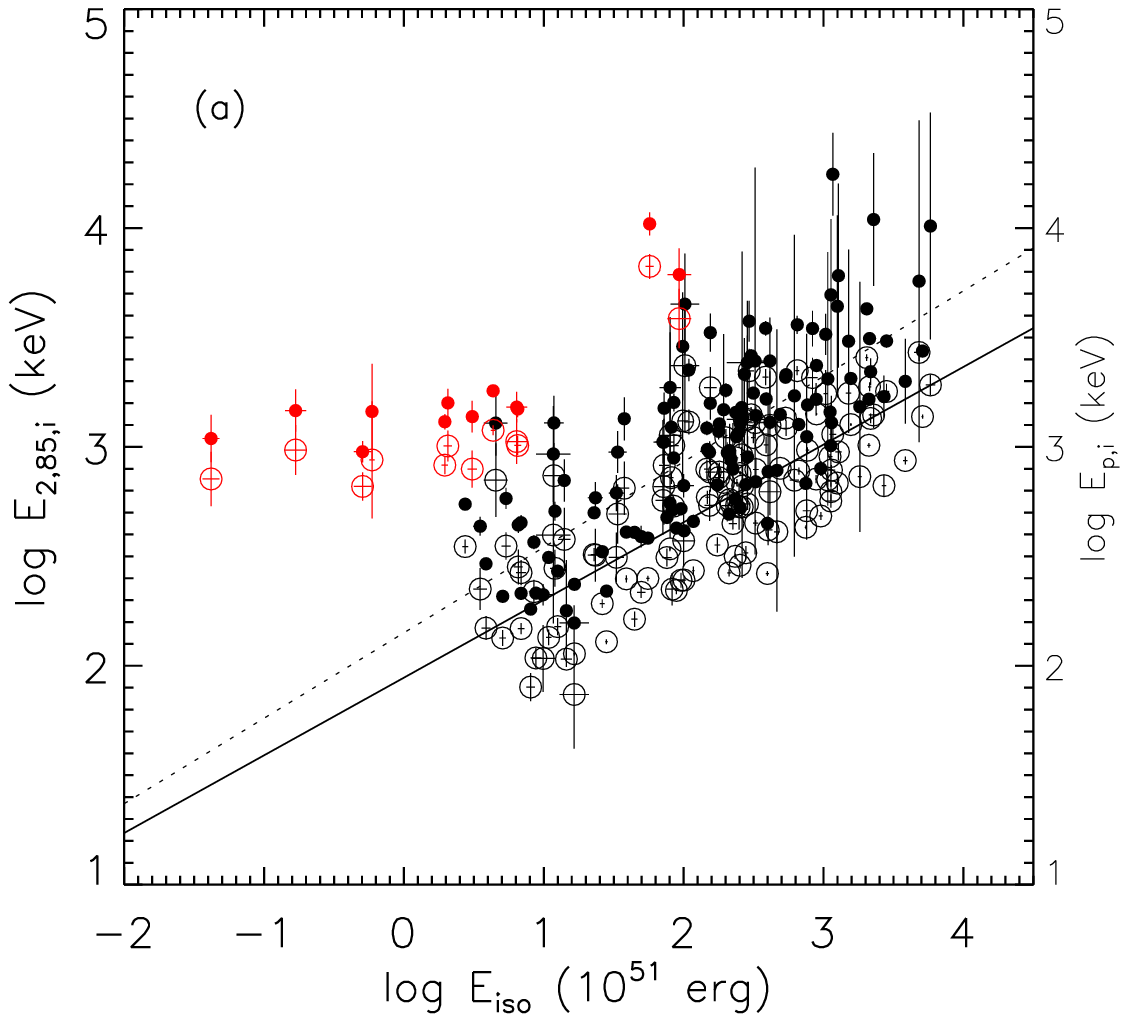}
  \includegraphics[width=0.48\textwidth]{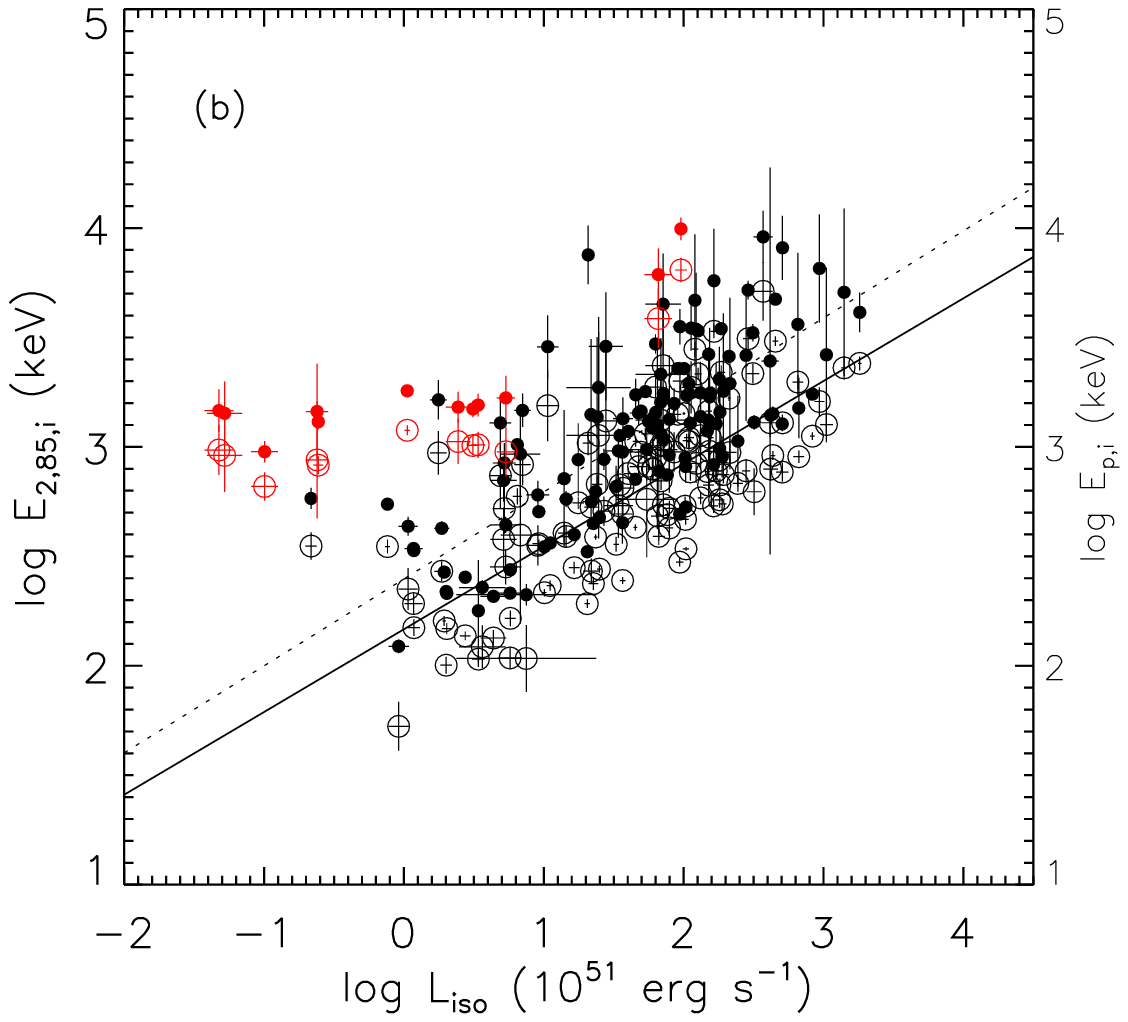}
  \caption{$E_{2,85,i}$ vs. $E_{iso}$ for the F spectra (a) and P spectra (b) and $E_{2,85,i}$ vs. $L_{iso}$ for the F spectra (c) and P spectra (d), along with the Amati and Yonetoku relations for the F and P spectra. All the symbols are same as Figure 7.}
\label{fig:example_figure}
\end{figure*}

\begin{figure*}
  \includegraphics[width=0.48\textwidth]{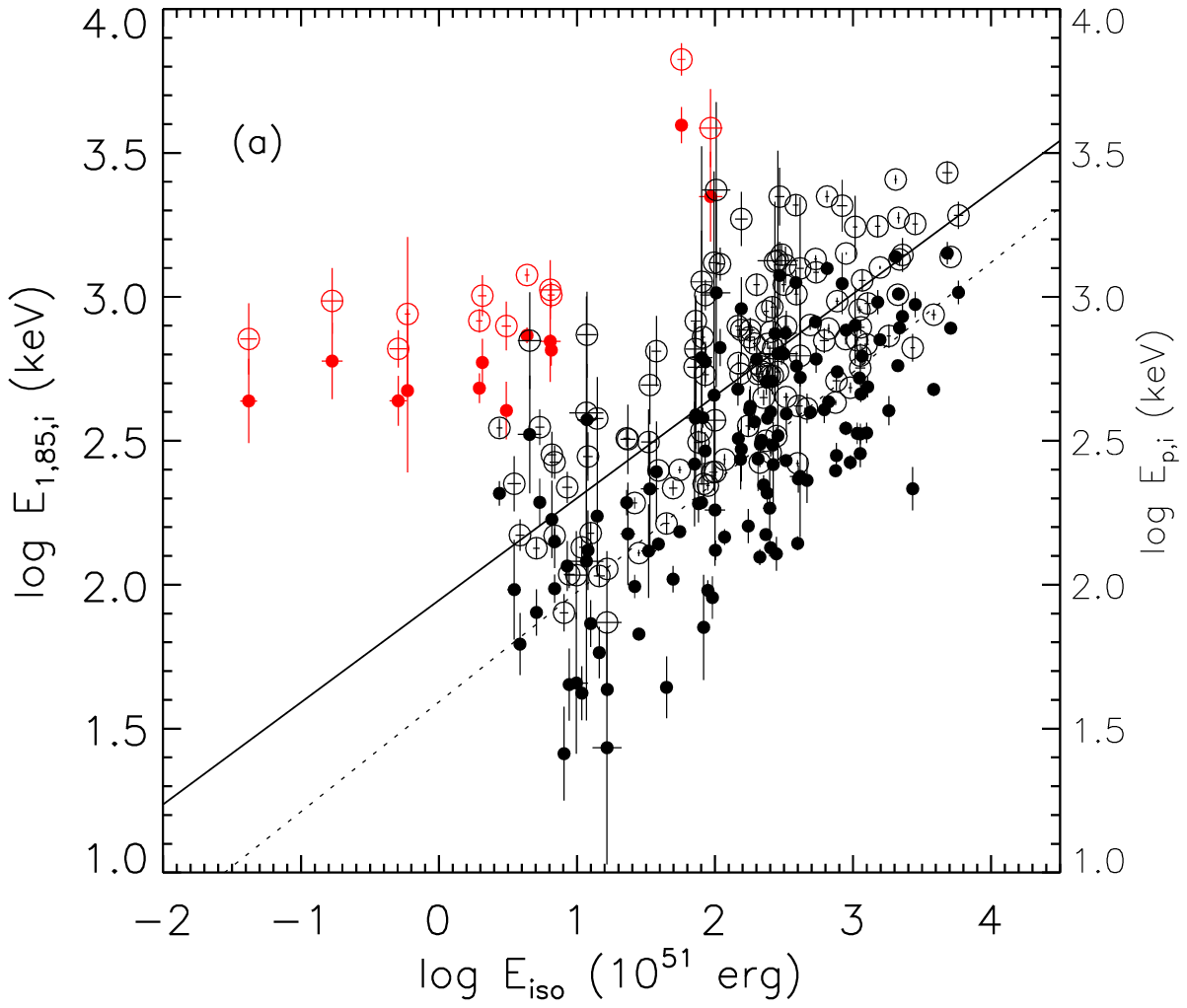}
  \includegraphics[width=0.48\textwidth]{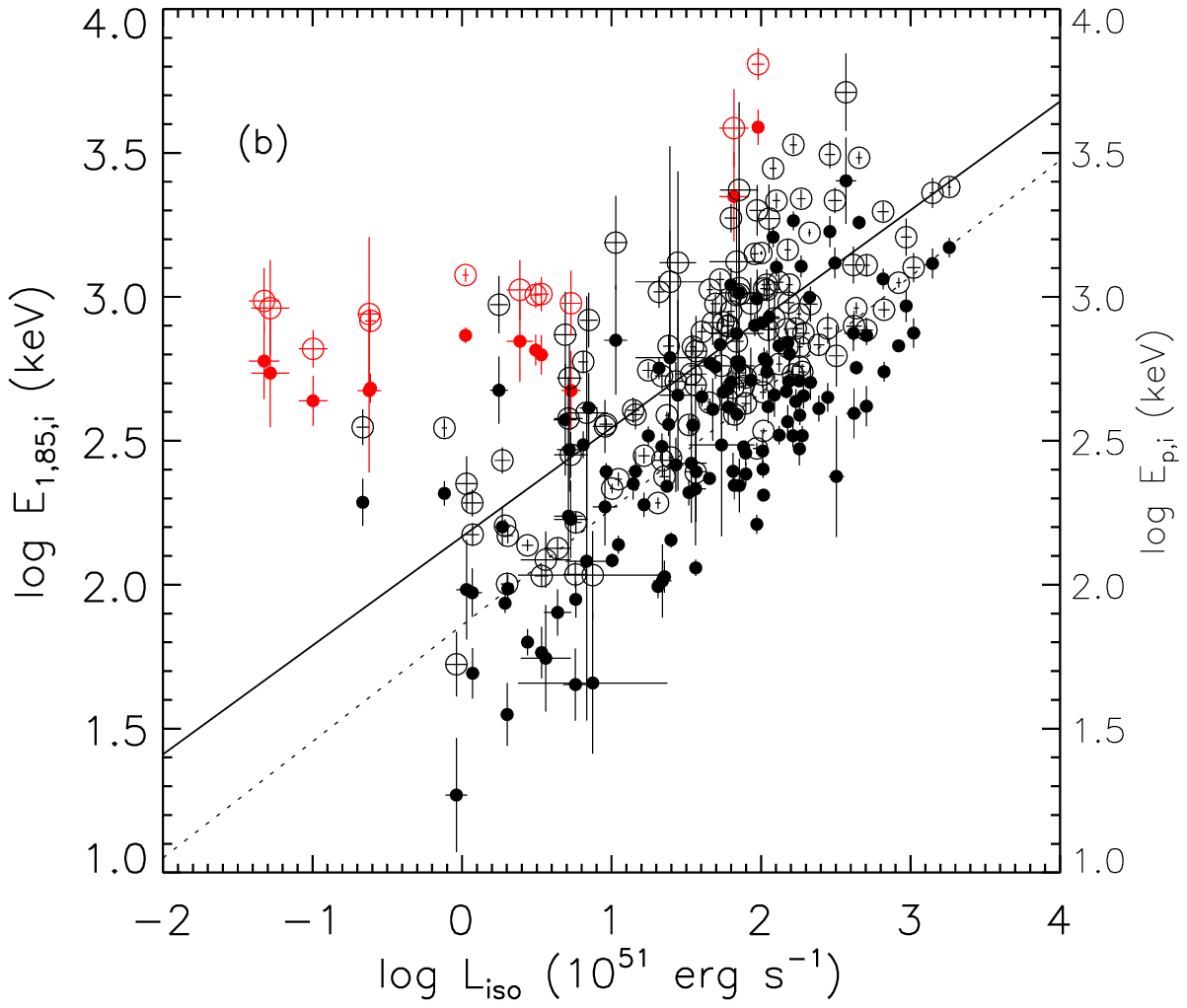}
  \caption{$E_{1,85,i}$ vs. $E_{iso}$ for the F spectra (a) and P spectra (b) and $E_{1,85,i}$ vs. $L_{iso}$ for the F spectra (c) and P spectra (d), along with the Amati and Yonetoku relations for the F and P spectra. All the symbols are same as Figure 7.}
\label{fig:example_figure}
\end{figure*}

\begin{figure*}
  \includegraphics[width=0.48\textwidth]{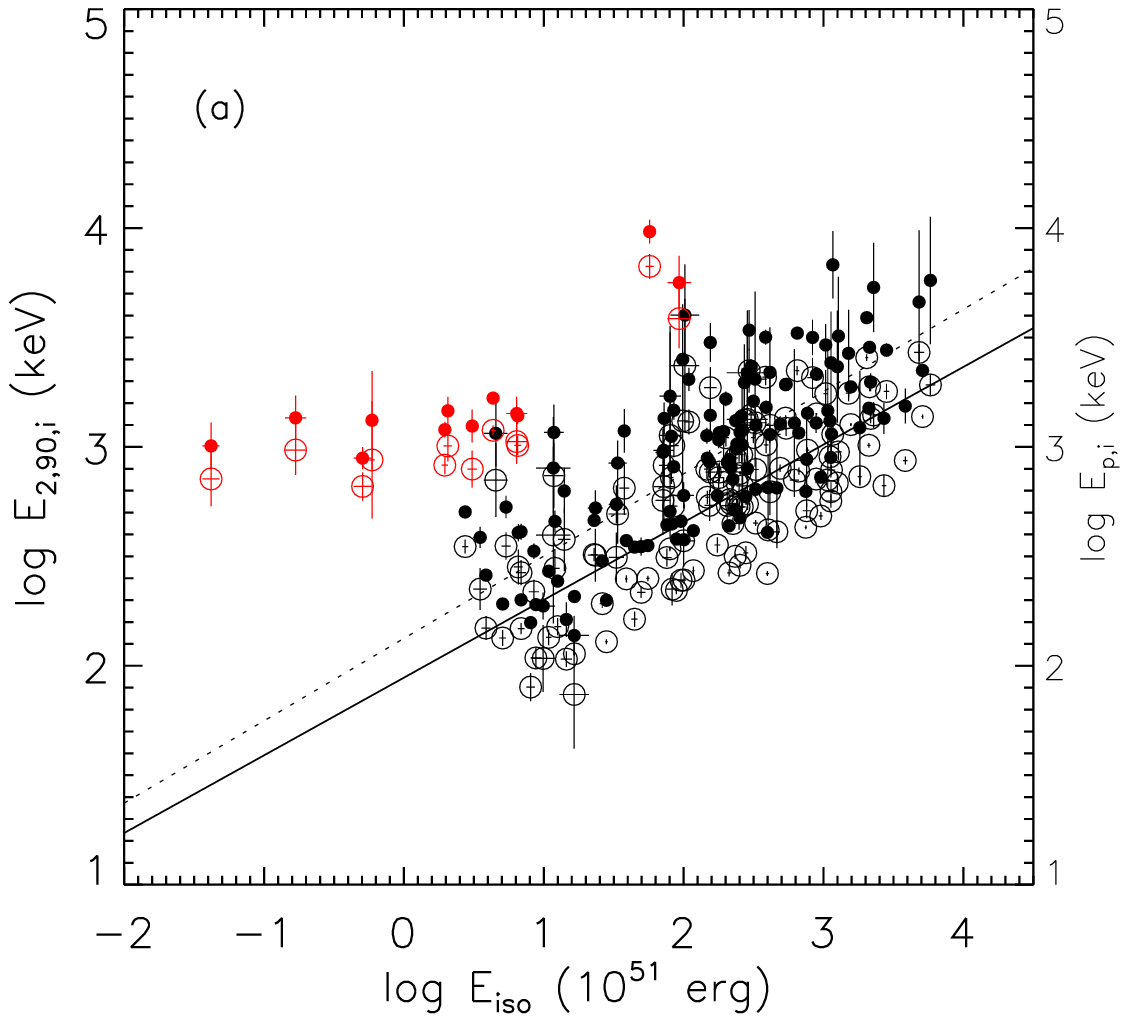}
  \includegraphics[width=0.48\textwidth]{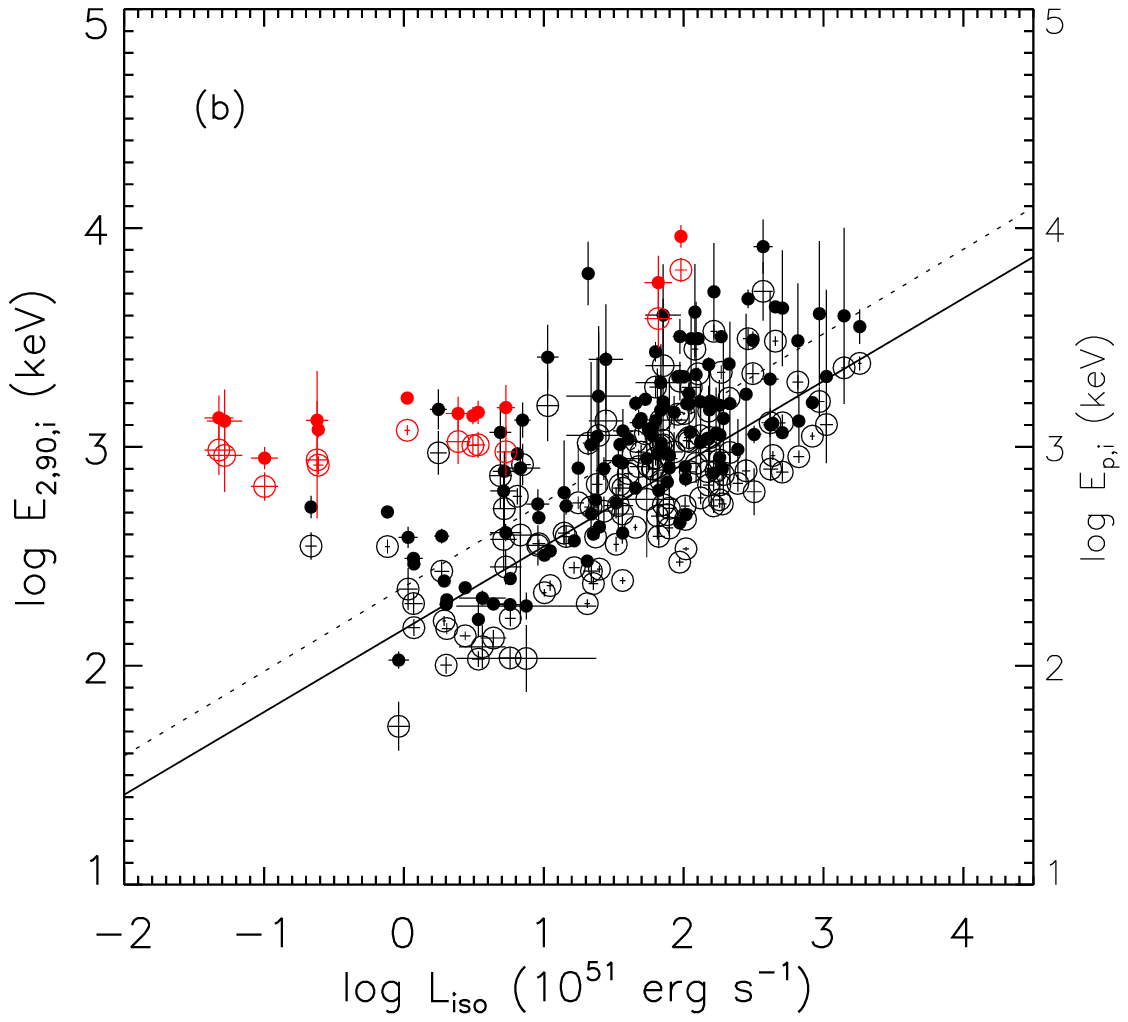}
  \caption{$E_{2,90,i}$ vs. $E_{iso}$ for the F spectra (a) and P spectra (b) and $E_{2,90,i}$ vs. $L_{iso}$ for the F spectra (c) and P spectra (d), along with the Amati and Yonetoku relations for the F and P spectra. All the symbols are same as Figure 7.}
\label{fig:example_figure}
\end{figure*}

\begin{figure*}
  \includegraphics[width=0.48\textwidth]{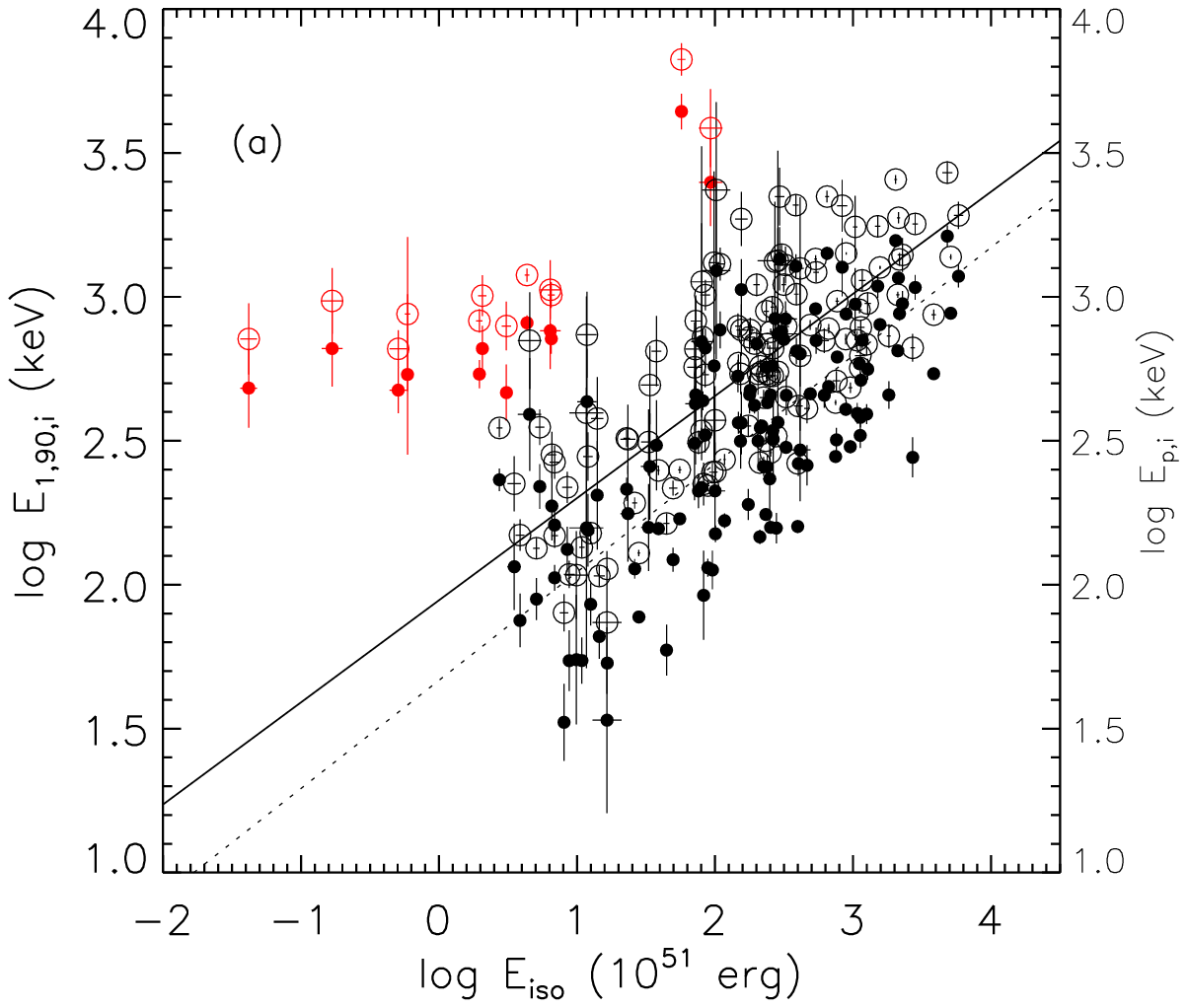}
  \includegraphics[width=0.48\textwidth]{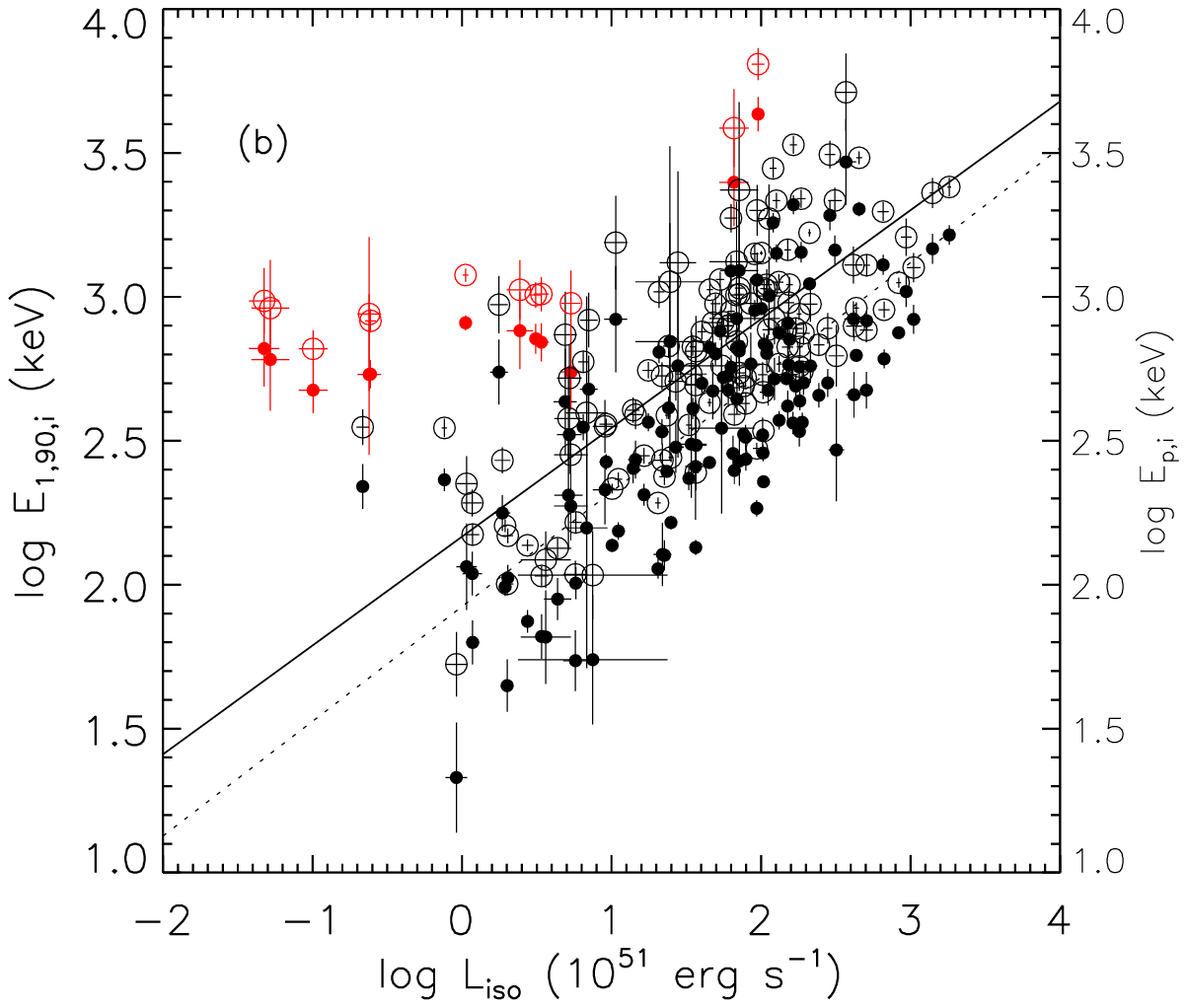}
  \caption{$E_{1,90,i}$ vs. $E_{iso}$ for the F spectra (a) and P spectra (b) and $E_{1,90,i}$ vs. $L_{iso}$ for the F spectra (c) and P spectra (d), along with the Amati and Yonetoku relations for the F and P spectra. All the symbols are same as Figure 7.}
\label{fig:example_figure}
\end{figure*}

\begin{figure*}
  \includegraphics[width=0.48\textwidth]{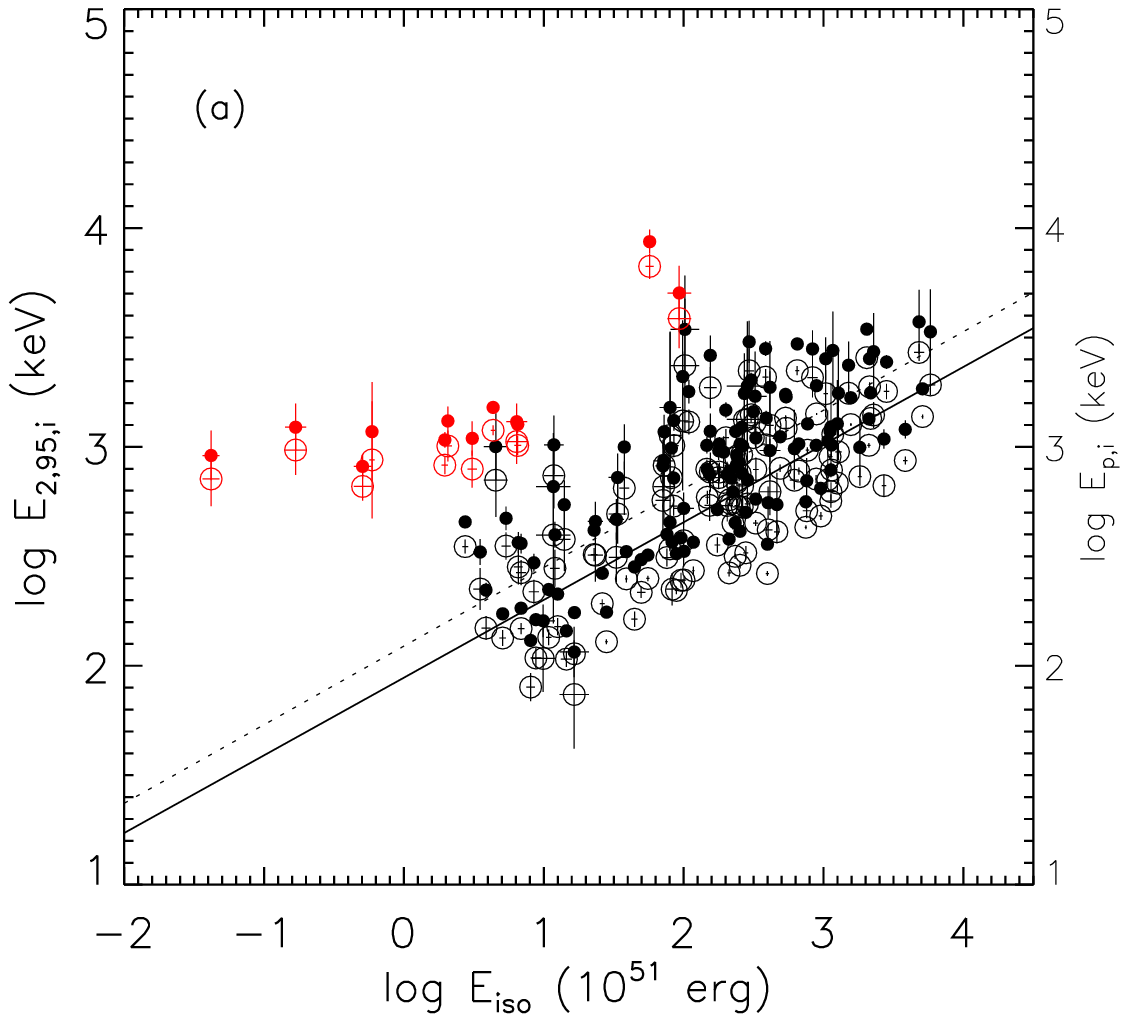}
  \includegraphics[width=0.48\textwidth]{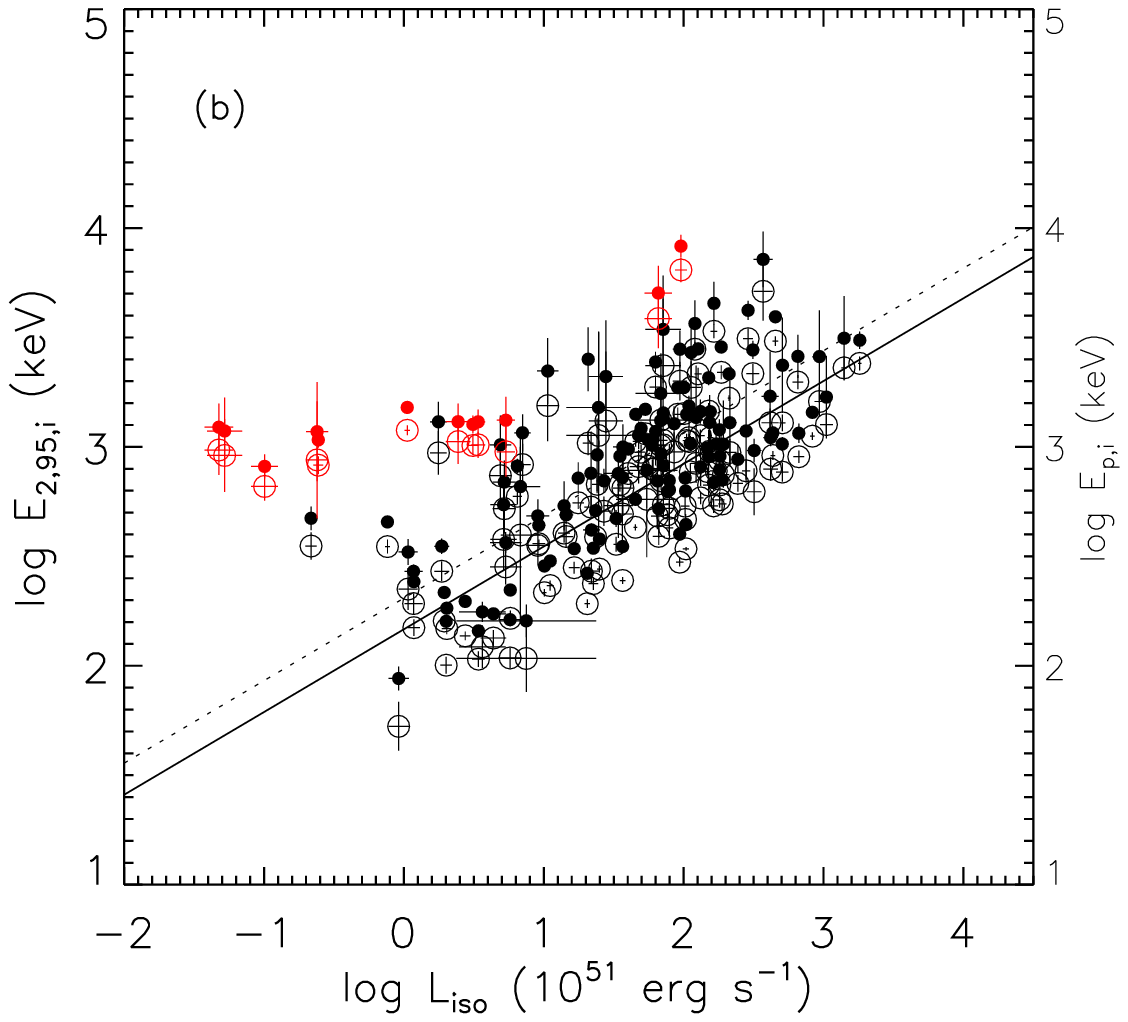}
 \caption{$E_{1,85,i}$ vs. $E_{iso}$ for the F spectra (a) and P spectra (b) and $E_{1,85,i}$ vs. $L_{iso}$ for the F spectra (c) and P spectra (d), along with the Amati and Yonetoku relations for the F and P spectra. All the symbols are same as Figure 7.}
\label{fig:example_figure}
\end{figure*}

\begin{figure*}
  \includegraphics[width=0.48\textwidth]{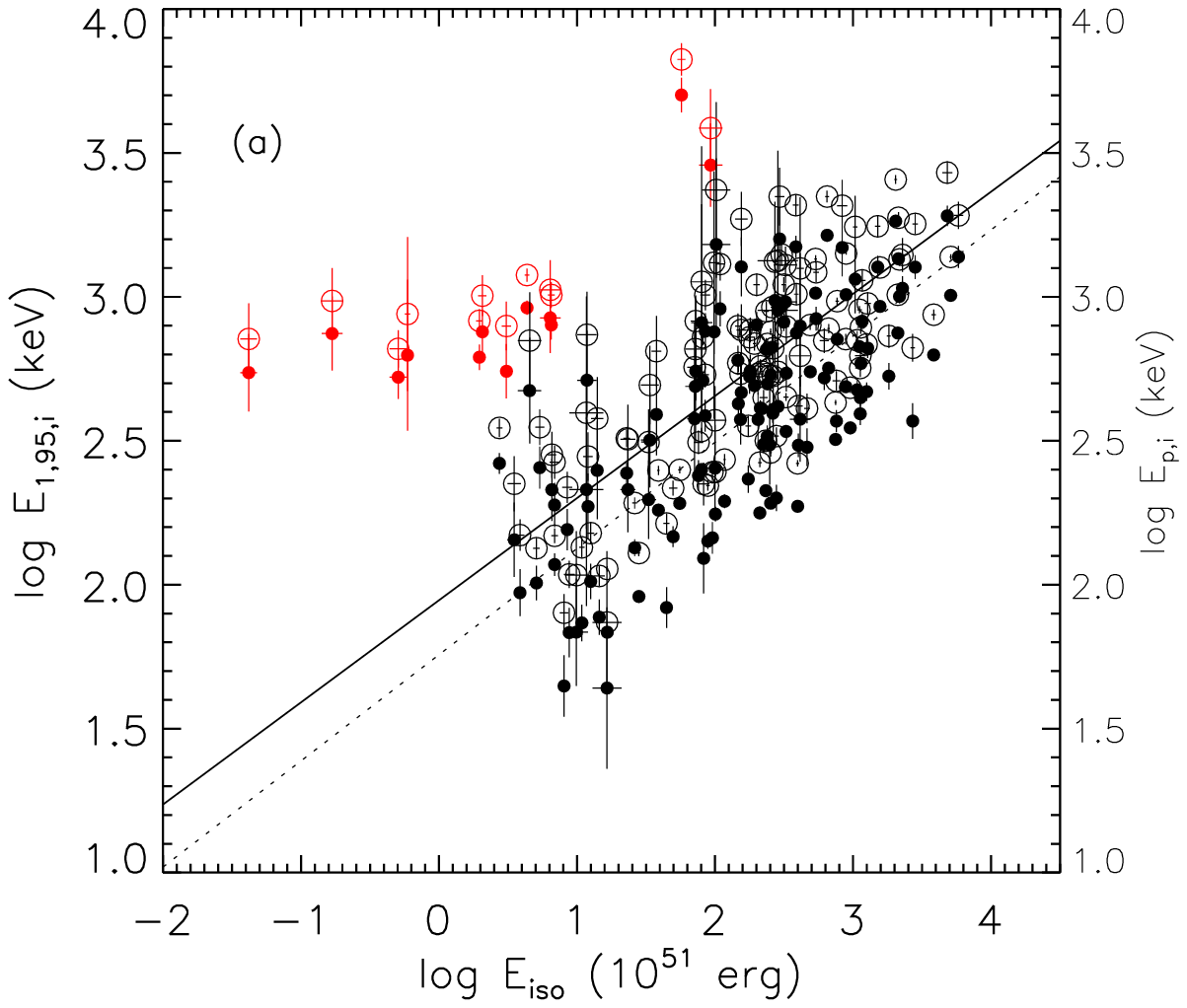}
  \includegraphics[width=0.48\textwidth]{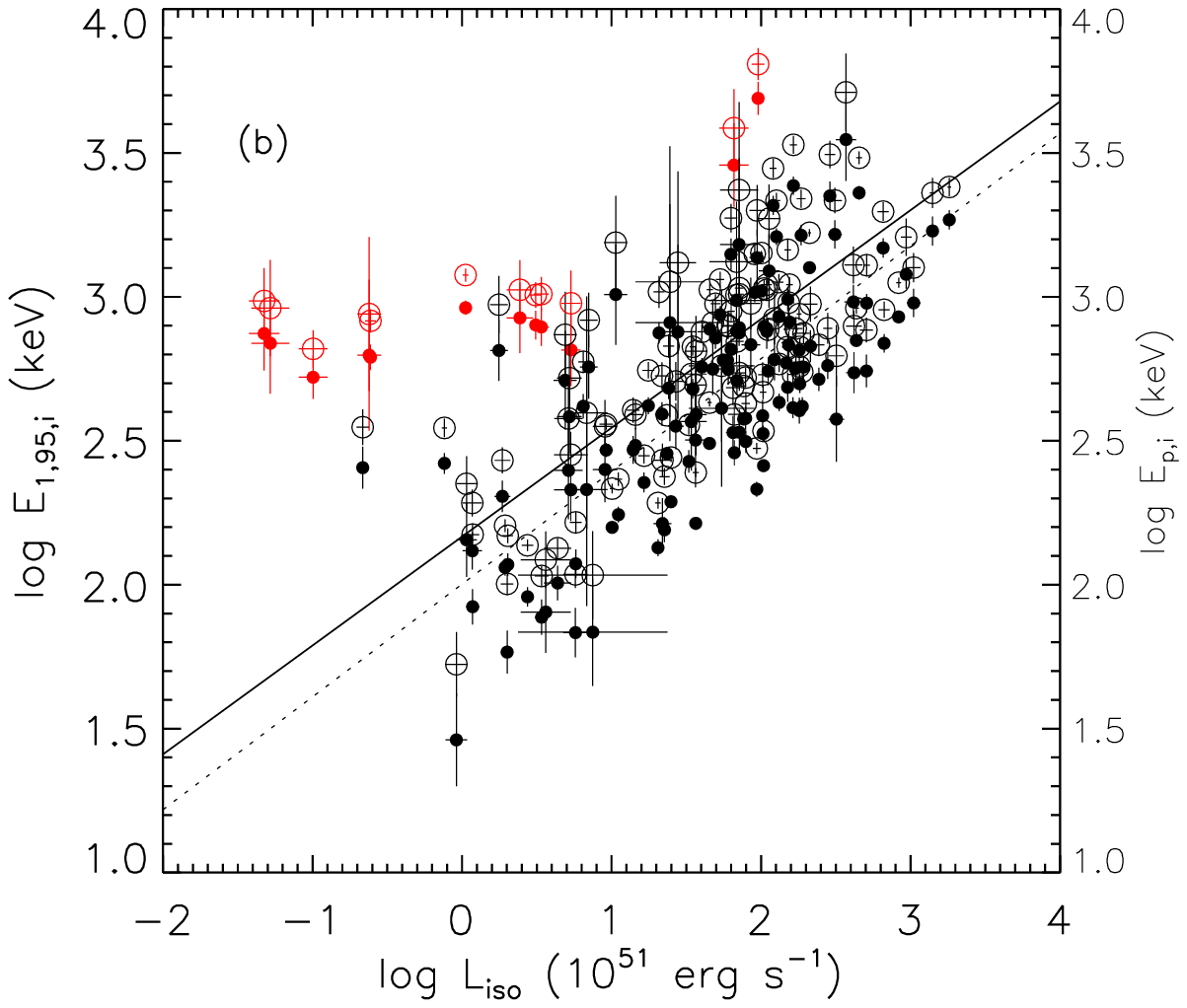}
  \caption{$E_{1,95,i}$ vs. $E_{iso}$ for the F spectra (a) and P spectra (b) and $E_{1,95,i}$ vs. $L_{iso}$ for the F spectra (c) and P spectra (d), along with the Amati and Yonetoku relations for the F and P spectra. All the symbols are same as Figure 7.}
\label{fig:example_figure}
\end{figure*}

\begin{figure*}
  \includegraphics[width=0.48\textwidth]{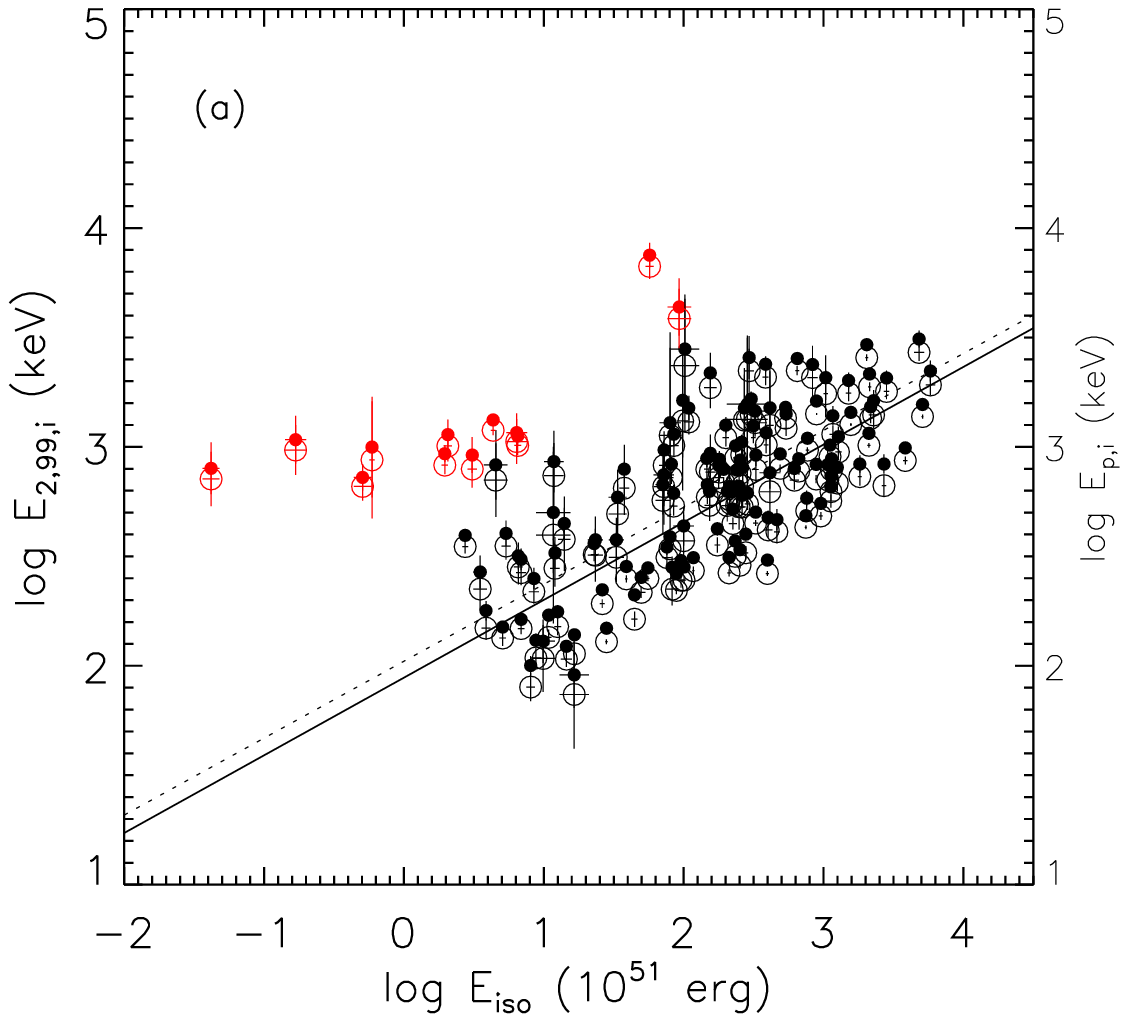}
  \includegraphics[width=0.48\textwidth]{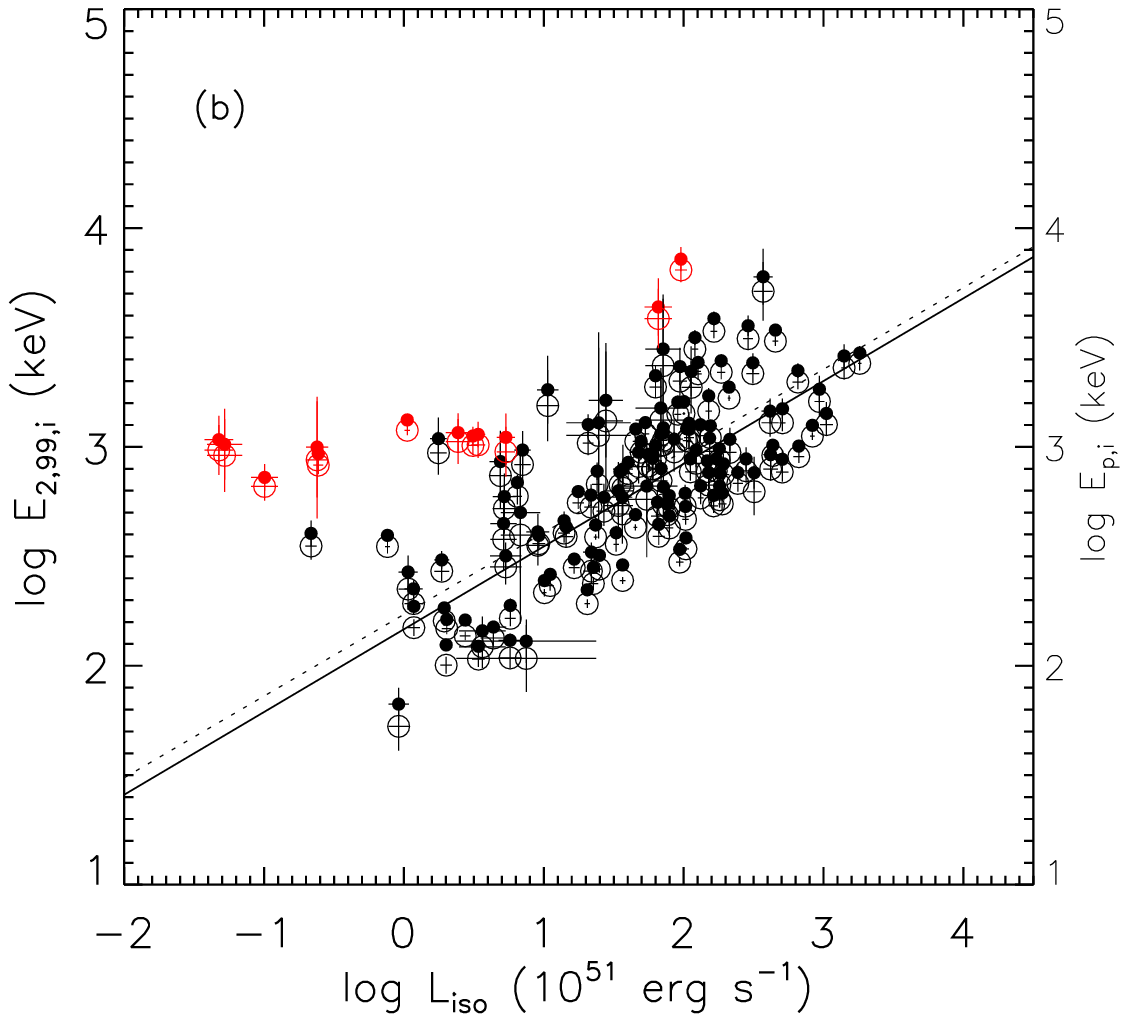}
  \caption{$E_{1,99,i}$ vs. $E_{iso}$ for the F spectra (a) and P spectra (b) and $E_{1,99,i}$ vs. $L_{iso}$ for the F spectra (c) and P spectra (d), along with the Amati and Yonetoku relations for the F and P spectra. All the symbols are same as Figure 7.}
\label{fig:example_figure}
\end{figure*}

\begin{figure*}
  \includegraphics[width=0.48\textwidth]{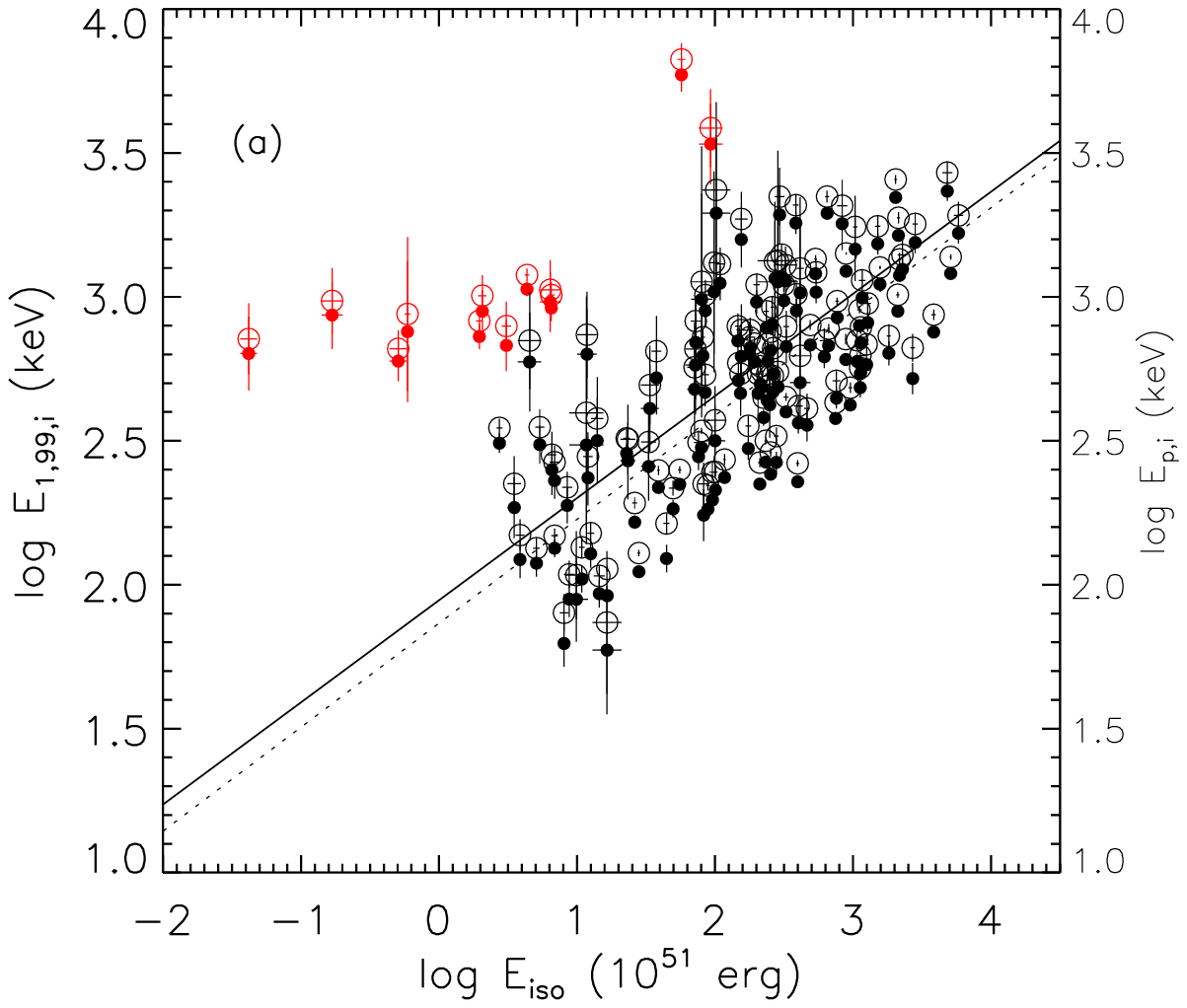}
  \includegraphics[width=0.48\textwidth]{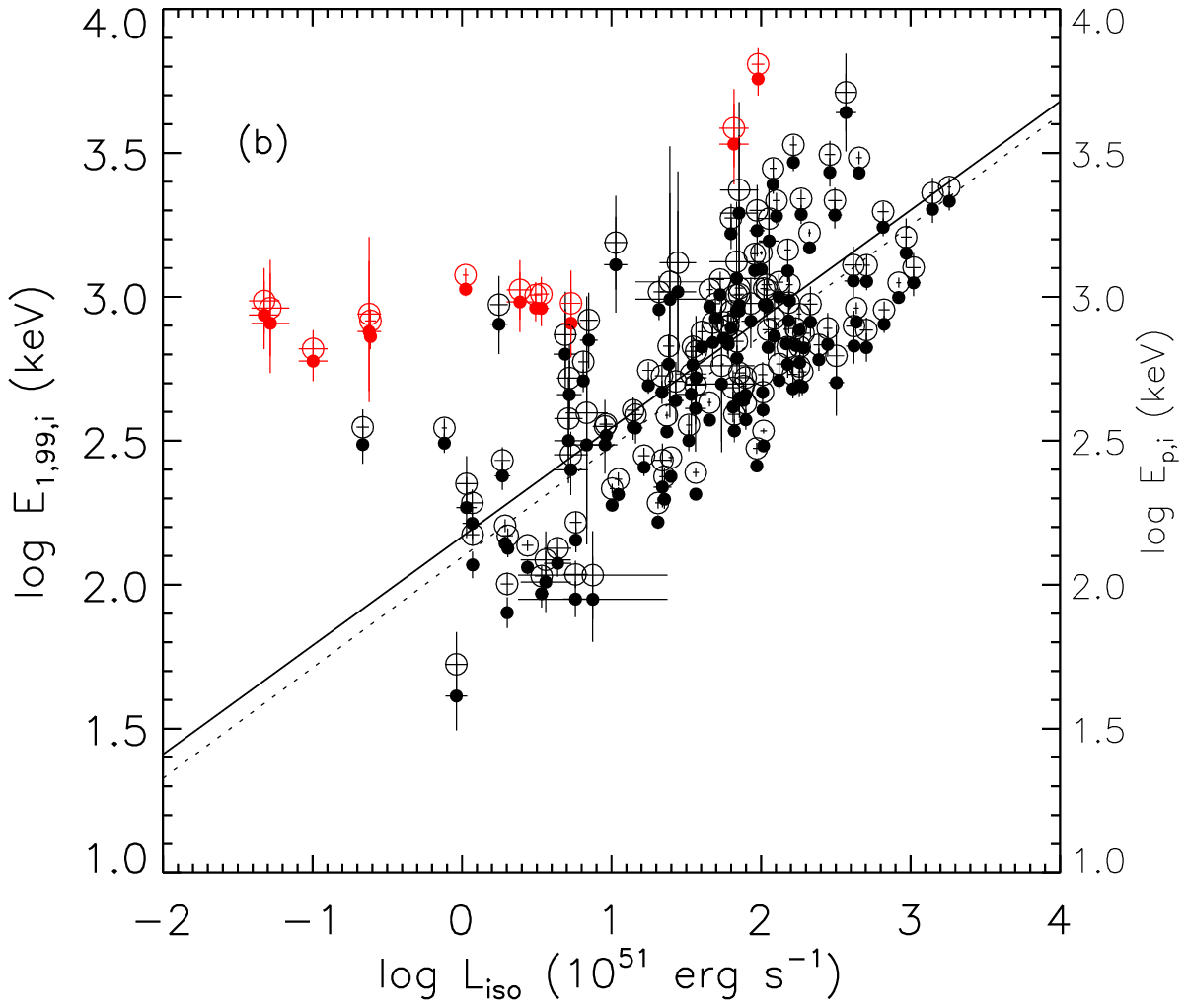}
  \caption{$E_{1,99,i}$ vs. $E_{iso}$ for the F spectra (a) and P spectra (b) and $E_{1,99,i}$ vs. $L_{iso}$ for the F spectra (c) and P spectra (d), along with the Amati and Yonetoku relations for the F and P spectra. All the symbols are same as Figure 7.}
\label{fig:example_figure}
\end{figure*}

\section{DISCUSSION AND CONCLUSIONS}
Spectrum-energy correlations of GRBs are empirical connections between the measurable properties of the prompt gamma-ray emission spectra and the energy or luminosity of GRBs. These correlations are very important because they were proposed to measure the universe and classify GRBs as useful probes (e.g. Dai et al. 2004; Wang et al. 2017, Minaev \& Pozanenko 2020). In this paper, we have analysed in detail the correlations between the absolute spectral width of gamma-ray burst spectra with BEST model parameters from KW and the equivalent isotropic energy ($E_{iso}$) as well as  the isotropic peak luminosity ($L_{iso}$) for time-integrated F and peak flux P spectra, respectively. Different from Paper I we mainly consider the absolute spectral width in the rest frame and the isotropic-equivalent radiated energy rather than the relative spectral width. The sample includes 141, 145 bursts for the F spectra and P spectra, respectively, in which contains 12 and 12 short burst. The majority of BEST spectral data are fitted by Compton model (93/141, 101/145 for the F and P spectra, respectively). The sample size with known-redshift we used in this paper is much larger than that of Paper I (86 and 75 for the F and P spectra). Using the best estimate of the observed spectral parameters of GRBs we compute six different absolute spectral widths for the F and P spectra based on the $EF_{E}$ spectra peak at 0.5, 0.75, 0.85, 0.90, 0.95, and 0.99 maximum to investigate the relationships of the rest-frame width-$E_{iso}$ and width-$L_{iso}$. Of course, the six different widths we defined are arbitrary. Our purpose is to check if the different spectral width are also correlated with $E_{iso}$ and $L_{iso}$ and compare with the well-known Amati and Yonetoku relation.

We first consider the popular rest-frame half maximum width $W_{ab,50,i}$ for the F spectra and P spectra. The $W_{ab,50,i}$ is strongly correlated with $E_{iso}$ as well as $L_{iso}$ for both F spectra and P spectra. Similar to the Amati relation relation it is found that the short bursts are also outliers of the long ones for $W_{ab,50,i}-E_{iso}$ relation (see, Table 3 and Figure 7). While for $W_{ab,50,i}-L_{iso}$ relation a few short bursts deviate from the long bursts even if most of the short bursts are consistent with long bursts. This property seems to shows that the long bursts have a different origin from the short bursts. Recently, Zhang et al. (2018) found that the correlations for short GRBs also established and they had the consistent power-law indices with long GRBs. Minaev \& Pozanenko (2020) further confirmed the correlation with a much larger short burst sample. Due to the huge difference of the number of the long and short burst we only check the correlations by comparing the $W_{ab,i}$ and $E_{iso}$ as well as $L_{iso}$ with the Amati relation and Yonetoku relation for the case of short bursts. The correlation properties are demonstrated in Table 8. It is surprise that there are almost no difference for the correlated properties, the slopes, and the scatters except some slight different intercepts for all of the spectral widths. Take the case of $W_{ab,50,i}$ for example, the relationships is shown in Figure 25 and the slopes of $W_{ab,50}-E_{iso}$ and $E_{p,i}-E_{iso}$ are 0.278$\pm0.069$ and 0.279$\pm0.065$, respectively. While the slopes of $W_{ab,50,i}-L_{iso}$ and $E_{p,i}-L_{iso}$ are 0.246$\pm0.052$ and 0.240$\pm0.048$. These show that the slopes of width-$E_{iso}$ and width-$L_{iso}$ are consistent with that of the Amati relation and the Yonetoku relation, respectively. Therefore, we deduce that the short burst slopes of $W_{ab,i}-E_{iso}$ and $W_{ab,i}-L_{iso}$ are also consistent with that of the long bursts since the short burst and the long burst of Amati relation and Yonetoku relation have consistent slope (Zhang et al. 2018). Of course, The validity of them needs further identify with a much larger short burst sample.

\begin{figure*}
  \includegraphics[width=0.45\textwidth]{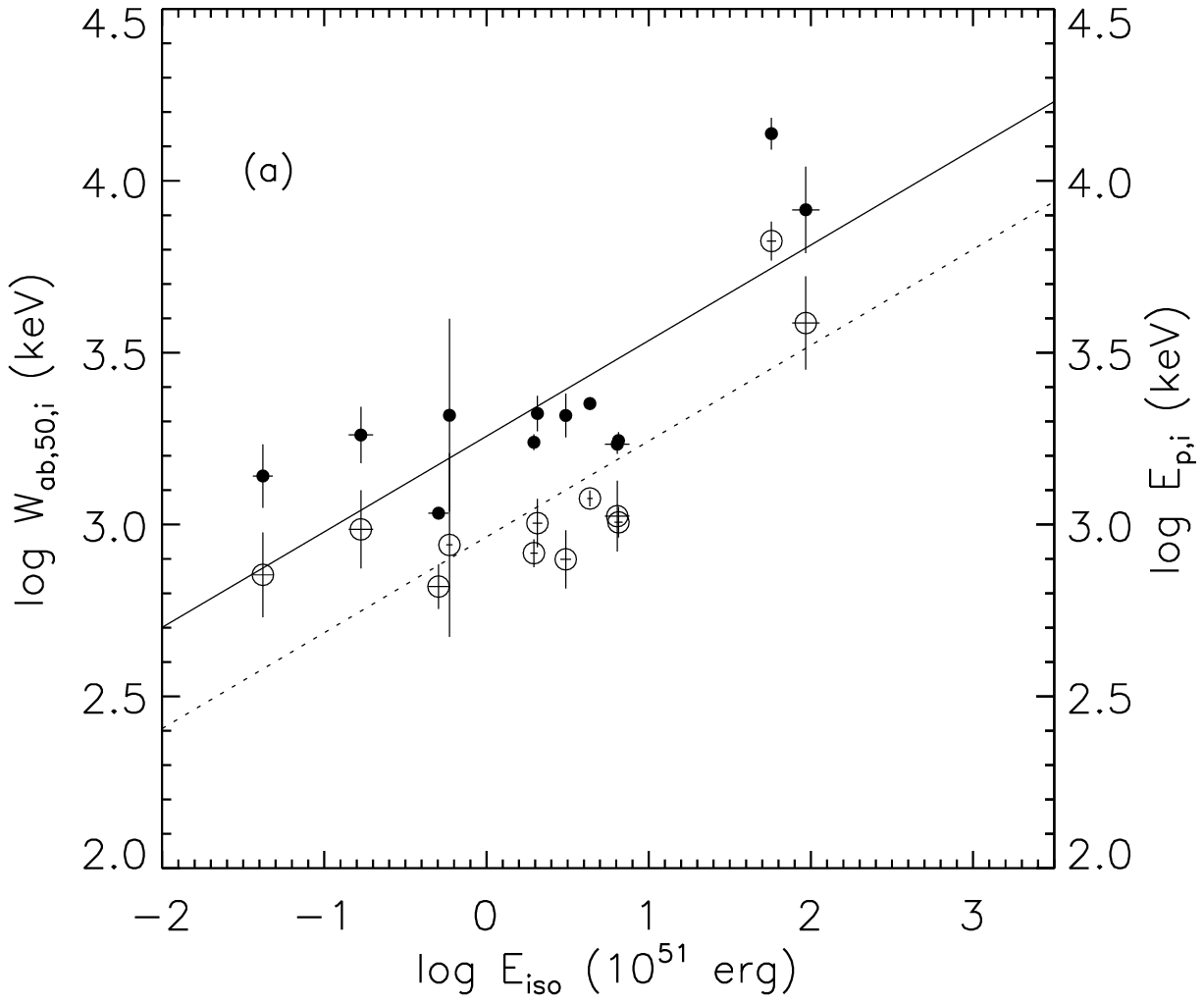}
  \includegraphics[width=0.45\textwidth]{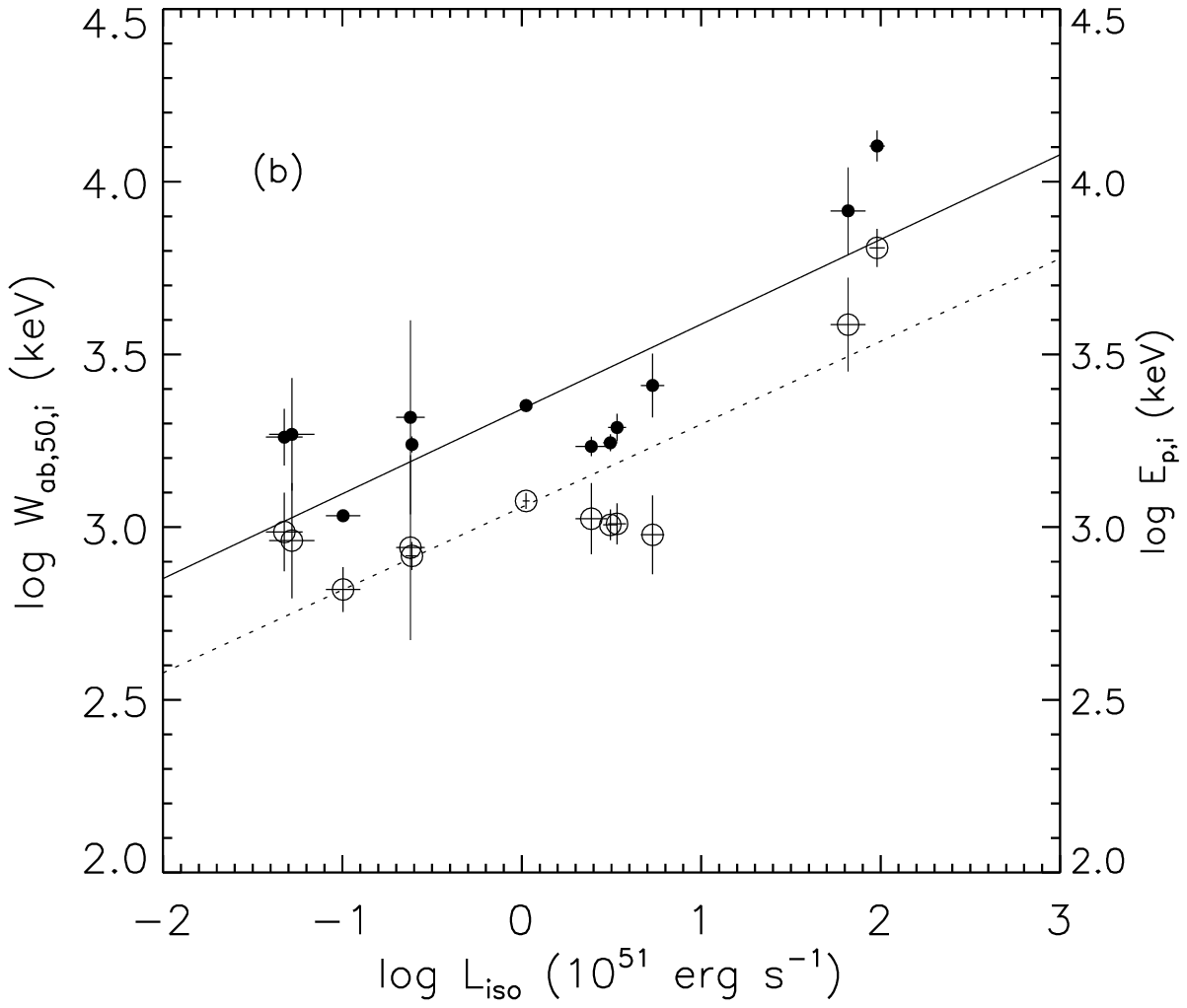}
  \caption{A plots of $W_{ab,50,i}$-$E_{iso}$ (a), $W_{ab,50,i}$-$L_{iso}$ (b) and $E_{p,i}$-$E_{iso}$ and $E_{p,i}$-$L_{iso}$ for the short burst sample. The black filled circles and black open circles represent the relations of $W_{ab,50,i}-E_{iso}$,  $W_{ab,50,i}-L_{iso}$, and $E_{p,i}-E_{iso}$, $E_{p,i}-L_{iso}$, respectively. The solid line are the best-fitting lines obtained by fitting $W_{ab,50,i}-E_{iso}$, $W_{ab,50,i}-L_{iso}$, while the dashed lines are the fitting lines of $E_{p,i}-E_{iso}$, $E_{p,i}-L_{iso}$, respectively.}
\label{fig:example_figure}
\end{figure*}

\startlongtable
\begin{deluxetable}{cccccccc}
\tablecaption{Correlation analysis results of six absolute spectral widths in the rest frame and $E_{iso}$ as well as $L_{iso}$ for the short GRBs.}
\tablehead{\colhead{Correlation}  & \colhead{number}  &\colhead{$\rho$} &\colhead{$P$} &\colhead{a} &\colhead{b} &\colhead{$\sigma$}}
\startdata
\hline
$W_{ab,50,i,F}-E_{iso}$   &   12  & 0.60  & 3.86 $\times 10^{-2}$     & 0.28$\pm$0.07   & 3.26 $\pm$0.07  & 0.22$\pm$ 0.04\\
$W_{ab,75,i,F}-E_{iso}$   &   12  & 0.60  & 3.86 $\times 10^{-2}$     & 0.28$\pm$0.07   & 3.06 $\pm$0.07  & 0.21$\pm$0.04 \\
$W_{ab,85,i,F}-E_{iso}$   &   12  & 0.60  & 3.86 $\times 10^{-2}$     & 0.28$\pm$0.07   & 2.93 $\pm$0.07  & 0.21$\pm$0.04\\
$W_{ab,90,i,F}-E_{iso}$   &   12  & 0.60  & 3.86 $\times 10^{-2}$     & 0.28$\pm$0.07   & 2.84 $\pm$0.07  & 0.21$\pm$0.04\\
$W_{ab,95,i,F}-E_{iso}$   &   12  & 0.60  & 3.86 $\times 10^{-2}$     & 0.28$\pm$0.07   & 2.68 $\pm$0.07  & 0.21$\pm$0.04\\
$W_{ab,99,i,F}-E_{iso}$   &   12  & 0.60  & 3.86 $\times 10^{-2}$     & 0.28$\pm$0.07   & 2.33 $\pm$0.07  & 0.21$\pm$0.04\\
$E_{p,i,F}-E_{iso}$       &   12  & 0.83  & 9.51 $\times 10^{-4}$     & 0.28$\pm$0.07   & 2.96 $\pm$0.07  & 0.19$\pm$0.04\\
$W_{ab,50,i,P}-L_{iso}$   &   12  & 0.62  & 3.32 $\times 10^{-2}$     & 0.25$\pm$0.05    & 3.34 $\pm$0.05  & 0.18$\pm$0.04\\
$W_{ab,75,i,P}-L_{iso}$   &   12  & 0.62  & 3.32 $\times 10^{-2}$     & 0.25$\pm$0.05    & 3.14 $\pm$0.05  & 0.18$\pm$0.04 \\
$W_{ab,85,i,P}-L_{iso}$   &   12  & 0.62  & 3.32 $\times 10^{-2}$     & 0.25$\pm$0.05    & 3.02 $\pm$0.05  & 0.18$\pm$0.05\\
$W_{ab,90,i,P}-L_{iso}$   &   12  & 0.62  & 3.32 $\times 10^{-2}$     & 0.25$\pm$0.05    & 2.92 $\pm$0.05  & 0.18$\pm$0.04\\
$W_{ab,95,i,P}-L_{iso}$   &   12  & 0.62  & 3.32 $\times 10^{-2}$     & 0.25$\pm$0.05    & 2.77 $\pm$0.05  & 0.18$\pm$0.04\\
$W_{ab,99,i,P}-L_{iso}$   &   12  & 0.62  & 3.32 $\times 10^{-2}$     & 0.25$\pm$0.05    & 2.41 $\pm$0.05  & 0.18$\pm$0.04\\
  $E_{p,i,P}-L_{iso}$     &   12  & 0.69  & 1.39 $\times 10^{-2}$     & 0.24$\pm$0.05    & 3.06 $\pm$0.05  & 0.16$\pm$0.04\\
\enddata
\tablecomments{The F and P correspond to the F spectra and P spectra, respectively.}
\end{deluxetable}

We then consider if the correlations of $width-E_{iso}$ and $width-L_{iso}$ for other narrower widths defined above, such as $W_{ab,75}$, $W_{ab,85}$, $W_{ab,90}$, $W_{ab,95}$, $W_{ab,99}$ exist and then compare the $width-E_{iso}$ and $width-L_{iso}$ with the Amati and Yonetoku relation. We find all of the correlations of $width-E_{iso}$ and $width-L_{iso}$ establish regardless of the F spectra and P spectra. In addition, the variabilities of the correlation the $width-E_{iso}$ and $width-L_{iso}$ for the F spectra much larger than those of the P spectra. All of the scatters of the $width-E_{iso}$ and $width-L_{iso}$ decrease with the decrease of the width regardless of the F and P spectra. The more stable correlations of width-$E_{iso}$ and width-$L_{iso}$ for the P spectra seem to reveal that the correlations of peak flux spectra can better represent the connections between width and $E_{iso}$ as well as $L_{iso}$. The F spectra are the average spectra and the GRB spectra evolute with time (e.g. Crider, et al. 1997, Kaneko et al. 2006; Peng,et al. 2009). Hence the peak flux spectra can better describe the observed spectra.

In fact the spectral width may be a good physics quantity to describe the spectral shape. The observed spectral width was used to compare with widths of spectra from fundamental emission processes (Axelsson \& Borgonovo 2015, Vurm \& Beloborodov 2016). As Paper I and Axelsson \& Borgonovo (2015) pointed out that most of the observed spectra are much narrower than the emission from an electron distribution and are also significantly wider than a blackbody spectrum. The discrepancy between the observed spectral width and the theory width can provide some evidences for the GRB radiation mechanism. Recently Bharali et al. (2017) put forward that if the system undergoes a rapid temperature evolution the observed spectral shape can be broadened. Particularly, if invoking thermal radiation a way must be found to broaden the spectrum (Vurm \& Beloborodov 2016). Beloborodov (2013) pointed out that several factors inevitably broaden the synchrotron peak in a more realistic model. Moreover, they thought observed spectra is hardly consistent with the synchrotron model, even if uncertainties in the observed spectrum due to the detector response and limited photon statistics are taken into account. Therefore, they thought that the narrow MeV spectral peak provided strong evidence for thermalization of radiation at early, opaque stages of the GRB explosion.

In fact, there are similar width-energy and width-luminosity relations in astrophysics field. Many studies have shown that the temporal profile also connect with the luminosity. For example, the well-known lag-luminosity found by Norris et al. (2000) as well as the variability-luminosity by Reichart et al. (2001). Norris et al. (2005) found the pulse width is strongly correlated with the pulse lag. Hakkila et al. (2008) derived a new pulse width in the rest frame versus peak luminosity correlation in GRB pulses, which appear to be nature from lag-luminosity and width-lag relation. Hashimoto et al. (2020) confirmed a positive time-integrated luminosity-duration relation of non-repeating fast radio bursts but repeating fast radio bursts do not indicate clear correlation between luminosity-duration. These results reveal different physical origins of the two types of fast radio bursts. Tu \& Wang (2019) also found a correlation between isotropic energy and intrinsic duration for the first time with a Swift GRB sample. In addition, the duration-energy correlation also established for solar flares and stellar superflares. It is very striking that the dependences of duration on the isotropic energy are very similar since they share similar power-law index of duration on energy for the GRB, solar flares and stellar superflares. Furthermore, many studies suggested there is dependence of the temporal profiles of GRBs on energy (e.g. Fenimore et al. 1995; Norris et al. 2005; Peng et al. 2006). The connection between temporal profiles on energy seems to support the width-$E_{iso}$ and width-$L_{iso}$ relations.

When we compare the relations of the six width-$E_{iso}$ and width-$L_{iso}$ with the well-known Amati and Yonetoku relation most of the correlations of width-$E_{iso}$ and width-$L_{iso}$ are much tighter than those of the Amati and Yonetoku relations for the F spectra. Whereas for the P spectra their correlations are very consistent. Moreover, width-$E_{iso}$ and width-$L_{iso}$ relations would approach the Amati and Yonetoku relation as the width decreases. The $W_{ab,90,i}-E_{iso}$ and $W_{ab,90,i}-L_{iso}$ relations almost overlap with the Amati and Yonetoku relations and their correlation properties, the slopes, and the scatters are also well consistent. When the widths are less than $W_{ab,90,i}$ the correlations of $width-E_{iso}$ and $width-L_{iso}$ deviate from the Amati and Yonetoku relations. As a consequent, the Amati and Yonetoku relations appear to be the special cases of width-$E_{iso}$ and width-$L_{iso}$ relations.

We further investigate the correlations between the upper ($E_{2}$) and lower energy ($E_{1}$) bounds of the $EF_{E}$ spectrum from six different widths with $E_{iso}$ as well as $L_{iso}$. It is found that all of the $E_{2}$ and $E_{1}$ are also correlated with $E_{iso}$ as well as $L_{iso}$. The $E_{2}$-$E_{iso}$, $E_{1}$-$E_{iso}$ relations and $E_{2}$-$L_{iso}$, $E_{1}$-$L_{iso}$ relations are close to the Amati relation and Yonetoku relation when $E_{2}$ and $E_{1}$ approach the peak energy $E_{p}$. Once more the Amati and Yonetoku relation seem to be the special cases of corresponding $E_{2}-E_{iso}$, $E_{2}-L_{iso}$ or $E_{2}-E_{iso}$, $E_{2}-L_{iso}$ correlations. Moreover, the the Amati and Yonetoku relation are not the best relationships by comparing the correlation parameters with the almost same scatters.

Both $E_{iso}$ and $L_{iso}$ are correlated with all of spectral widths as well as all of the location energy ($E_{2}$ and $E_{1}$), which further suggest that the $E_{iso}$ and $L_{iso}$ are related to the shape of GRB spectra. At present we can not find out the reasons of causing the correlated relationships. These issues motivate us suspect the $E_{iso}$ and $L_{iso}$ may be also related to the area of the $EF_{E}$ spectrum or the spectrum centroid energy, which deserve the further investigations.

Paper I considered the correlations between the spectral widths and the photon indices $\alpha$ and $\beta$, and $T_{90}$. We also inspect these relations with KW sample for the popular half width $W_{ab,50}$. Because the sample consists of BAND and COMP the absolute spectral widths also consist of the two models. For the Band model, the widths are strongly correlated with the high-energy index $\beta$ ($\rho$= 0.82, p $< 10^{-4}$ for the F spectra and $\rho$ = 0.87, $p < 10^{-4}$  for the P spectra) and less correlated with $\alpha-\beta$ ($\rho$ = -0.57, $p < 10^{-4}$ for the F spectra and $\rho$ = -0.68, $p < 10^{-4}$ for the P spectra). However, there are no correlation between the spectral width and the lower energy index $\alpha$  ($\rho$ = -0.02, p = 0.93 for the F spectra and $\rho$ = -0.06, p = 0.71 for the P spectra). Likewise, we do not find that there have any correlations between the spectral width and $\alpha$ ($\rho$=0.04, p=0.65 for the F spectra and $\rho$=0.04, p=0.68 for the P spectra).

We also check the relationships between the absolute spectral width ($W_{ab,50}$) and the duration ($T_{90}$). We first examine the relationships for the case of the F spectra and the P spectra in the observed frame. The Spearman rank-order correlation coefficients and the P values ($\rho$=0.24, $p =4.35\times10^{-3}$ for the F spectra and $\rho$=0.23, $p =6.69\times10^{-3}$ for the P spectra) show that there are weak correlations between them. While for the intrinsic case the corresponding correlation coefficients and the P values ($\rho$=0.21, $p =1.25\times10^{-2}$ for the F spectra and $\rho$=0.22, $p =8.50\times10^{-3}$ for the P spectra) also show the correlations between $W_{ab,50,i}$ and $T_{90,i}$ are not strong for both the F spectra and the P spectra.

In conclusion our analysis further confirm the tight width-energy and width-luminosity relations in GRBs and the relations are also important correlation relationships to understand the physics of GRB. The Amati and Yonetoku relations are not necessarily the best correlated relationships to relate the location energy to isotropic energy as well as isotropic peak luminosity. These results may provide some help to understand the origins of the Amati and Yonetoku relations.

\acknowledgments
We would like to thank the anonymous referee for constructive suggestions to improve the manuscript. This work was supported by the National Natural Science Foundation of China (grant 11763009, 11263006), the Natural Science Fund of the Education Department of Guizhou Province(KY2015455), the Natural Science Fund of the Liupanshui Normal College (LPSSY201401).

\end{sloppypar}
\bibliography{reference}

\begin{thebibliography}{}
\bibitem[Ahlgren(1989)]{Ahlgren2012}Ahlgren, B., Larsson, J., Valan, V., et al. \ 2019, \apj, 880, 76
\bibitem[Ahlgren(1989)]{Ahlgren2012}Ahlgren, B., Larsson, J., Ahlberg, E., et al. \ 2019, \mnras, 485, 474
\bibitem[Amati(2002)]{Amati2002}Amati, L., Frontera1, F., Tavani, M., et al. \ 2002,\aap, 390, 81
\bibitem[Axelsson(2015)]{Axelsson2015}Axelsson, M., \&  Borgonovo, L., \ 2015,\mnras, 447, 3150
\bibitem[Band(1993)]{Band1993}Band, D. L., Matteson, J., Ford, L., et al. \ 1993, \apj, 413, 281
\bibitem[Band(2005)]{Band2005}Band, D. L., \& Preece, R. D., \ 2005, \apj, 627, 319
\bibitem[Beloborodov(2013)]{Beloborodov213}Beloborodov, A., \ 2013, \apj, 764, 157
\bibitem[Burgess(2017)]{Burgess2020}Burgess, J., B\'{e}gu\'{e}, D.,  Greiner, J., et al. \ 2020, \ NatAs, 4, 174
\bibitem[Crider(1997)]{Crider1997}Crider, A., Liang, E. P., Smith, I. A., et al. \ 1997, \apj, 479, L39
\bibitem[Dai(2004)]{Dai2004}Dai, Z. G., Liang, E. W., \& Xu, D. \ 2004, \apj, 612, L101
\bibitem[Eichler(1989)]{Eichler2012}Fenimore, E. E., in't Zand, J. J. M., Norris J. P., Bonnell J. T., Nemiroff R. J., \ 1995, \apj, 448, L101
\bibitem[Ghirlanda(2004)]{Ghirlanda2004}Ghirlanda, G. W.,  Ghisellini, G., Lazzati, D., et al. \ 2004, \apj, 616, 331
\bibitem[Hakkila(1993)]{Horv1998}Hakkila, J., Giblin, W., Norris, J., et al.  \ 2008, \apj, 677, L81
\bibitem[Hashimoto(1993)]{Hashimoto1998}Hashimoto, T., Goto, T., Wang, T., et al.  \ 2020, arXiv:2004.02079
\bibitem[Horv\'{a}th(2010)]{Horv2010}Horv\'{a}th I., Bagoly, Z., Bal\'{a}zs, L. G., et al. \ 2010, \apj, 713, 552
\bibitem[Kaneko(2006)]{Others2013}Kaneko, Y., Preece, R. D., Briggs, M. S., et al. \ 2006, \apjs, 166, 298
\bibitem[Kouveliotou(1993)]{Kouveliotou1993}Kouveliotou C. et al., \ 1993, \apj, 413, L101
\bibitem[Minaev(2020)]{Minaev2020}Minaev, P. Y.; \& Pozanenko, A. S., \ 2020, \mnras, 492, 1919
\bibitem[Norris(2000)]{Norris2000}Norris, J. P., MARANI, G. F., BONNELL, J. T., \ 2000, \apj, 534, 248
\bibitem[Norris(2005)]{Norris2005}Norris, J. P., Bonnell, J. T., Kazanas, D., et al. \ 2005, \apj, 627, 324

\bibitem[Oganesyan(2019)]{Oganesyan2019}Oganesyan, G., Nava, L., Ghirlanda, G., et al. \ 2019, \aap, 628, 59
\bibitem[Peng(2006)]{Peng2006}Peng, Z. Y., Qin, Y. P., Zhang, B. B., et al., \ 2006, \mnras, 368, 1351
\bibitem[Peng(2009)]{Peng2009}Peng, Z. Y., Ma, L., Zhao, X. H., et al.,\ 2009, \apj, 698, 417
\bibitem[Peng(2019)]{Peng2019}Peng, Z. Y., Zhao, X. H., Yin, Y., Wang, D. Z.,\ 2019, \apj, 881, 51 (Paper I)
\bibitem[Qin(1993)]{Qin2013}Qin, Y, \& Chen, Z., \ 2013, \mnras, 430, 163
\bibitem[Robotham(2015)]{Robotham2015}Robotham, A. S. G.; Obreschkow, D.  \ 2015, \pasa, 32, 33
\bibitem[Reichart(2001)]{Reichart2001}Reichart, D. E.; Lamb, D. Q.; Fenimore, E. E., et al. \ 2001, \apj, 552,57
\bibitem[Svinkin(2000)]{Svinkin2000}Svinkin, D., Frederiks, D., Aptekar, R., et al. \ 2016, \apjs, 224, 10
\bibitem[Titarchuk(2012)]{Titarchuk2012}Titarchuk, L., Farinelli, Ruben., Frontera, Filippo., et al. \ 2012, \apj, 752, 116
\bibitem[Tsvetkova(2017)]{Tsvetkova2017}Tsvetkova, A., Frederiks, D., Golenetskii, S., et al. \ 2017, \apj, 850, 161

\bibitem[Tu(2018)]{Tu2018}Tu, Z., \& Wang, F., \ 2018, \apj, 869, L23
\bibitem[Uhm(1993)]{Uhm2014}Vurm, I., \&  Beloborodov, A., \ 2016, \apj, 831, 175
\bibitem[Wang(2017)]{Wang2017}Guo-Jian Wang, G, J., Yu, H., Li, Z. X., Jun-Qing Xia, J. Q., Zhu, Z. H., \ 2017, \apj, 836, 103
\bibitem[Woosley(1993)]{Yonetoku2004}Yonetoku, D., Murakami, T., Nakamura, T., et al. \ 2004, \apj, 609, 935
\bibitem[Zhang(2018)]{Zhang2018}Zhang, Z. B.,  Zhang, C. T., Zhao, Y. X., et al. \ 2018, \pasp, 130, 4202
\end{thebibliography}
\end{document}